%% file: P03a_CO_v4.tex
\documentclass[a4paper,traditabstract,longauth]{aa} 
\usepackage{graphicx}
\usepackage{txfonts}
\usepackage{color}
\usepackage[breaklinks, colorlinks, citecolor=blue]{hyperref}
\usepackage{url}
\usepackage{longtable,lscape}
\usepackage{amsfonts,amssymb}
\usepackage{longtable,lscape}
\usepackage{fixltx2e}   
\usepackage{natbib} 
\newcommand{\typeone}{{\sc{type~1}}}\newcommand{\typetwo}{{\sc{type~2}}}
\newcommand{\typethree}{{\sc{type~3}}}
\newcommand{\joz}{$J$=1$\rightarrow$0}
\newcommand{\jto}{$J$=2$\rightarrow$1}
\newcommand{\jtt}{$J$=3$\rightarrow$2}
\newcommand{\cooz}{CO(1$\rightarrow$0)}
\newcommand{\coto}{CO(2$\rightarrow$1)}
\newcommand{\cott}{CO(3$\rightarrow$2)}

\begin{document}

\include{Planck}

\title{\textit{Planck} 2013 results. XIII. Galactic CO emission}

\input{AuthorList_P03a_CO_authors_and_institutes.tex}


 
   \abstract
  {Rotational transition lines of CO play a major role in molecular
    radio astronomy and in particular in the study of star formation and the
    Galactic structure. Although a wealth of data exists in the Galactic plane
    and some well-known molecular clouds, there is no available CO high sensitivity all-sky
    survey to date.  
  Such all-sky surveys can be constructed using the \Planck\ HFI data because
    the three lowest CO rotational transition lines at 115, 230 and 345~GHz
    significantly contribute to the signal of the 100, 217 and 353~GHz HFI
    channels respectively.  
  Two different component separation methods are used to extract the CO maps
    from \Planck\ HFI data. The maps obtained are then compared to one another
    and to existing external CO surveys. From these quality checks the best CO
    maps in terms of signal to noise and/or residual 
   foreground contamination
    are selected.
  Three sets of velocity-integrated CO emission maps are produced: \typeone\
    maps of the CO $J$=1$\rightarrow$0, $J$=2$\rightarrow$1, and $J$=3$\rightarrow$2 rotational transitions with low
    foreground contamination but moderate signal-to-noise ratio; \typetwo\ maps
    for the $J$=1$\rightarrow$0 and $J$=2$\rightarrow$1 transitions with a better signal-to-noise ratio;
    and one \typethree\ map, a line composite map with the best signal-to-noise
    ratio in order to locate the faintest molecular
    regions. The maps are described in detail. They are shown to be fully
    compatible with previous surveys of parts of the Galactic Plane and also
    of fainter regions out of the Galactic plane.
  The \Planck\ HFI velocity-integrated CO maps for the $J$=1$\rightarrow$0, $J$=2$\rightarrow$1, and
    $J$=3$\rightarrow$2 rotational transitions provide an unprecedented all-sky CO view of
    the Galaxy. These maps are also of great interest to monitor potential CO
    contamination on CMB \Planck\ studies.}

\titlerunning{Galactic CO emission in \Planck}
 \authorrunning{\Planck\ collaboration}

   \maketitle

\section{Introduction}
\label{sec:introduction}

This paper, one of a set associated with the 2013 release of data from the \Planck\footnote{\Planck\ (\url{http://www.esa.int/Planck}) is a project of the European Space Agency (ESA) with instruments provided by two scientific consortia funded by ESA member states (in particular the lead countries France and Italy), with contributions from NASA (USA) and telescope reflectors provided by a collaboration between ESA and a scientific consortium led and funded by Denmark.} mission \citep{planck2013-p01}, describes the
construction and validation of full-sky carbon monoxyde (CO) maps from
\Planck\ data. 

The interstellar medium (ISM) represents about $10-15\%$ of the
  total mass of the Milky Way. The ISM is a mixture of atomic and
  molecular gas, the latter containing around $50$\% of its mass whilst
  filling only a tiny fraction of the volume \citep[see][for a
  general introduction and review]{2001RvMP...73.1031F,2005ARA&A..43..337C}. The cold
  neutral gas is confined close the Galactic disk mid-plane (about $100$~pc scale height), in clouds with varying molecular-to-atomic
  ratios. Molecular clouds, where hydrogen is molecular, are the sites
  of star formation, and as such, play a pivotal role in the
  interstellar matter cycle. Giant molecular clouds (GMCs) are the
  largest self-gravitating structures in spiral galaxies such as the
  Milky Way, and are also the most massive entities, reaching several times
  $10^6$~M$_{\sun}$, placing them at the top of the mass
  spectrum. Molecular clouds were discovered via the
  rotational emission line $J$=1$\rightarrow$0 of carbon monoxide in its
  fundamental electronic and vibrational levels \citep{1970ApJ...161L..43W,
    1972ApJ...174L..43P}. Contrary to the atomic component of the neutral ISM,
  which is directly observable via the spin-flip HI
  $\lambda$21~cm line, the bulk of molecular hydrogen is not directly
  observable in molecular clouds. Because CO is abundant, easily
  excited by collisions with H$_2$, and easily observable from the
  ground, it is considered a good tracer of the molecular component
  of the ISM. It is also a dominant coolant of molecular gas. The ability to detect the CO $J$=1$\rightarrow$0 line from the
  ground allows large surveys to be performed \citep{2001ApJ...547..792D},
  leading to the observational evidence that the molecular gas is
  structured in clouds that are turbulent and harbour the formation of
  all stars in the Galaxy. The astrophysical importance of carbon
  monoxide could hardly be over-appreciated.
 
Large-scale surveys the \joz\ line of $^{12}$CO, but also of $^{13}$CO and
  C$^{18}$O isotopologues, have been carried out with meter-sized
  radio telescopes, mostly through the fundamental rotational
  transitions. The most complete \cooz\ survey is that of
  \cite{2001ApJ...547..792D}, which covers the Milky Way at Galactic latitudes
  $|b|\le 30\deg$, with an effective spatial resolution of 0.5\deg. CO
  and isotopologues were observed with the 4-m NANTEN telescope
  providing spectral maps at slightly higher spatial resolution,
  towards specific GMCs \citep{2004ASPC..317...59M}. In addition,
  there exist a wealth of smaller \cooz\ line surveys, such as for
  example in Orion and Monoceros
  \citep{2005A&A...430..523W}. Observations of CO, for the Orion and Monoceros regions, by
  \cite{1985ApJ...295..402M,1998ApJ...492..205H,2000ApJ...535..167M} have revealed the existence of molecular clouds
  at Galactic latitudes up to 55\deg. However, these high-latitude
  observations provide only a limited view of the $|b|>30\deg$ sky.

  High-resolution full-sky surveys of higher $J$ CO transitions have never been
  carried out, essentially because these are much more time consuming
  than \joz\ observations (atmospheric transmission is poorer, amd
  higher-spatial resolution requires finer spatial sampling). High-$J$
  lines are expected to probe molecular gas with stronger
  excitation conditions (high density and/or warmer) better
  than does the \joz\ line. Combined with \cooz, observations of
  $J\ge2$ lines would therefore provide global constraints on the
  physical conditions in the molecular ISM. Only specific regions have
  been mapped in the \jto\ CO line, e.g., the Galactic Centre region has
  been observed by \citet{2001ApJS..136..189S}. Maps of W3 and W5
  \citep{2010ApJS..191..232B, 2011ApJS..196...18B} are also available.
  For the \jtt\ line, the Galactic centre has been observed by
  \citet{2012ApJS..201...14O}, Orion by \citet{1999ApJ...527L..59I}
  and W3 by \citet{2010ApJS..191..232B}.

The ESA \Planck\ satellite\footnote{\Planck\ (\url{http://www.esa.int/Planck}) is a project of the European Space
Agency (ESA) with instruments provided by two scientific consortia funded by ESA member
states (in particular the lead countries France and Italy), with contributions from NASA
(USA) and telescope reflectors provided by a collaboration between ESA and a scientific
consortium led and funded by Denmark.}
was launched on the 14 May 2009 in order to (primarily) 
measure with unprecedented precision the temperature and polarization anisotropies of
the cosmological microwave background (CMB). It observes the sky in
nine frequency bands covering 30--857 GHz with high sensitivity
and angular resolution from 31\arcm\  to 5\arcm. The Low Frequency
Instrument (LFI; \citealp{2010A&A...520A...3M,2010A&A...520A...4B,2011A&A...536A...3M}) 
covers the 30, 44, and 70~GHz bands with amplifiers cooled to 20~K. The High Frequency Instrument
(HFI;  \citealp{2010A&A...520A...9L,2011A&A...536A...4P}) covers
the 100, 143, 217, 353, 545, and 857 GHz bands with bolometers cooled down to 0.1~K. \Planck's
sensitivity, angular resolution, and frequency coverage make it a
powerful instrument for Galactic and extragalactic astrophysics as well as cosmology \citep{planck2013-p01}.

The first seven CO rotational transition lines lie within the spectral
bands of the HFI instrument. Of these, the first three,
\joz, \jto, and \jtt\ at 115, 230 and 345~GHz, respectively,
present the largest transmission coefficients making them a significant
foreground component in the \Planck\ intensity maps. 
In this paper, we extract full-sky CO maps for these three lines from the LFI and HFI data using
component separation methods. The \Planck\ intensity maps are presented in
Sect.~\ref{sec:planckdata}. In Sect.~\ref{sec:co_trans}, we
provide a brief description of HFI bandpasses and estimate the CO transmission
coefficients for the most important rotational lines.
Specifically tailored component separation methods for CO extraction in the \Planck\ maps are detailed in
Sect.~\ref{sec:co_extract}.  The \Planck\ CO maps obtained using the above
methods are presented in Sect.~\ref{sec:COmaps}. Uncertainties and foreground contamination on those
maps are discussed in Sect.~\ref{sec:errors}. The internal validation of the \Planck\ CO maps is presented
in Sect.~\ref{sec:co_char}. Detailed comparison to existing external CO surveys is presented in Sect.~\ref{sec:ext_comp}. 
Finally, we discuss the results and draw conclusions in Sect.~\ref{sec:conclusion}.

\begin{figure}
\begin{center}
\includegraphics[width=\columnwidth]{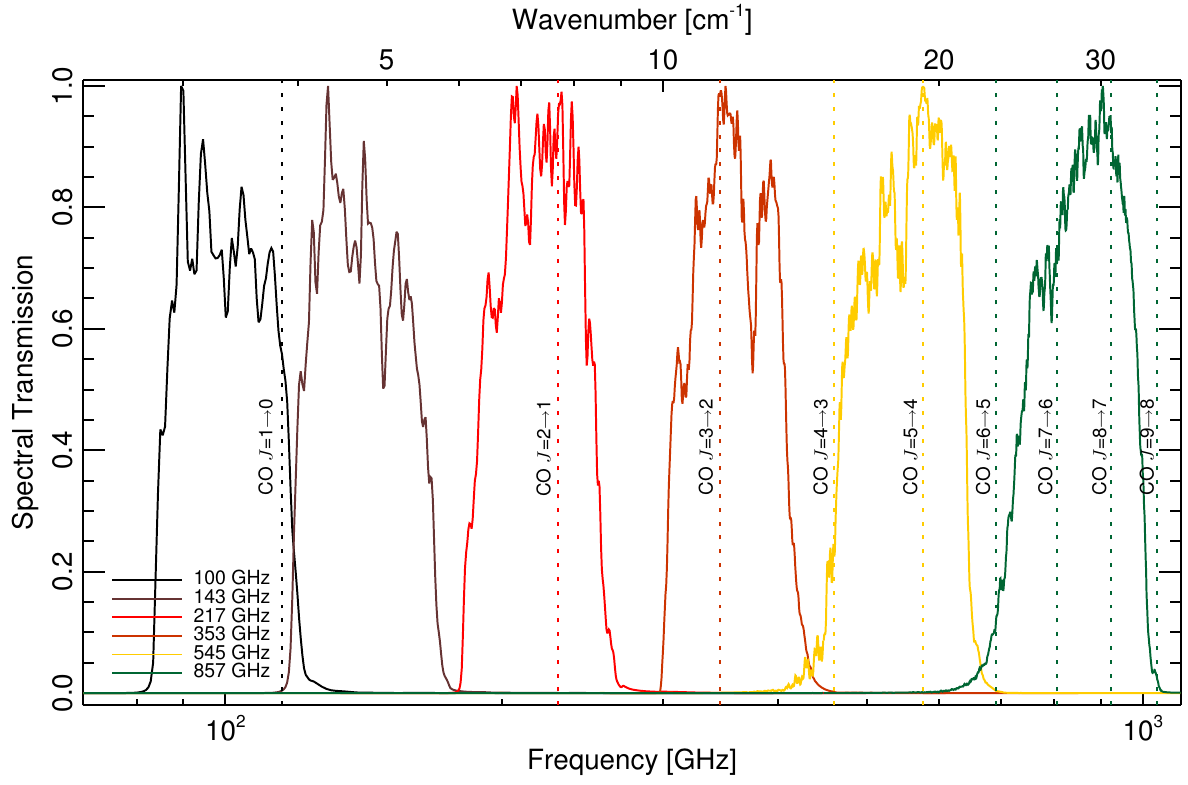}
\caption{\label{fig:AvgBPs} The average spectral response for each of the HFI frequency bands.  The vertical bars represent the CO rotational transitions.}
\end{center}
\end{figure}

\section{\textit{Planck} data}
\label{sec:planckdata}

This paper is based on the \Planck\ first 15.5 month survey mission (two full-sky surveys)
and uses the full-sky maps of the nine \Planck\ frequency bands, and also the 100, 217 and 353~GHz
full-sky bolometer maps. These maps are provided in {\tt HEALPix} pixelization \citep{2005ApJ...622..759G} 
with $N_{\rm side}=2048$ at full resolution. We refer to (\cite{planck2013-p02,planck2013-p02b,planck2013-p03,planck2013-p03b,2011A&A...536A...6P,2011A&A...536A...5Z}
for the generic scheme of TOI processing and map-making, as well as for calibration. 
These \Planck\ maps are given in $K_{\mathrm{CMB}}$ units, i.e. in temperature units referred to the CMB blackbody spectrum.
The scanning strategy of \Planck\ consists of circles on the sky that correspond to different positions of the satellite spin axis.
The latter is changed by 2.5\arcm\ every 40 to 60 minutes in order to cover the full sky in about seven months.
The data set acquired for each position of the satellite spin axis is called a ring and consist of 40 to 60
observations of the same circle on the sky. Thus, a noise map can be obtained for each frequency band or bolometer map from the difference of the maps of the first and second
half of the rings. The resulting noise maps are basically free from
astrophysical emission and thus a good representation of the statistical instrumental noise and systematic errors.
In the following we assume that the beam pattern of the maps can be well represented by effective circular Gaussians
with FWHM of  32\parcm 24, 27\parcm 01, 13\parcm 25, 9\parcm 65,
7\parcm 25, 4\parcm 99, 4\parcm 82, 4\parcm 68, and 4\parcm33
at 30, 44, 70, 100, 143, 217, 353, 545, and 857~GHz, respectively
\citep{planck2013-p02d,planck2013-p03c}.
 
 \begin{figure}
\centering
\includegraphics[width=\columnwidth]{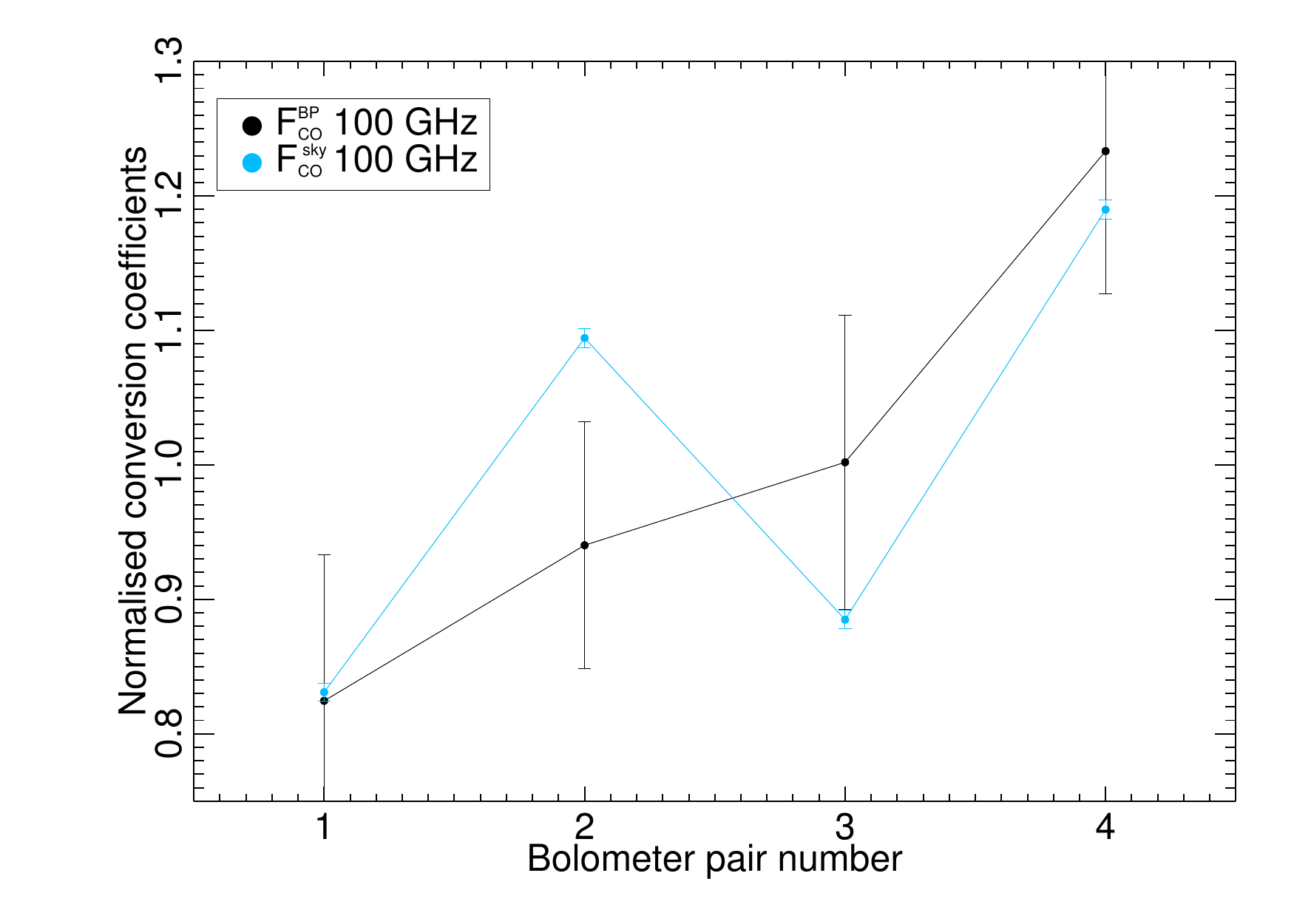}
\includegraphics[width=\columnwidth]{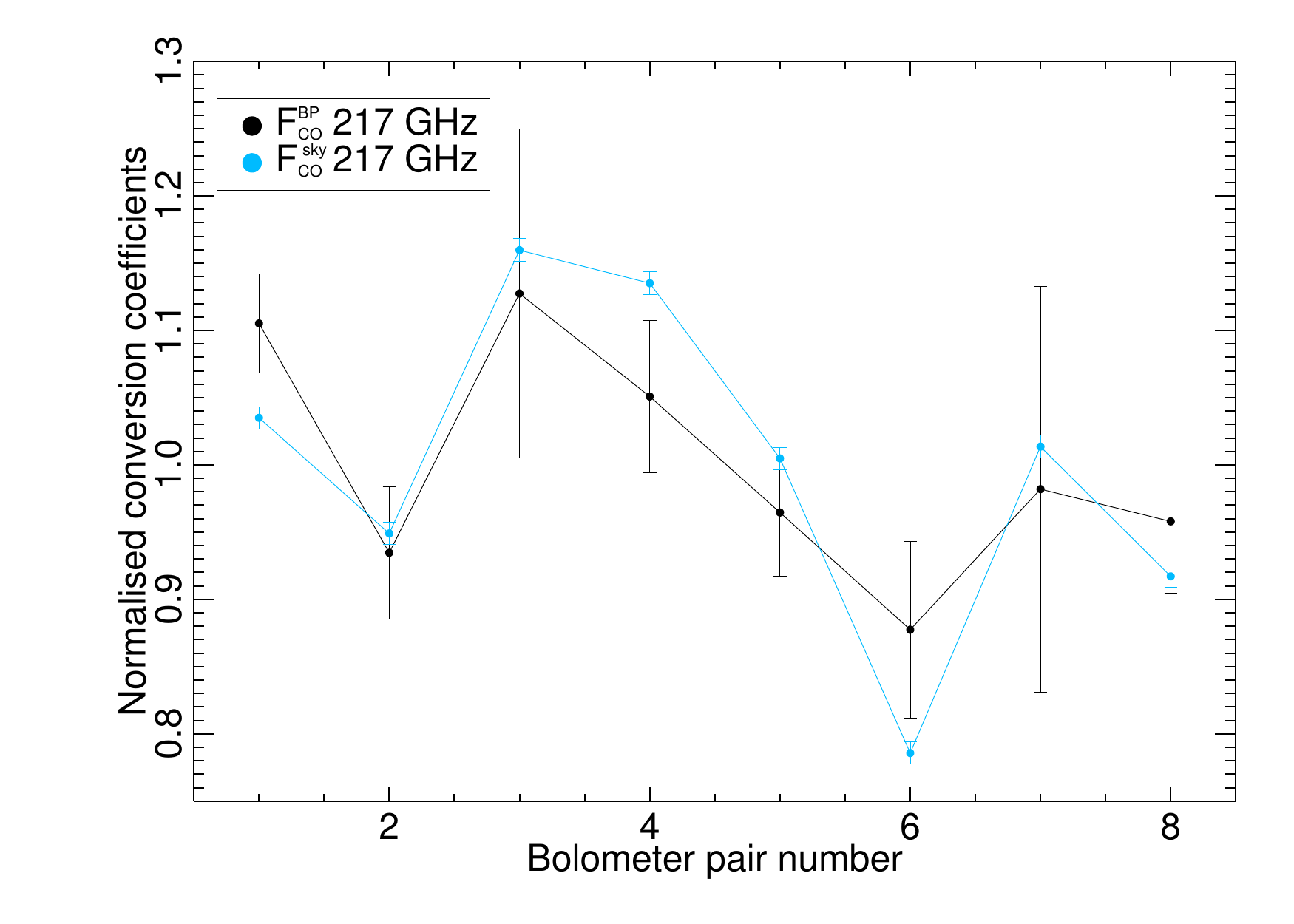}
\includegraphics[width=\columnwidth]{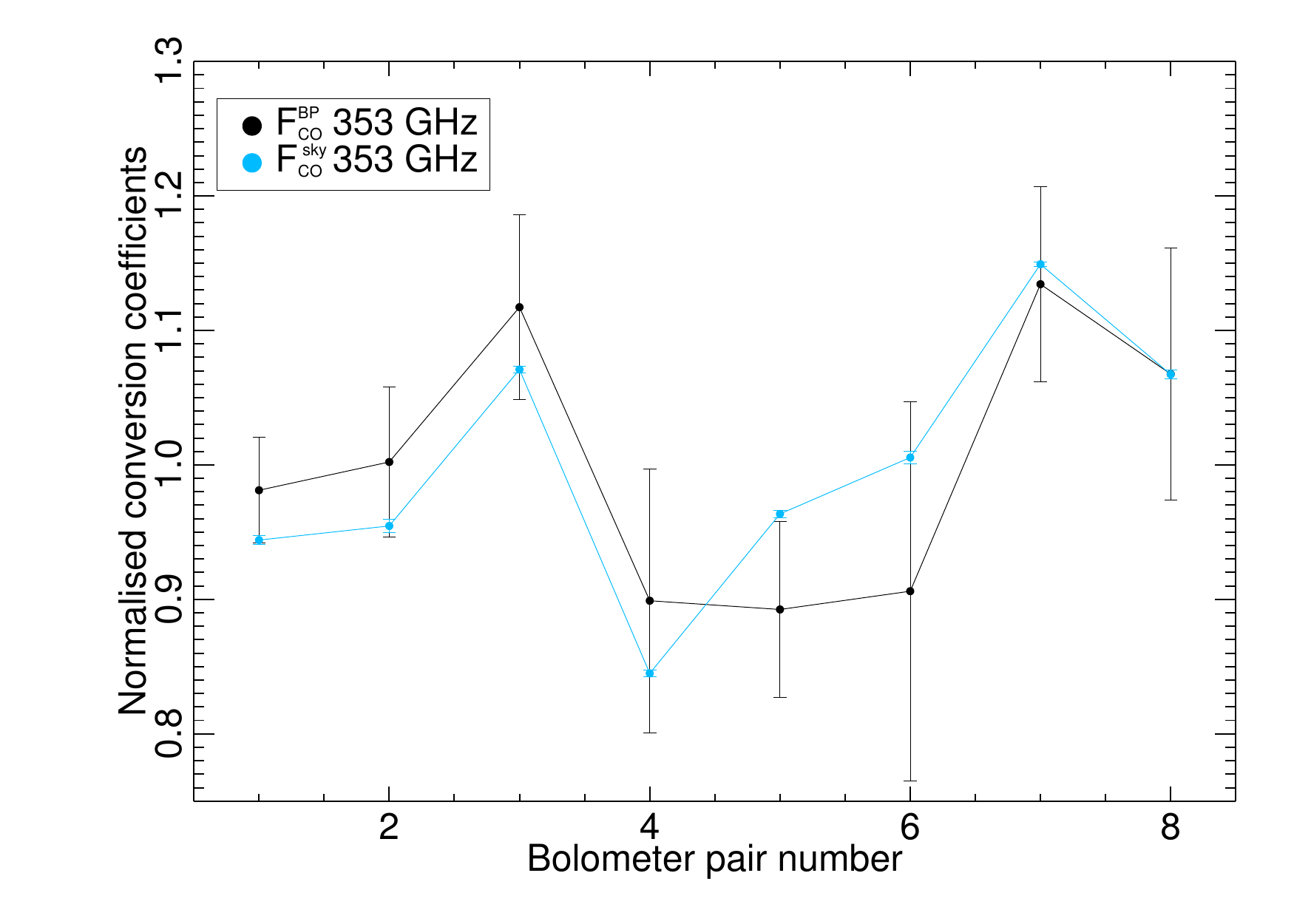}
\caption{Comparison of normalised CO conversion coefficients
  computed from the bandpass measurements (black) or estimated on the
  sky (blue) for the 100 (top), 217 (middle) and 353~GHz (bottom)
  channels. See Table~\ref{tab:co_coeff}.
\label{fig:compare_coeff}}
\end {figure}
\section{CO contribution in the \textit{Planck} HFI channels}
\label{sec:co_trans}

\subsection{HFI Spectral Response and CO Emission}
\label{sec:LS_CO}

In order to isolate the narrow CO features from the remainder of the components (e.g., CMB, dust, etc.), 
precise knowledge of the instrument spectral response as a function of frequency is required.  The original 
spectral resolution requirement for the spectral response of a given HFI detector was about 3 GHz; this corresponds 
to a velocity resolution of around 8000 km/s for the \cooz\ line.  As HFI does not have the ability to measure 
spectral response within a frequency band during flight, the ground-based Fourier Transform Spectrometer (FTS)
measurements provide the authoritative 
data on the HFI spectral transmission.  With good S/N, spectral information can be inferred to a fraction (i.e.,
$\sim$1/10th) of a spectral resolution element \citep{Spencer2010_MST}.  Thus, the pre-flight
calibration FTS measurements \citep{2010A&A...520A..10P}, which were carried out at a spectral resolution 
of about 0.6~GHz, may be used to estimate the spectral response at a resolution equivalent to 50 km s$^{-1}$ 
for the \cooz\ line. For spectral regions near CO rotational transitions, therefore, the spectral response was 
oversampled by a factor of around 10 using an interpolation based on the instrument line shape of the FTS\@.  
Table \ref{tab:CO} lists the relevant CO transitions, and the frequency ranges over which the HFI spectral response 
was oversampled \citep{planck2013-p03d}. The oversampling ranges were extended to 
include all of the common CO isotopologues.  The vertical bars above the spectral response curves of Fig. 
\ref{fig:AvgBPs} illustrate these oversampled regions.  

\begin{table}[tmb]
\begingroup
\newdimen\tblskip \tblskip=5pt
\caption{\label{tab:CO}Rotational $^{12}$CO transitions within the HFI bands and over-sampled regions.}                          
\nointerlineskip
\vskip -6mm
\footnotesize
\setbox\tablebox=\vbox{
   \newdimen\digitwidth 
   \setbox0=\hbox{\rm 0} 
   \digitwidth=\wd0 
   \catcode`*=\active 
   \def*{\kern\digitwidth}
   \newdimen\dpwidth 
   \setbox0=\hbox{.} 
   \dpwidth=\wd0 
   \catcode`!=\active 
   \def!{\kern\dpwidth}
\halign{\hbox to 1.5cm{\hfil#\hfil}\tabskip 1em&
     \hfil#\hfil \tabskip 1em&
     \hfil#\hfil \tabskip 1em&
     \hfil#\hfil \tabskip 0em\cr 
\noalign{\doubleline}
\noalign{\vskip 5pt}
\omit Band & CO transition & $\nu_{0}$ & Oversampling\cr
\omit $[$GHz$]$\hfil&$[J_{\mbox{\tiny{upper}}} \rightarrow
      J_{\mbox{\tiny{lower}}}]$ & $[$GHz$]$&$[$GHz$]$\cr
\noalign{\vskip 3pt\hrule\vskip 5pt}
 \noalign{\vskip 4pt}
\omit 100 & 1 $\rightarrow$ 0 & *115.271 & 109.6 --  *115.4 \cr
\noalign{\vskip 5pt}
\omit 217 & 2 $\rightarrow$ 1 & *230.538 & 219.3 --  *230.8 \cr
\noalign{\vskip 5pt}
\omit 353 & 3 $\rightarrow$ 2 & *345.796 & 329.0 --  *346.2 \cr
\noalign{\vskip 5pt}
\omit 545 & 4 $\rightarrow$ 3 & *461.041 & 438.6 --  *461.5 \cr
\omit 545 & 5 $\rightarrow$ 4 & *576.268 & 548.3 --  *576.8 \cr
\noalign{\vskip 5pt}
\omit 857 & 6 $\rightarrow$ 5 & *691.473 & 657.9 --  *692.2 \cr
\omit 857 & 7 $\rightarrow$ 6 & *806.652 & 767.5 --  *807.5 \cr
\omit 857 & 8 $\rightarrow$ 7 & *921.799 & 877.0 --  *922.7 \cr
\omit 857 & 9 $\rightarrow$ 8 &1036.912 & 986.6 -- 1037.9 \cr
\noalign{\vskip 3pt\hrule\vskip 5pt}
}
}
\endPlancktable 
\endgroup
\end{table}

\subsection{Spectral Band CO conversion coefficients}
\label{subsec:coeff_BP}

Unit conversion coefficients between CO brightness temperature and CMB temperature are determined in much the same manner as the other HFI 
unit conversion coefficients \citep[see][]{planck2013-p03d}:
\begin{equation}
F^{\mathrm{BP}}_{\mathrm{CO},b} = \frac{\int d\nu \ H^{b}_{\nu} \  I^{\mathrm{CO}}_{\nu} }{\int d\nu \ H^{b}_{\nu} \ I^{\mathrm{CMB}}_{\nu} }
\label{coconvbpequ}
\end{equation}
where $H^{b}_{\nu}$ is the spectral transmission of bolometer $b$ at frequency $\nu$, and $I^{\mathrm{CO}}_{\nu}$ and  $I^{\mathrm{CMB}}_{\nu}$ are
the CO and CMB intensities, respectively.
The CO velocity-integrated emission (VIE) is expressed as the product of Rayleigh-Jeans temperature and 
spectral line width in velocity units, i.e., K$_{\mbox{\tiny{RJ}}}$~km~s$^{-1}$.  For  the lowest rotational CO transitions a Doppler shifted line profile may be 
assumed with $\nu = \nu_{\rm CO}( 1 + \mathrm{v}/c)^{-1} \approx \nu_{\rm CO}(1 - \mathrm{v}/c)$ for $\mathrm{v} \ll c$. As the CO transitions occur at discrete frequencies, 
with a Doppler line-width much less than the transition frequency (about $10^3~$Hz cf.\ $10^{11}~$Hz), and much narrower than the 
available knowledge of the HFI detector spectral response ($\sim 10^8~$Hz), the velocity integration may be approximated by a delta function 
distribution at $\nu_{\rm CO}$, i.e.\ $\delta_{\nu_{\rm CO}}$. 

Bandpass CO conversion coefficients for the first three CO transitions were determined 
using the above relation \citep[see][]{planck2013-p03d}) ; similar data are available for the individual 
HFI detectors and for the other transitions within HFI bands. Table
\ref{tab:co_coeff} (first column) lists the relative values of these CO conversion
coefficients for the $^{12}$CO isotopologue (averaged for each $a$ and $b$ pair of polarized
bolometers) when normalized to the average transmission of all bolometers
in the band (this format is chosen to ease comparison to the
sky-calibrated conversion coefficients, see next section). To recover the
physical conversion factor, values in the table must be multiplied by the average of $^{12}$CO conversion coefficients,
namely, $1.42\times10^{-5}$K$_{\rm CMB}$~km~s$^{-1}$ at 100~GHz, $4.50\times10^{-5}$K$_{\rm CMB}$~km~s$^{-1}$ at 217~GHz and $17.37\times10^{-5}$K$_{\rm CMB}$~km~s$^{-1}$ at 353~GHz.

As will be described in Sect.~\ref{subsubsec:13co}, the $^{13}$CO
isotopologue may contribute to the \Planck\ CO maps. The conversion coefficients of
$^{13}$CO are computed in the same way as described for
$^{12}$CO, but for the $^{13}$CO transitions at $\nu_0^{[1-0]}=110.2$~GHz,
$\nu_0^{[2-1]}=220.40$~GHz, and $\nu_0^{[3-2]}=330.6$~GHz. The average
conversion factor over all bolometers in the 100, 217, and 353~GHz channel
are $1.62\times10^{-5}$K$_{\rm CMB}$~km~s$^{-1}$, 
$3.63\times10^{-5}$K$_{\rm CMB}$~km~s$^{-1}$, and $13.0\times10^{-5}$K$_{\rm CMB}$~km~s$^{-1}$, respectively.

\subsection{Sky-calibrated CO conversion coefficients}
\label{subsec:coeff_sky}

In the following sections we will use CO conversion coefficients to extract maps
of the CO emission from the \Planck\ frequency maps. For this purpose, we need
an accurate estimation of the relative CO conversion coefficients between bolometers (see appendix~\ref{sec:simuvalidation}
for further discussions). The bandpass CO conversion coefficients present two main problems.
First, the estimation of the spectral band transmissions of each of the bolometers may be affected by systematic
errors as we are dealing with narrow lines. Indeed, HFI bandpasses are not sampled 
to a sufficient resolution to allow for a satisfactory CO
extraction using these coefficients and then need to be interpolated as discussed above. 
Second, both $^{12}$CO and $^{13}$CO (and other isotopologues) are present
in the \Planck\ maps. The emissions from $^{12}$CO and $^{13}$CO are spatially
correlated with a varying ratio across the sky and we can not discriminate between the two.
Thus, an accurate determination of the relative CO conversion coefficients between bolometers
of the same \Planck\ channel is difficult when using bandpass information only. 

A way around these issues has been found in estimating the CO conversion coefficients 
directly from sky measurements of well-known molecular clouds.
For these regions, we can assume -- to first order -- that the map of the sky
emission for a detector is a linear combination of CO and thermal dust emissions weighted by
the CO and thermal dust conversion coefficients of the detector. 
Thus, using an external CO emission template, $P_{\rm CO}$, obtained
from the \citet{2001ApJ...547..792D} $^{12}$CO \joz\
survey and a dust emission template, the 545~GHz \Planck-HFI channel
map (in K$_{{\rm CMB}}$ units), $P_{\mathrm{d}}$, it is possible to determine sky-calibrated CO and dust transmission
coefficients $F_{{\rm CO},b}^{\rm sky}$ and $F_{{\rm dust},b}^{\rm sky}$ of bolometer $b$. 
Notice that we implicitly assume here a perfect spatial correlation between the $^{12}$CO
and $^{12}$CO components.

Neglecting other astrophysical components (i.e., CMB, free-free, synchrotron, AME and point sources),
we perform a simple linear fit between each bolometer map $M_{b}$ in the native
units (K$_{\rm CMB}$) and
the CO and dust templates: 
\begin{equation}
M_{b} = F_{{\rm CO},b}^{\rm fit} \ P_{\rm  CO} +  F_{{\rm dust},b}^{\rm fit} \ P_{\mathrm{d}}\;,$$
\label{eq:skycoeff}
\end{equation}
where $F_{{\rm CO},b}^{\rm fit}$ and $F_{{\rm dust},b}^{\rm fit}$ are
the results of the linear regression.
To avoid contamination from dust polarization emission we combine the maps of pairs of polarized bolometers
into a single map and a single CO conversion coefficient is computed for the pair of detectors.

Due to the spatial correlation between the CO and dust emission, we expect
the CO conversion coefficients, $F_{{\rm CO},b}^{\rm fit}$, to be affected by dust
contamination and so biased.
We simply assume here that the estimated CO conversion factor can be expressed as $F_{{\rm
    CO},b}^{\rm fit} = \alpha \ F_{{\rm CO},b}^{\rm sky} + \beta$
corresponding to an offset and a shift in total amplitude.
However, we will consider that thermal dust conversion coefficients
are not affected by CO contamination significantly
and thus $F_{{\rm dust},b}^{\rm sky}= F_{{\rm dust},b}^{\rm fit}$.

To compute $\alpha$ and $\beta$ we solve the following system of equations.
First, we  construct dust weighted bolometer (pairs of bolometers) maps. 
Then, we obtain a first set equations by imposing that the difference of two dust weighted bolometer (pairs of bolometer) maps, 
$$\frac{M^{\nu}_{b}}{F_{{\rm dust},b}^{\rm fit} P_{\mathrm{d}}}-\frac{M^{\nu}_{b^\prime}}{F_{{\rm dust},b^\prime}^{\rm fit} P_{\mathrm{d}}}$$ 
must be correlated with the CO template (weighted by the dust map). A second set of equations is obtained by
searching for the factor $\gamma$ that minimizes the correlation of  the difference
$$\frac{M^{\nu}_{b}}{F_{{\rm dust},b}^{\rm fit} P_{\mathrm{d}}}- \gamma \times \frac{M^{\nu}_{b^\prime}}{F_{{\rm dust},b^\prime}^{\rm fit} P_{\mathrm{d}}}$$
with the CO dust weighted template, $P_{{\rm CO}}/P_{\mathrm{d}}$.
Uncertainties in the final CO conversion coefficients are obtained from the dispersion of the coefficients found for different sky regions.
For the \cooz\ line we use sky regions for which the
\citet{2001ApJ...547..792D} map is above 2~K$_{\rm
  RJ}$~km~s$^{-1}$. For the \jto\ and \jtt\ lines we consider only the
nine brightest CO clouds in the \citet{2001ApJ...547..792D} map. 

The sky CO conversion coefficients\footnote{Compatible results within statistical errors are obtained when performing a similar, but independent, analysis with {\tt Commander}. This results shows these conversion coefficients are robust against some possible systematic effects on the method.} (normalized to their mean value)
obtained from this analysis are presented in the right column
of Table~\ref{tab:co_coeff} and compared to $^{12}$CO bandpass conversion
coefficients in Fig.~\ref{fig:compare_coeff}. 
The discrepancy between the relative values of the bandpass and sky determined
CO conversion can be explained from the quoted uncertainties.
A more detailed comparison of sky and bandpass determined CO conversion
coefficients, and a description of the systematic uncertainties is given in the \citet{planck2013-p03d} companion paper.
However, it is important to notice that they are not expected to be equal as
the sky-based CO conversion coefficients account also for the contribution of other spatially 
correlated CO isotopologues, mainly $^{13}$CO and other molecular
lines. This is extensively discussed in Sect.~\ref{subsubsec:13co}.

Finally, note that for the \jto\ and \jtt\ lines the
maps obtained from the sky-calibrated CO conversion
coefficients are calibrated in the Dame et al. \cooz\ line units and 
need to be re-calibrated to their actual frequencies; this is the purpose of the next section.

\begin{table}[tmb]
\begingroup
\newdimen\tblskip \tblskip=5pt
\caption{Relative CO conversion coefficients of HFI bolometers (pairs of bolometers) in the 100,
  217 and 353~GHz channels normalized to the average conversion coefficient for
  each channel. \label{tab:co_coeff}}                         
\nointerlineskip
\vskip -6mm
\footnotesize
\setbox\tablebox=\vbox{
   \newdimen\digitwidth 
   \setbox0=\hbox{\rm 0} 
   \digitwidth=\wd0 
   \catcode`*=\active 
   \def*{\kern\digitwidth}
   \newdimen\dpwidth 
   \setbox0=\hbox{.} 
   \dpwidth=\wd0 
   \catcode`!=\active 
   \def!{\kern\dpwidth}
\halign{\hbox to 1.5cm{\hfil#\hfil}\tabskip 2em&
     \hfil#\hfil \tabskip 1em&
     \hfil#\hfil \tabskip 1em&
     \hfil#\hfil \tabskip 0em\cr 
\noalign{\doubleline}
\omit Bolo ID & $F_{^{12}\mathrm{CO}}^{\rm BP}$ & $F_{^{13}\mathrm{CO}}^{\rm BP}$ & $F_{\rm CO}^{\rm sky}$ \cr
\noalign{\vskip 3pt\hrule\vskip 5pt}
   {\joz} & & & \cr
\noalign{\vskip 3pt\hrule\vskip 5pt}
\omit 100-1 (a+b)/2 &0.82 $\pm$ 0.10 & 1.03 $\pm$ 0.12  & 0.83 $\pm$ 0.01 \cr
\omit 100-2 (a+b)/2 &0.94 $\pm$ 0.09 & 0.97 $\pm$ 0.10  & 1.09 $\pm$ 0.01 \cr
\omit 100-3 (a+b)/2 &0.99 $\pm$ 0.11 & 0.87 $\pm$ 0.14 & 0.88 $\pm$ 0.01  \cr
\omit 100-4 (a+b)/2 &1.24 $\pm$ 0.10 & 1.13 $\pm$ 0.24 & 1.19 $\pm$ 0.01  \cr
\noalign{\vskip 3pt\hrule\vskip 5pt}
   {\jto} & & & \cr
\noalign{\vskip 3pt\hrule\vskip 5pt}
\omit 217-1               &  1.10 $\pm$ 0.03   &  0.98 $\pm$ 0.06 & 1.03  $\pm$ 0.01 \cr
\omit 217-2               &  0.94 $\pm$ 0.05   &  0.94 $\pm$ 0.13 & 0.95  $\pm$ 0.01\cr
\omit 217-3               &  1.13 $\pm$ 0.12   &  1.07 $\pm$ 0.06 &1.16  $\pm$ 0.01\cr
\omit 217-4               &  1.05 $\pm$ 0.06   &  0.88 $\pm$ 0.20 &1.14  $\pm$ 0.01\cr
\omit 217-5 (a+b)/2 &  0.97 $\pm$ 0.05   &  0.89 $\pm$ 0.06 & 1.00  $\pm$ 0.01 \cr
\omit 217-6 (a+b)/2 &  0.88 $\pm$ 0.06   &  1.06 $\pm$ 0.12 & 0.79  $\pm$ 0.01  \cr
\omit 217-7 (a+b)/2 &  0.97 $\pm$ 0.15   &  1.08 $\pm$ 0.05 & 1.01  $\pm$ 0.01  \cr
\omit 217-8 (a+b)/2 &  0.96 $\pm$ 0.05   &  1.08 $\pm$ 0.04 & 0.92  $\pm$ 0.01\cr
\noalign{\vskip 3pt\hrule\vskip 5pt}
   {\jtt} & & & \cr
\noalign{\vskip 3pt\hrule\vskip 5pt}
\omit 353-1               & 0.98 $\pm$ 0.04 & 0.66 $\pm$ 0.07 &  0.94  $\pm$ 0.01    \cr
\omit 353-2               & 1.00 $\pm$ 0.05 & 1.05 $\pm$ 0.05 &  0.95 $\pm$ 0.01      \cr
\omit 353-3 (a+b)/2 & 1.11 $\pm$ 0.07 & 1.22 $\pm$ 0.11 & 1.07 $\pm$ 0.01 \cr
\omit 353-4 (a+b)/2 & 0.90 $\pm$ 0.09 & 1.13 $\pm$ 0.08 & 0.84 $\pm$ 0.01 \cr
\omit 353-5 (a+b)/2 & 0.89 $\pm$ 0.06 & 1.24 $\pm$ 0.09 & 0.96 $\pm$ 0.01  \cr
\omit 353-6 (a+b)/2 & 0.91 $\pm$ 0.14 & 0.89 $\pm$ 0.11 & 1.01 $\pm$ 0.01  \cr
\omit 353-7               & 1.14 $\pm$ 0.07 & 0.12 $\pm$ 0.04 & 1.15 $\pm$ 0.01     \cr
\omit 353-8               & 1.06 $\pm$ 0.09 & 0.80 $\pm$ 0.08 & 1.07 $\pm$ 0.01    \cr
\noalign{\vskip 3pt\hrule\vskip 5pt}
}
}
\endPlancktable 
\endgroup
\end{table}

\subsection{Unit convention}
In the following the \Planck\ CO maps are extracted using the sky determined CO conversion coefficients.
As discussed above these coefficients convert from the units of the
\citet{2001ApJ...547..792D} survey that it is used as a CO template to
the original $K_{\rm CMB}$ units of the \Planck\ HFI maps. 
To be useful to scientific purpose these maps must be converted into the emission
of the transition line. This is done by recalibrating the \Planck\ CO
maps so that they are in units of the expected $^{12}$CO contribution at the
transition frequency: 

\begin{equation}
M^{\rm final}_{\rm CO}  = M^{\rm sky}_{\rm CO} \times\frac{\langle F^{\rm
    sky}_{\rm CO}\rangle}{\langle F^{\rm BP}_{\rm CO}\rangle}.
\end{equation}
Here $M^{\rm sky}_{\rm CO}$ is the CO map of any transition calibrated on the
\citet{2001ApJ...547..792D} data, using the sky CO conversion coefficients. The
quantities $\langle F^{\rm BP}_{\rm CO}\rangle$ and  $\langle F^{\rm sky}_{\rm CO}\rangle$
are the average across detectors of the bandpass and sky $^{12}$CO conversion
coefficients. 

\begin{figure*}[!th]
\centering
\includegraphics[width=\columnwidth]{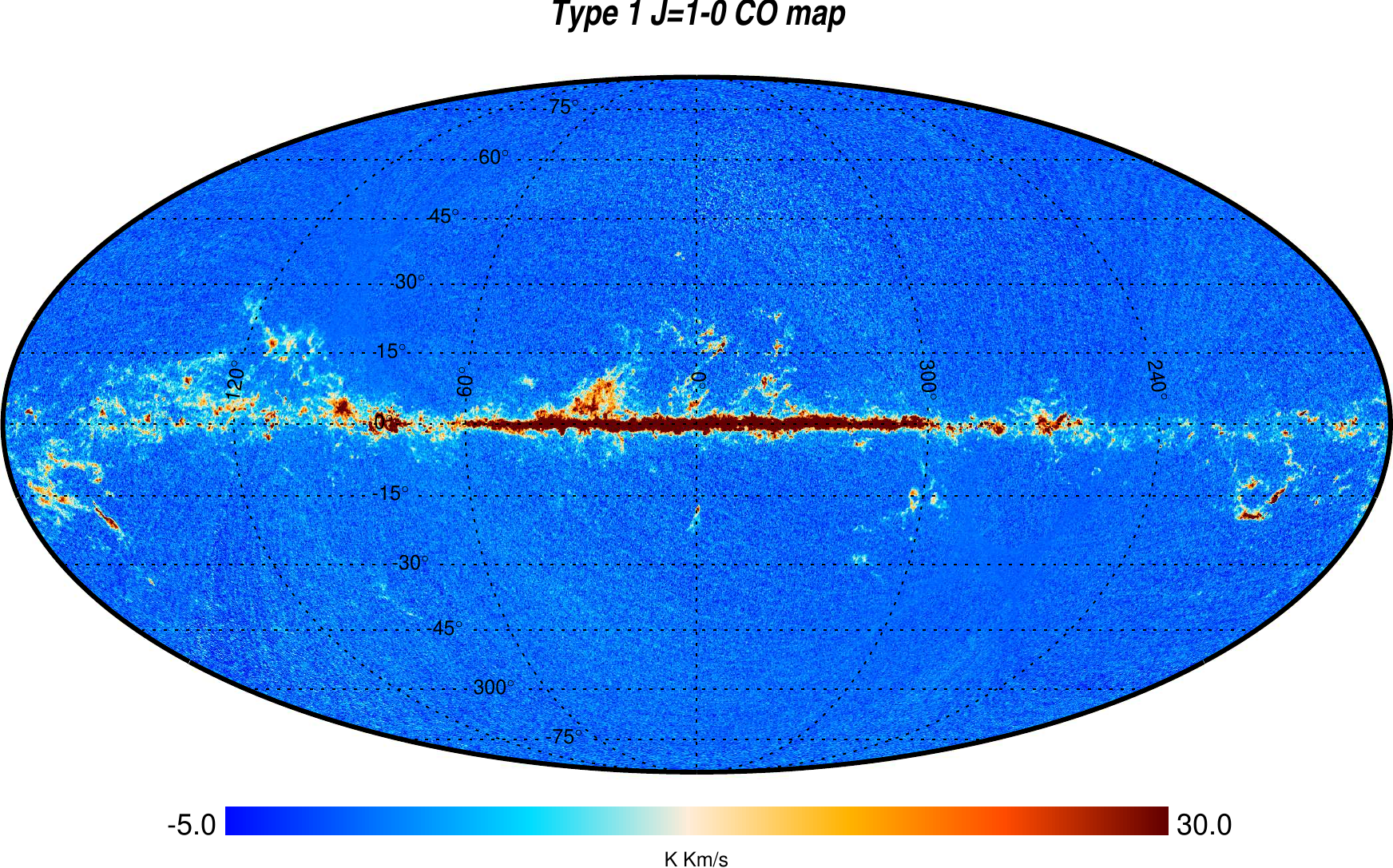}
\includegraphics[width=\columnwidth]{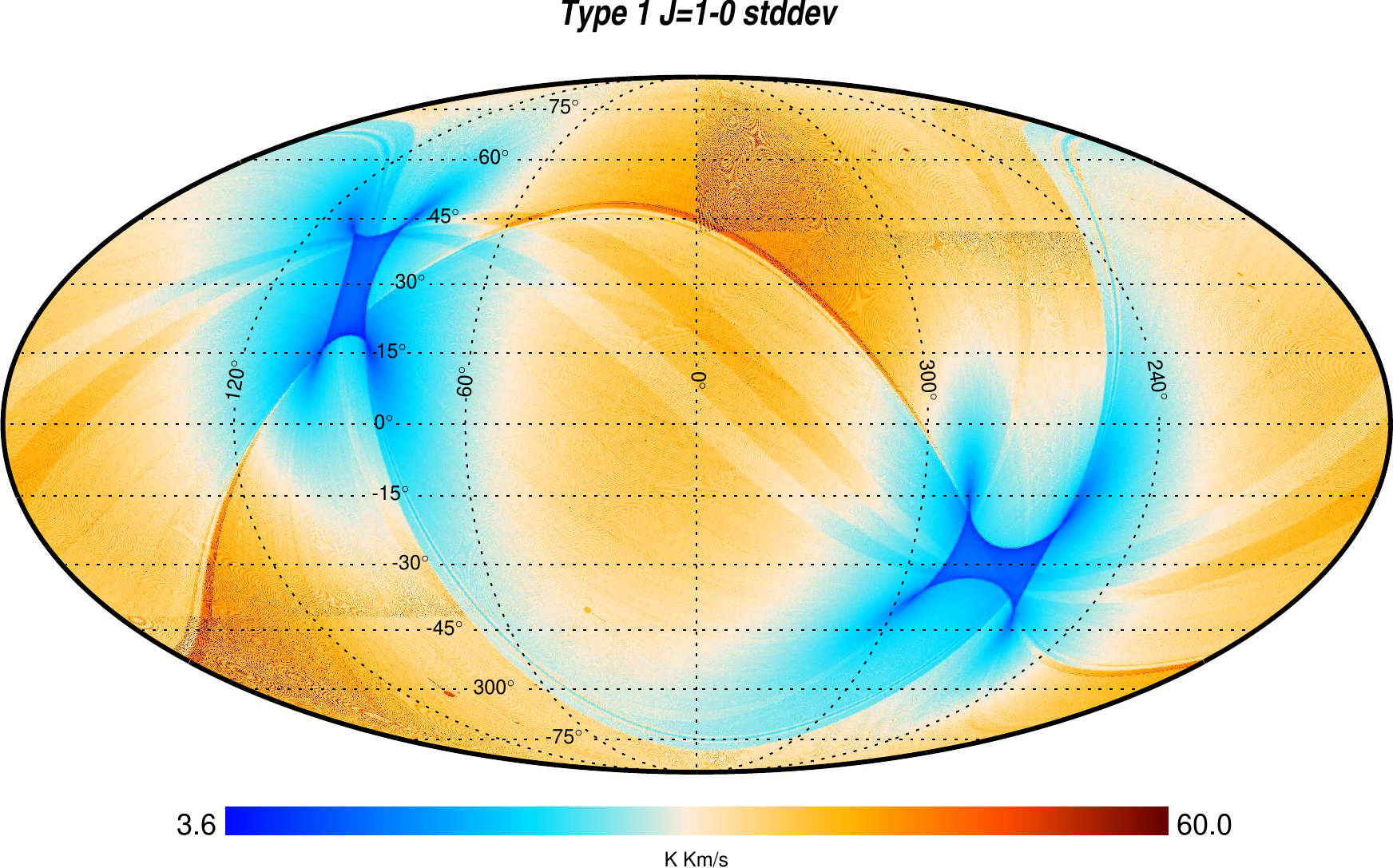}
\includegraphics[width=\columnwidth]{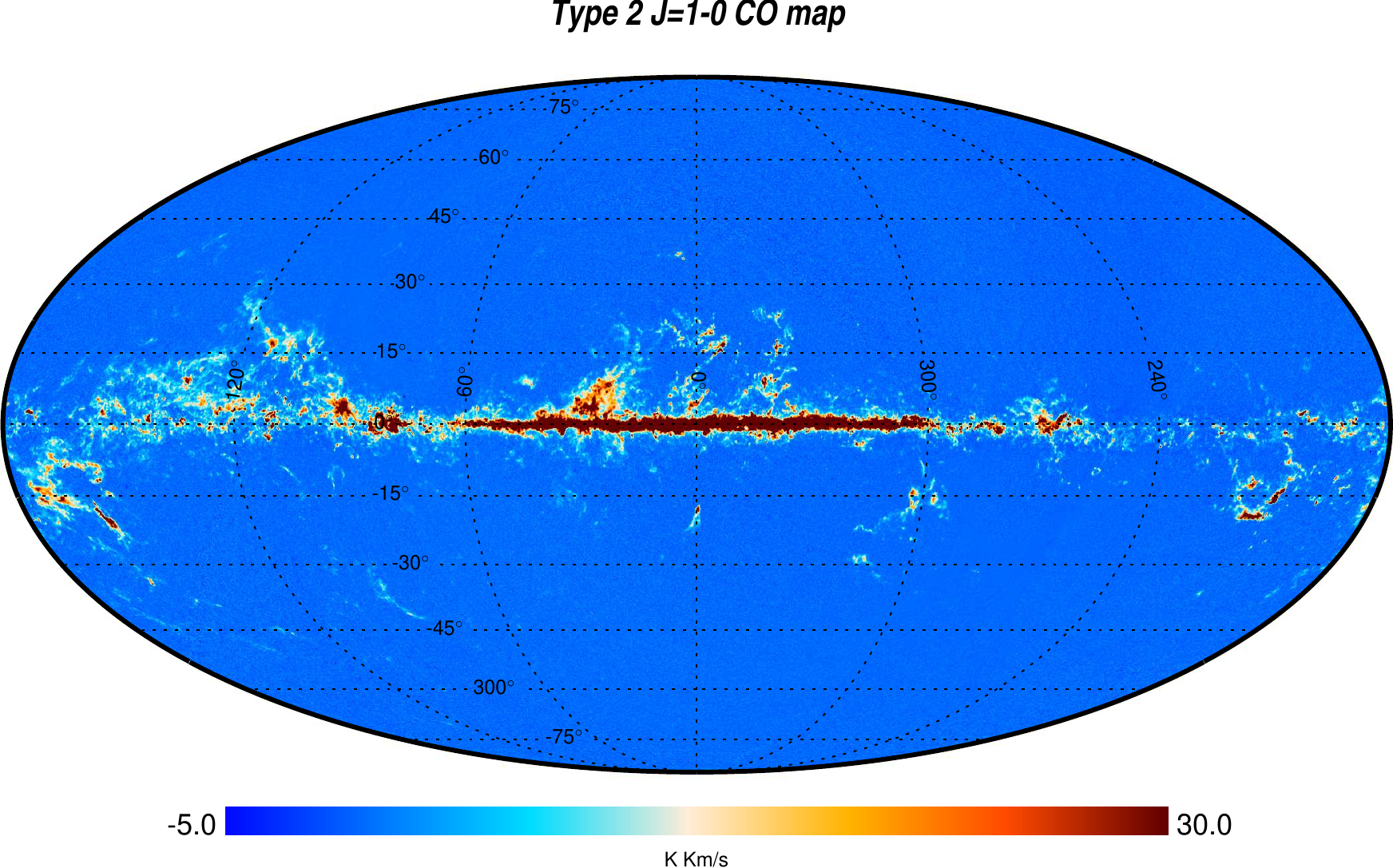}
\includegraphics[width=\columnwidth]{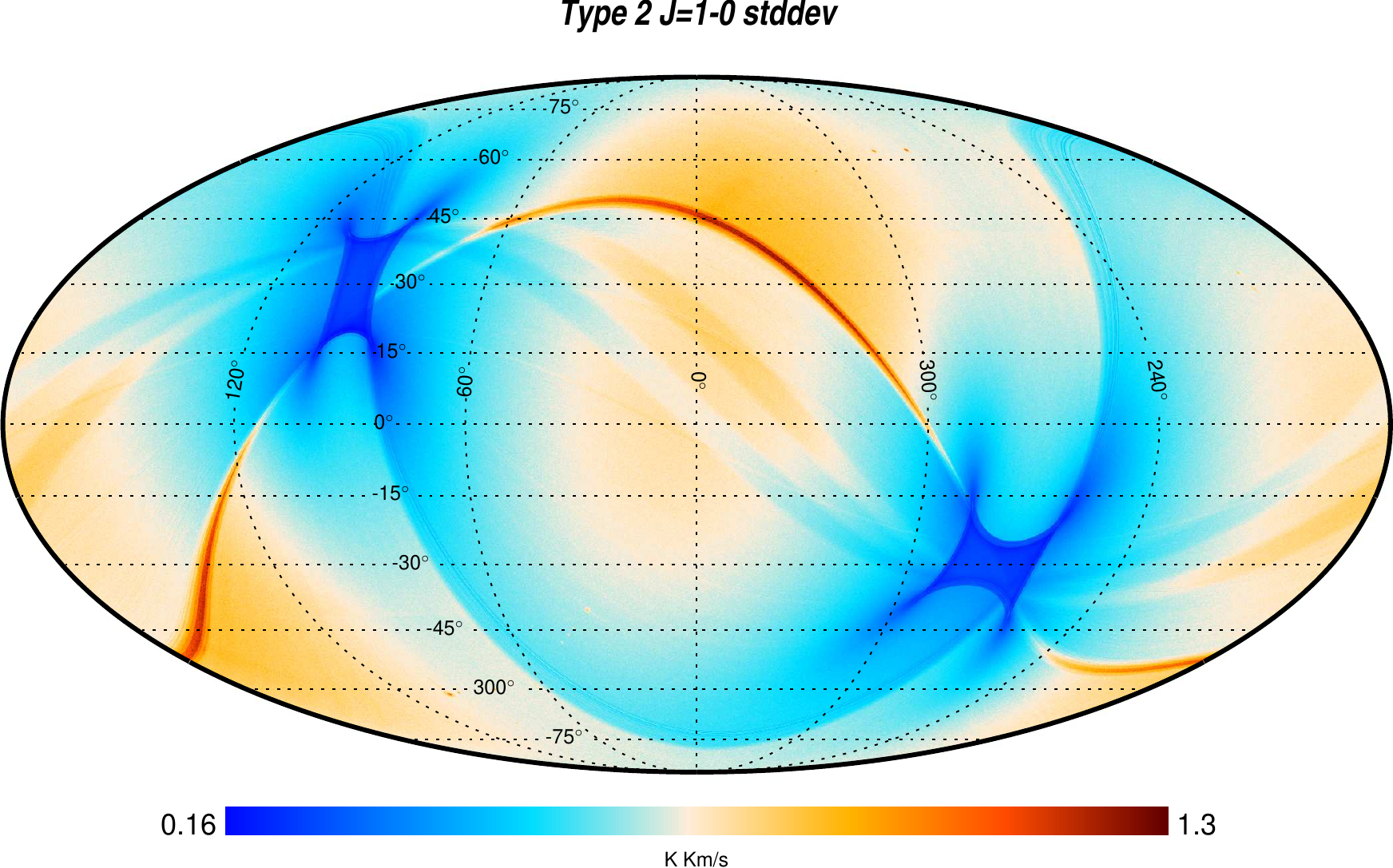}
\includegraphics[width=\columnwidth]{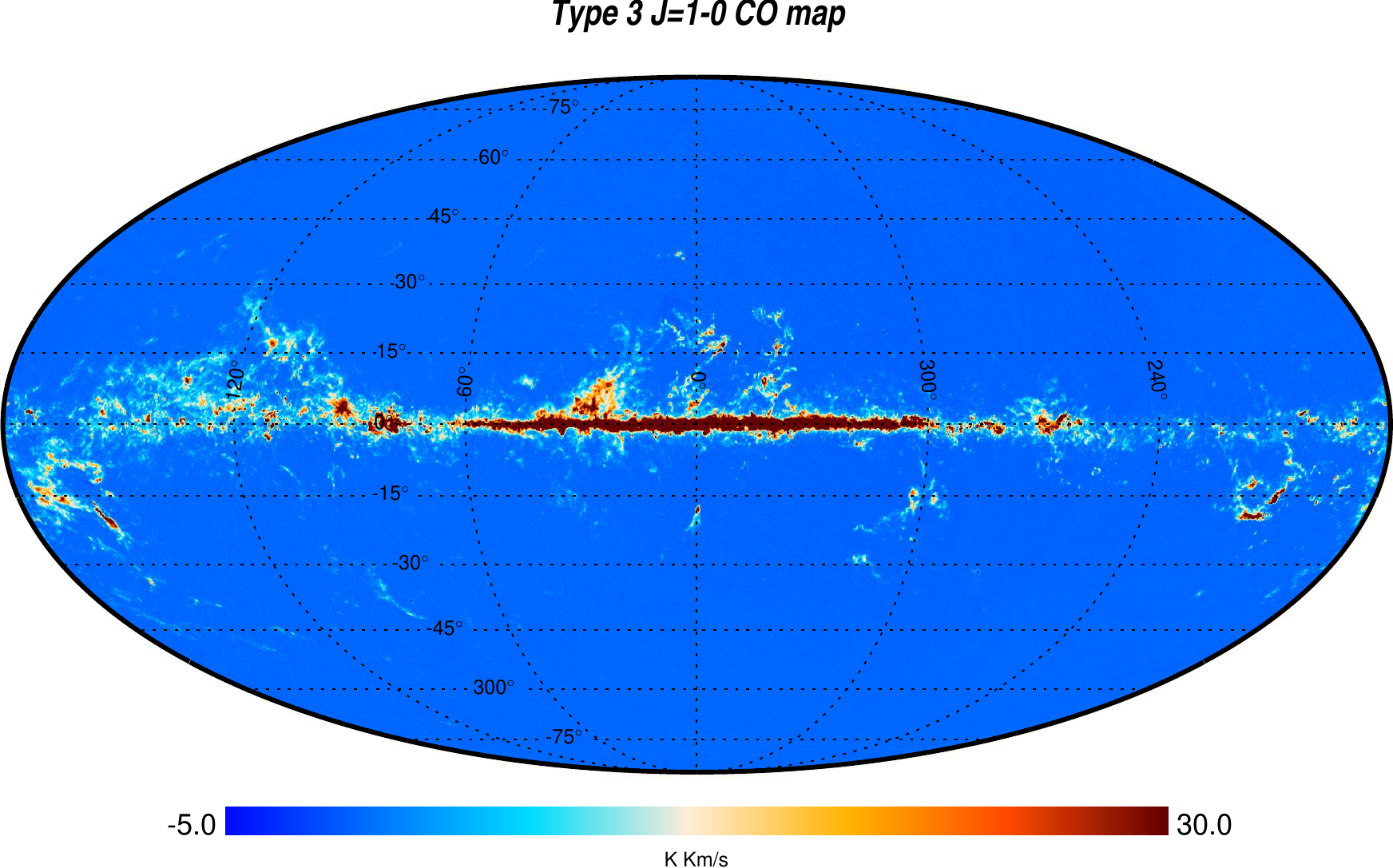}
\includegraphics[width=\columnwidth]{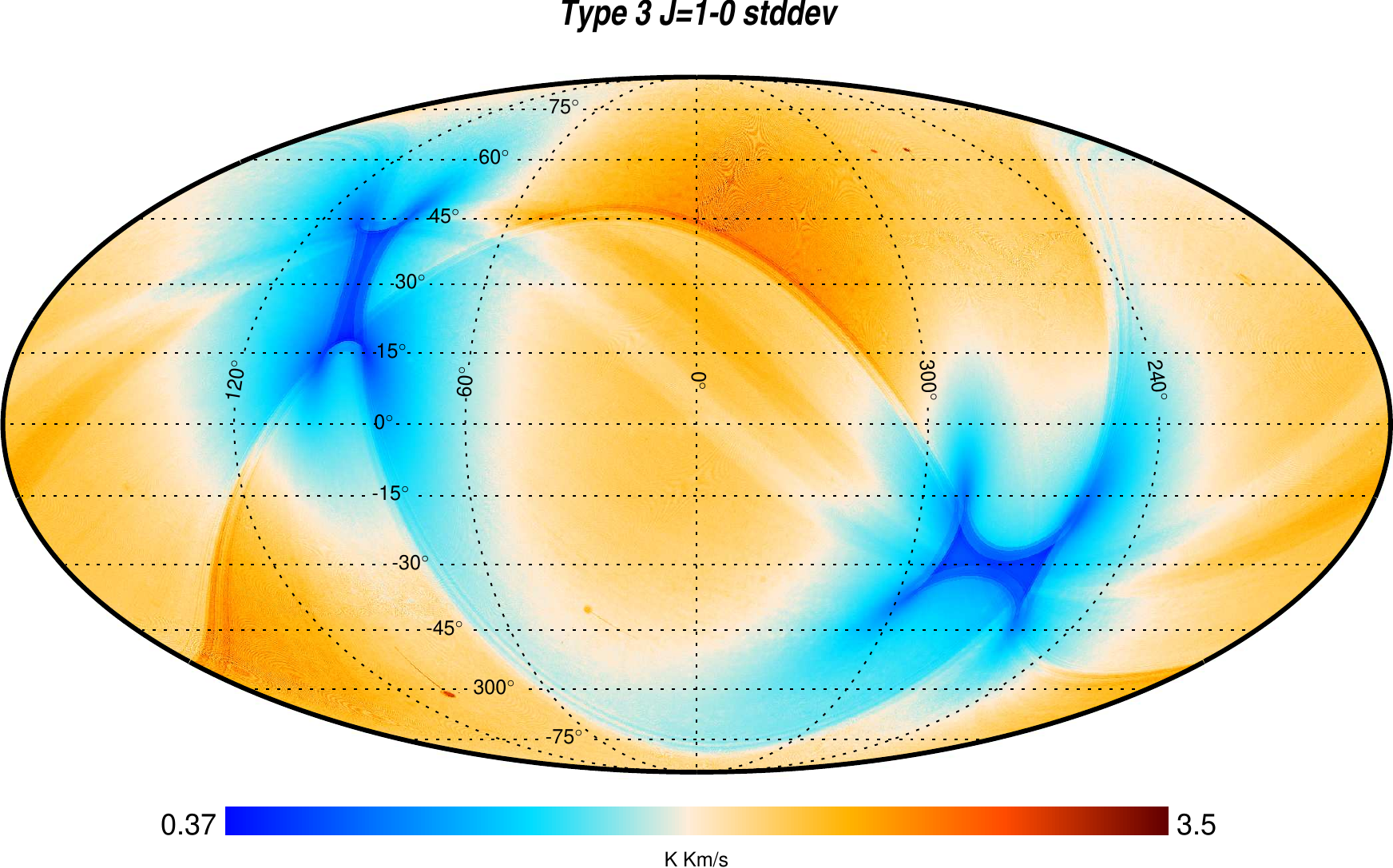}

\caption{Full-sky CO \joz\ maps (left) and their respective statistical
  error maps (right) for the three types of \Planck\ CO products.  For
display purposes only, the \typeone\ and \typethree\ maps are
smoothed to 15\arcm\ to ease the comparison to the \typetwo\ maps. The error
maps correspond to the non-smoothed product. The maps are in Galactic coordinates and
follow the {\tt HEALPix} pixelization scheme for $N_{\mathrm{side}}=2048$. 
\label{fig:fullskyCO10}}
\end {figure*}

\begin{figure*}[!th]
\centering
\includegraphics[width=\columnwidth]{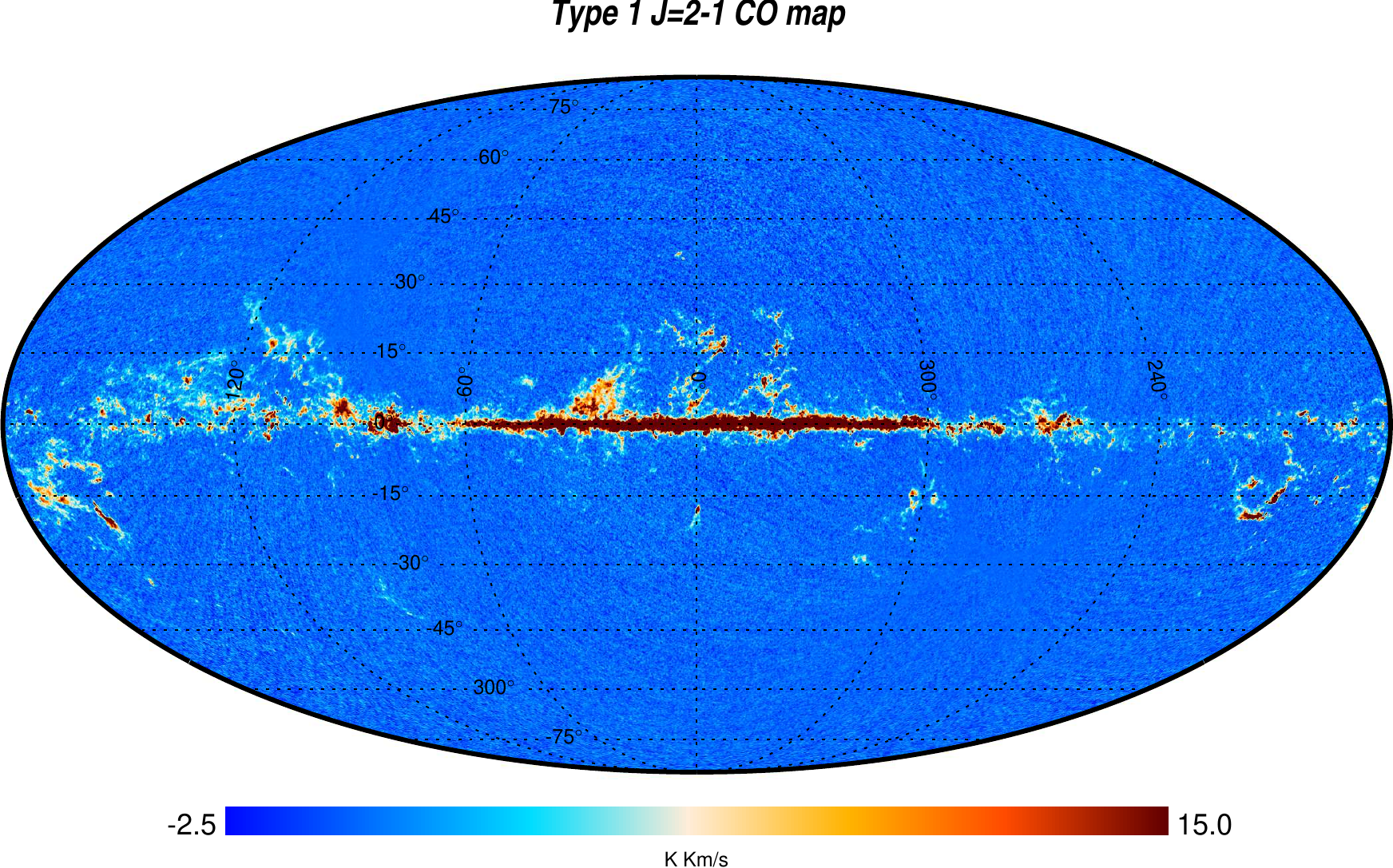}
\includegraphics[width=\columnwidth]{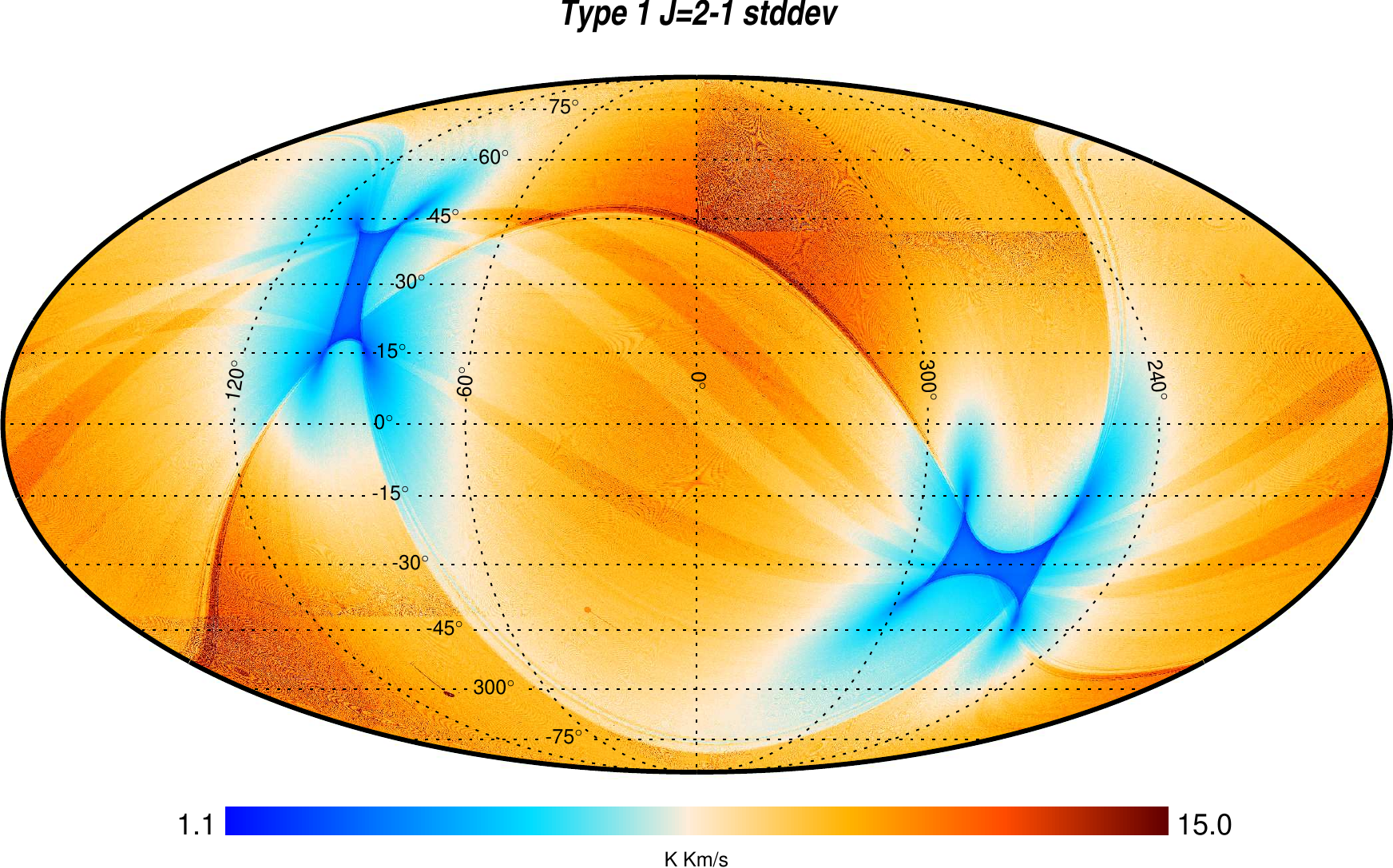}
\includegraphics[width=\columnwidth]{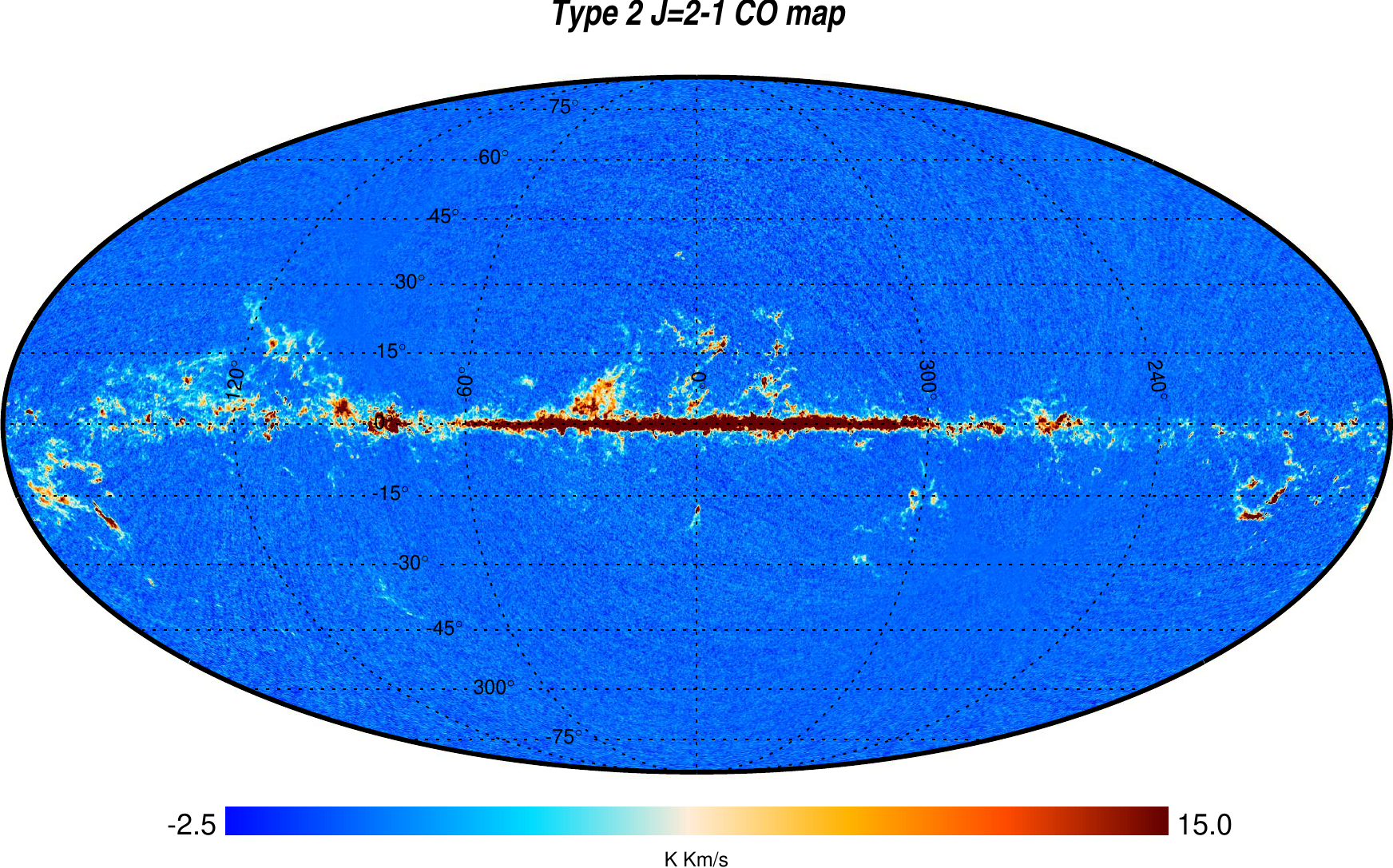}
\includegraphics[width=\columnwidth]{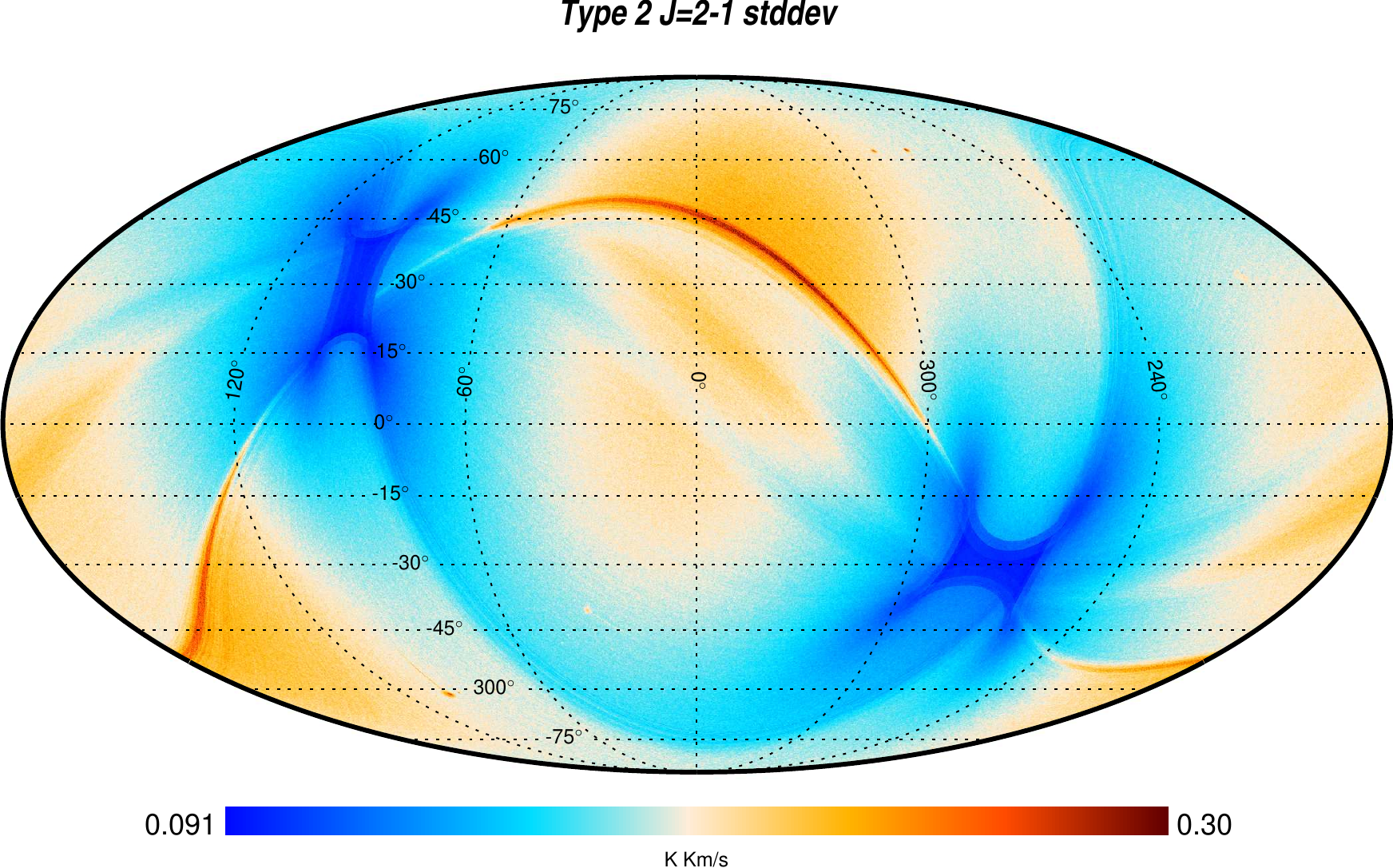}
\caption{Full-sky CO \jto\ maps (left) and their respective statistical
  error maps (right) for \typeone\ (top) and \typetwo\ (bottom) \Planck\ CO products. The \typeone\ map
  has been smoothed to 15\arcm\ for display purposes but the error
  map corresponds to the non smoothed product. The maps are in Galactic coordinates and
follow the {\tt HEALPix} pixelization scheme for $N_{\mathrm{side}}=2048$. 
  \label{fig:fullskyCO21}}
\end {figure*}

\begin{figure*}[!th]
\centering
\includegraphics[width=\columnwidth]{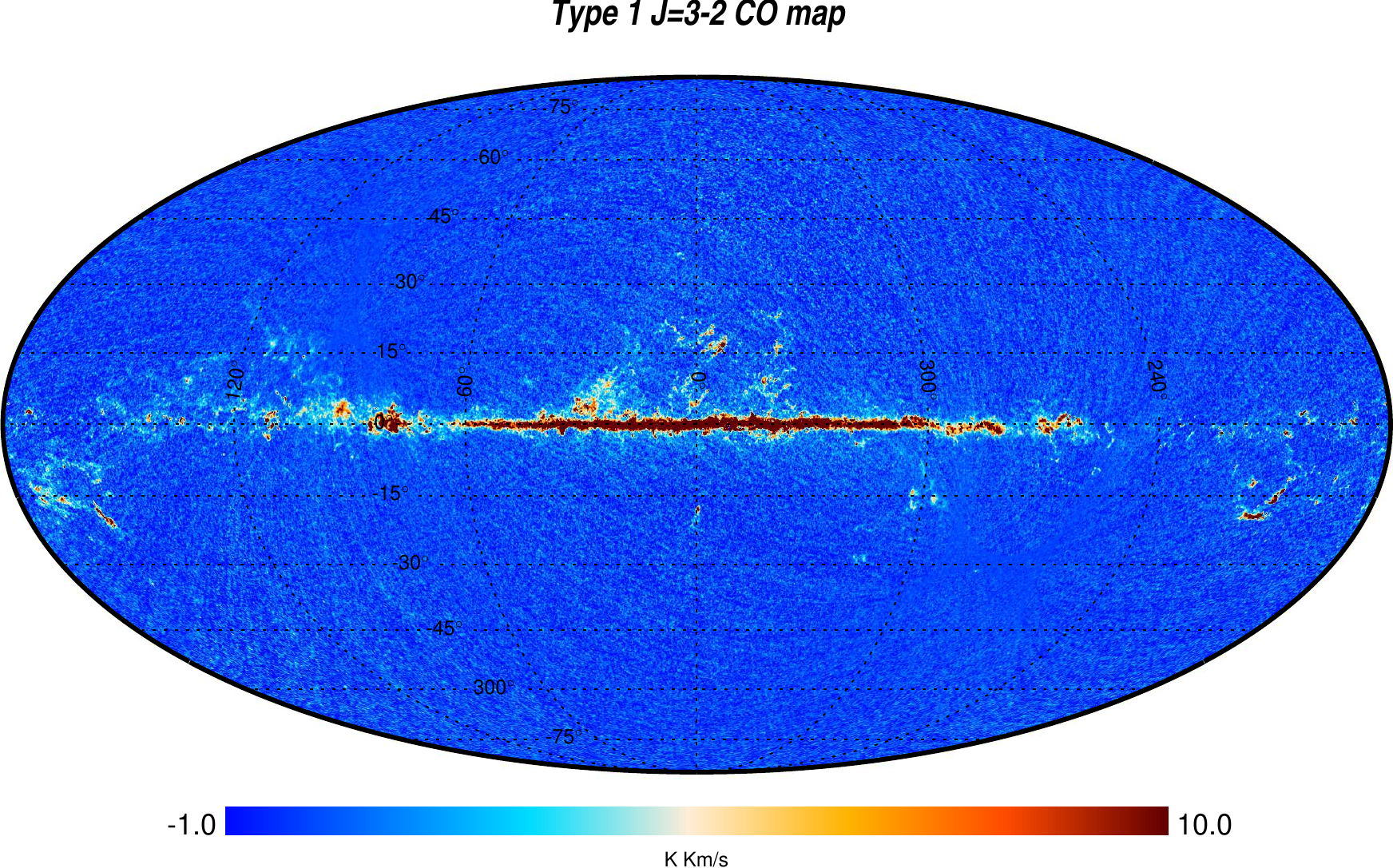}
\includegraphics[width=\columnwidth]{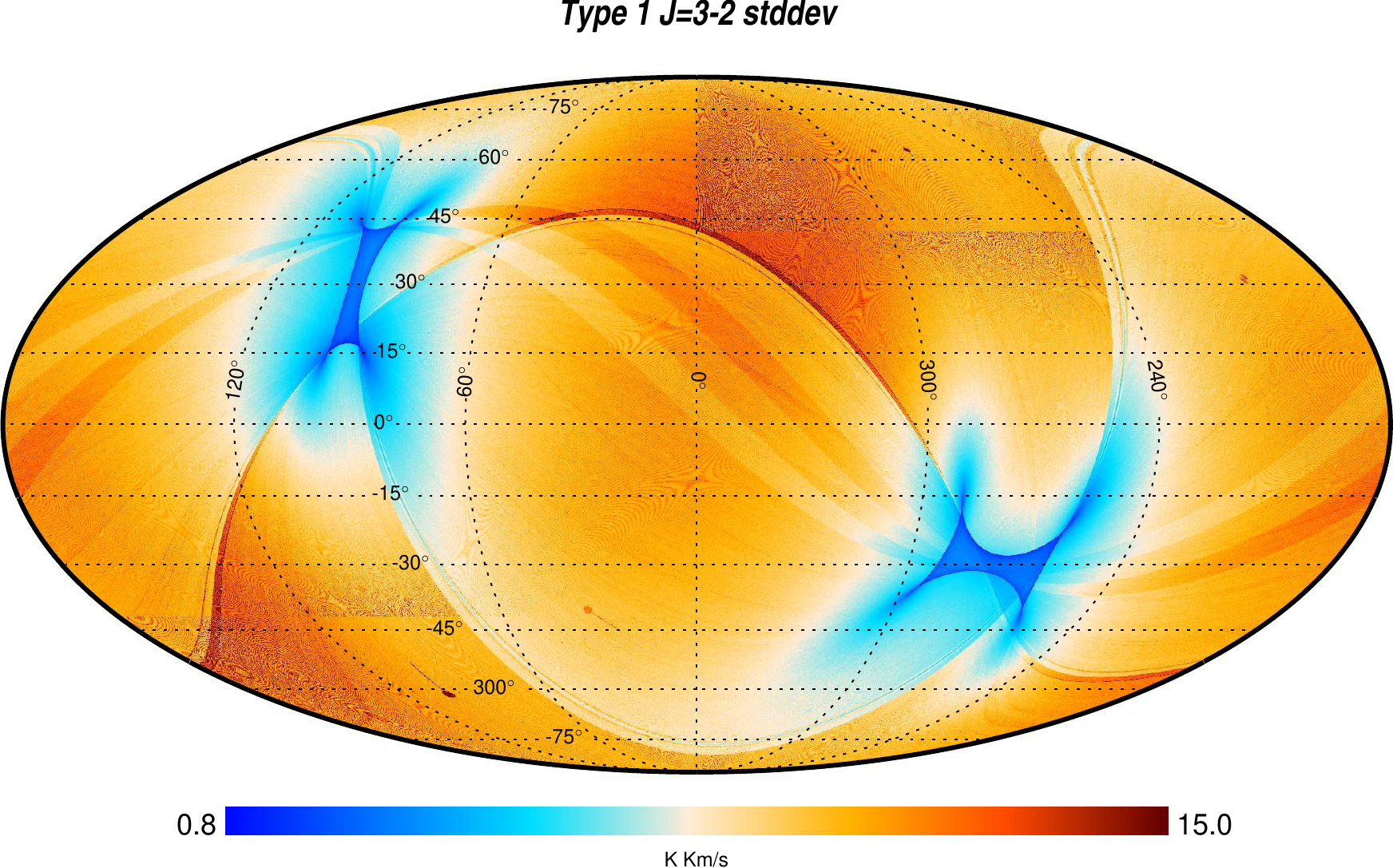}
\caption{\typeone\ full-sky CO \jtt\ maps (left) and its statistical error
  map (right). The map has been smoothed to 15\arcm\ for display purpose but the error
  map corresponds to the non smoothed product. The maps are in Galactic coordinates and
follow the {\tt HEALPix} pixelization scheme for $N_{\mathrm{side}}=2048$. 
\label{fig:fullskyCO32}}
\end {figure*}

\section{Extracting CO from \textit{Planck} data}
\label{sec:co_extract}


To extract the CO emission from the \Planck\ maps, three main approaches have been
considered: i) a single channel analysis (Sect.~\ref{subsec:multibolo}), ii)
a  multi-channel approach (Sect.~\ref{subsec:multifreq}) and finally iii) a multi-line
approach (Sect.~\ref{subsec:combined}) using fixed CO line ratios. While the
two first methods allow us to reconstruct specific 
CO transition lines, the third one does not discriminate between CO
transitions, but yields the best signal-to-noise ratio. 
Before detailing the specifics of these approaches, we present
in Sect.~\ref{subsec:compsep} the various component separation methods
that have been used and adapted to the specifics of CO observations with the \Planck\ satellite. 

\begin{table*}[tmb]
\begingroup
\newdimen\tblskip \tblskip=5pt
\caption{Main characteristics of the \Planck\ CO products.\label{tab:prod_summary}}                         
\nointerlineskip
\vskip -6mm
\footnotesize
\setbox\tablebox=\vbox{
   \newdimen\digitwidth 
   \setbox0=\hbox{\rm 0} 
   \digitwidth=\wd0 
   \catcode`*=\active 
   \def*{\kern\digitwidth}
   \newdimen\dpwidth 
   \setbox0=\hbox{.} 
   \dpwidth=\wd0 
   \catcode`!=\active 
   \def!{\kern\dpwidth}
\halign{\hbox to 1.5cm{\hfil#\hfil}\tabskip 2em&
     \hfil#\hfil \tabskip 1em&
     \hfil#\hfil \tabskip 1em&
     \hfil#\hfil \tabskip 1em&
     \hfil#\hfil \tabskip 1em&
     \hfil#\hfil \tabskip 1em&
     \hfil#\hfil \tabskip 1em&
     \hfil#\hfil \tabskip 0em\cr 
\noalign{\doubleline}
\omit  Type & Line & Resolution & Noise / 15\arcm pixel & $N_{\rm side}$ &Method & Components considered & Data used \cr
     &     &      [\arcm]     &    [K\,km\,s$^{-1}$]&    &  &  & Freq [GHz] \cr
\noalign{\vskip 3pt\hrule\vskip 5pt}
\noalign{\vskip 2pt}
\omit \typeone &\cooz & 9.65 & 1.77 & 2048 & {\tt MILCA} & CO, CMB  & 100 (bolo. maps) \cr
\omit \typeone &\coto & 4.99 & 0.74 & 2048 & {\tt MILCA} & CO, CMB, dust & 217 (bolo. maps) \cr
\omit \typeone &\cott & 4.82 & 0.73 &2048 & {\tt MILCA} & CO, dust & 353 (bolo. maps) \cr
 \noalign{\vskip 4pt} 
\omit \typetwo &\cooz & 15 & 0.45 & 2048 & {\tt Ruler} & CO, CMB, dust, free-free & 70, 100,143,353 \cr
\omit \typetwo &\coto & 15 & 0.12 & 2048 & {\tt Ruler} & CO, CMB, dust, free-free & 70,143,217,353 \cr
\noalign{\vskip 4pt}
\omit \typethree & \dots & 5.5 & 0.16& 2048 & {\tt Commander-Ruler} & CO, CMB, dust,  & 30-857 \cr
                                                                                                   &  &  &  &   &   & power-law sync/free-free  & \cr
\noalign{\vskip 3pt\hrule\vskip 5pt}
}
}
\endPlancktable 
\endgroup
\end{table*}

\subsection{Component separation methods}
\label{subsec:compsep}

Component separation algorithms as presented in~\citet{planck2013-p06}
are mainly specifically tailored for CMB extraction and its statistical analysis.
Thus, for this paper we have adapted and tested several algorithms for CO extraction
from which we have selected two. Here we give a
brief overview of them and of their main characteristics, while a more
detailed description can be found in the references given below.

\subsubsection{{\tt MILCA}}
The {\tt{MILCA}} (Modified Internal Linear
Combination Algorithm) method  \citep{thesegh} was specifically developed within the \Planck\ collaboration 
or the reconstruction of thermal Sunyaev-Zeldovich effect and CO contributions on the \Planck\ maps.
{\tt MILCA} is an extension of the standard ILC algorithm originally
aiming at CMB extraction \citep{2003ApJS..148...97B,2004ApJ...612..633E}. {\tt{MILCA}} 
provides a flexible way of selecting wanted and un-wanted spectral components \citep{milca, thesegh}
and corrects for the noise bias in standard ILC algorithms.
Furthermore, the weights of the internal linear combination can be computed both in real and harmonic
space to improve the component separation efficiency.
{\tt MILCA} was tested against CO-oriented simulations, 
showing no bias in the reconstruction when the CO
conversion coefficients are perfectly known (see Appendix~\ref{sec:simuvalidation}).

\subsubsection{\tt Ruler}
 {\tt Ruler} is an inversion algorithm, which provides the generalized
 least squares solution (GLSS), given a parametric model of the
 Galactic emission. In a nutshell, input data, either bolometer maps
 or frequency maps, are linearly combined, on a pixel-by-pixel basis,
 according to weights that account for both the foreground spectral
 properties and the \Planck\ channel specifications, namely
 instrumental noise and bandpass. A comprehensive description of the
 method is given in \citet{planck2013-p06}. Validation on simulations
 are presented in Appendix~\ref{sec:simuvalidation}.

Note that {\tt Ruler} and
 {\tt MILCA} have both been tested in the single- and multi-channel
 configurations described in Sect.~\ref{subsec:extract_methods} and were
 found to produce compatible results. This brings confidence in the
 robustness of these independent algorithms.
 
\subsubsection{\tt Commander-Ruler}
The {\tt Commander-Ruler} component separation pipeline consists of two
steps: i) the \Planck\ channel maps are brought to a common resolution
and the likelihoods of the non linear degrees of freedom of a chosen
parametric foreground model are jointly sampled, through a Gibbs
sampling Monte-Carlo algorithm \citep[{\tt Commander},][]{Eriksen2008ApJ676}; ii) for
each sample of the derived distribution, the linear degrees of
freedom, namely the amplitude maps of the components, are computed at
the full \Planck\ resolution via a generalized least squared solution
({\tt Ruler}, described above). The first and second moments of the
posterior distribution of a parameter define its mean value and
uncertainty, which accounts for both the instrumental and foreground
modelling error.
{\tt Commander-Ruler} has been extensively tested using the {\tt FFP6} simulations \citep[see][]{planck2013-p28} as part of \Planck\
component separation, a thorough description of which is provided in \citep{planck2013-p06}.

\subsection{CO extraction strategies}
\label{subsec:extract_methods}
\subsubsection{Single-channel approach -- \typeone\ maps}
\label{subsec:multibolo}
In the single-channel approach we exploit differences in the spectral
transmission of a given CO line among the bolometers (pairs of bolometers) of the same frequency channel.
The main advantages of this solution are to give access to the first
three transitions of CO at the native resolution of the \Planck\ maps and to avoid
contamination from other channels. However, this type of CO extraction results in
a lower signal-to-noise ratio due to the use of individual
bolometer (pair of bolometers) maps.

The CO $J$=1$\rightarrow$0, $J$=2$\rightarrow$1 and
$J$=3$\rightarrow$2 maps are obtained with {\tt MILCA}, using all
bolometers in the 100, 217 and 353~GHz channel, respectively, and
are denoted the \Planck\ \typeone\ CO product.
Different constraints are applied depending on
the line under scrutiny:
\begin{itemize}
\item The $J$=1$\rightarrow$0 map is
  obtained by requesting cancellation of a flat spectrum
  (i.e. the CMB) while preserving CO (using the sky-calibrated
  conversion coefficients given in Table~\ref{tab:CO}). 
  Notice that in this case the diffuse Galactic foreground contamination
  can be well approximated by a flat spectrum within the band
  and there is no need for extra constraints.
\item At 217 GHz, dust becomes more of a major foreground and should
  be dealt with. The $J$=2$\rightarrow$1 map is therefore extracted using an
  extra constraint on the dust transmission in the different
  bolometers, computed from a grey-body spectrum defined as $I_{\nu}^{\rm dust}\propto
  \nu^\beta B_\nu(T_d)$ where $T=17K$, $\beta=1.6$ and $B_\nu(T)$ is
  the Planck function. The choice of this dust spectrum is discussed
  in Sect.~\ref{subsubsec:dust}.
\item At 353~GHz, the CMB becomes sub-dominant compared to the dust
  and we only request keeping the CO and removing the dust. We also
  found that the dust spectrum used for the $J$=2$\rightarrow$1 line is not optimal
  at 353~GHz and thus we used the sky-based dust
  conversion coefficients $F_{\rm dust}^{\rm sky}$ which were also
  fitted in the linear regression of Eq.(\ref{eq:skycoeff}) and for which
  we obtained better results. 
\end{itemize}

\subsubsection{Multi-channel approach -- \typetwo\ maps}
\label{subsec:multifreq}
 The multi-channel approach makes use of the intensity maps in
 several channels to isolate the CO contribution from the transmission
 of CO, CMB, dust and free-free emission in the different
 channels. The {\tt Ruler} method was selected to be used in this
 configuration, to extract the \joz\ and \jto\ lines. 

In order to construct
 the \cooz\ line, we use LFI 70~GHz, HFI-100, 143 and 353~GHz channels,
 while \coto\ is obtained using LFI 70~GHz, 143, 217 and 353~GHz
channels. All maps are smoothed to 15\arcm. To solve for the two CO transitions, we assume that the CO contribution to the 353~GHz channel
is negligible\footnote{While this assumption eases the CO extraction, it also
leads to an overall systematic calibration error that can be corrected
for afterwards (see Sect.~\ref{subsubsec:foregrounds}).}. The \cooz\ and \coto\ maps
obtained using the multi-channel method constitute the
\Planck\ \typetwo\ CO product.

For both lines, the requirement is to extract CO while fitting at the same time for 
CMB, dust and free-free emission. By construction, the CMB spectrum is flat
across channels. For the dust, the same grey-body spectrum as for the
single-channel $J$=2$\rightarrow$1 line is assumed. Free-free emission is modelled as a power-law
spectrum $\propto \nu^{-2.15}$. The dust and free-free transmission
coefficients are obtained by integrating their spectra over \Planck's
bandpasses.

\subsubsection{Multi-line approach -- \typethree\ map}
\label{subsec:combined}

While the multi-channel approach provides single line maps at relatively high
signal-to-noise ratio, it is possible to use a multi-line approach to increase even further the S/N ratio, potentially allowing the discovery of new faint
molecular clouds at high Galactic latitudes. For this, we need to
assume that the line ratios (\coto/\cooz\ and  \cott/\cooz) are
constant across the sky. This approach is used as part of
the {\tt Commander}-{\tt Ruler} pipeline employed for the component separation. 

The \typethree\ CO map results from a seven-band run of this pipeline, including
30 to 353 GHz channel maps, with the sky modelled as a
superposition of CMB, CO, dust -- treated as a modified blackbldy --
and a power law to describe the low-frequency Galactic emission. The
dust optical depth and the dust temperature, as well as the low-frequency
component spectral index are fitted at every pixel. Since the total number of parameters would exceed the number of
frequencies considered, the three CO line emissions are assumed to be
perfectly correlated. A single CO map is solved for, the so-called
\typethree\ map, using the average
CO bandpass transmission (see Sect.~\ref{subsec:coeff_BP}) in each channel, whereas the CO line ratios are given by
the posterior average of the distribution obtained from a dedicated
{\tt Commander} run on small bright CO regions (Taurus, Orion,
Polaris, etc.). The line ratios found for
\coto/\cooz\ and  \cott/\cooz\ are 0.595 and 0.297, respectively.
Notice that there are significant spatial variations on these ratios as discussed in Sect.~\ref{subsubsec:13co}.
For the analysis we have selected bright CO regions with large ratios.

\section{\textit{Planck} CO maps}
\label{sec:COmaps}
The \Planck\ CO delivery consists of three types of products,
corresponding to the three extraction methods described above: 
\begin{itemize}
\item \typeone\ maps are extracted using the single-channel approach
   described in Sect.~\ref{subsec:multibolo} and
  come at the native resolution of the corresponding \Planck-HFI channel;
\item \typetwo\ maps come from the multi-channel method given
  in Sect.~\ref{subsec:multifreq} and have a resolution of 15\arcm.;
\item The \typethree\ map comes from the multi-line approach described in
  Sect.~\ref{subsec:combined}, and note that it assumes fixed CO line
ratios in order to obtain the highest possible signal-to-noise ratio, with
is delivery at a resolution of $\sim 5\arcm.5$.
\end{itemize}
The main characteristics of all \Planck\ CO maps are gathered in
in Table~\ref{tab:prod_summary} while details of the
\Planck\ CO product is given in Appendix~\ref{app:coprod}.
The \typeone, \typetwo\ and \typethree\ terminology will be used throughout the
paper.
All maps discussed below are in units of
K$_{\rm RJ}\,{\rm km} \,{\rm s}^{-1}$ at the transition line they represent.

\subsection{The CO $J$=1$\rightarrow$0 line at 115~GHz}
The \typeone\ (top), \typetwo\ (middle) and \typethree\ (bottom)
\Planck\ \cooz\ maps are given in the left column of
Fig.~\ref{fig:fullskyCO10}. The right column corresponds their statistical 
error maps and will be discussed further in Sect.~\ref{sec:errors}.

The resolution of the \typeone\ product is 9.65\arcm, i.e. the native resolution of the
Planck-HFI 100~GHz channel. The \typetwo\ product has a 15\arcm\ beam
due to additional smoothing required to combine several \Planck\ channels: in particular the use of Planck-LFI 70~GHz channel is essential to remove the free-free emission (see Sect.~\ref{subsubsec:foregrounds}).  Finally, the \typethree\
has a varying resolution across the sky, resulting from the {\tt
  Ruler} solution, and a beam profile is computed from {\tt FFP6}
simulations. A good Gaussian approximation can be achieved with a
FWHM of 5.5\arcm. A simple eye inspection finds a good overall agreement between
the maps; an in-depth comparison is conducted in Sect.\ref{sec:co_char}.

\subsection{The CO $J$=2$\rightarrow$1 line at 230~GHz}

The \coto\ line can only be extracted using the single-channel or the
multi-channel approach. Therefore the 230~GHz \Planck\ CO product
consists of \typeone\ (5\arcm\ resolution) and \typetwo\ (15\arcm\ resolution) maps only, which are displayed in the
left column of Fig.~\ref{fig:fullskyCO21}. The dust emission
increases with frequency and becomes more of an issue for the $J$=2$\rightarrow$1
line extraction. This will be discussed at length in
Sect.~\ref{subsubsec:foregrounds}, but this issue is already clearly
visible when comparing the two products: the \typetwo\ CO presents a
diffuse dust emission throughout the Galactic plane that is not
present in the \typeone\ map.

\subsection{The CO $J$=3$\rightarrow$2 line at 345~GHz}
As mentioned previously, the single-channel approach is the only way
to extract this higher $J$ CO line and the corresponding \typeone\ CO
map is shown in Fig.~\ref{fig:fullskyCO32}. This map has a resolution of 4\arcm.82.

\section{Uncertainties and foregrounds}
\label{sec:errors}
\subsection{Statistical errors}
Statistical uncertainties in the CO maps may be obtained using so-called
half-ring differences\footnote{First and last ring sets are
  independent data sets built using the first and second half of
the stable pointing periods (see Sect.~\ref{sec:planckdata}).} \citep{2011A&A...536A...6P}. For the
\typeone\ product, standard deviation ($\sigma-$)maps are generated at the level of individual
bolometer maps, starting from the half-ring differences noise map that is 
whitened using the number of hits
in each pixels. A given bolometer $\sigma-$map, $\sigma_b$, is then obtained by
dividing the standard deviation of the noise map by the square root
of the number of hits in each pixel, namely
\begin{equation}
\sigma_b=\frac{1}{\sqrt{N_b}}{\rm stddev}\left[\frac{\left(M_b^F - M_b^L\right)\sqrt{N_b}}{2}\right]\,,
\end{equation}
where $N_b$ is the hit number map and $M_b^F$ and $M_b^L$ are the
first and last half-ring bolometer maps respectively. 
This standard deviation of each bolometer map $\sigma_b$ is then propagated
quadratically to the CO map level using the weights found by {\tt MILCA} for the linear
combination. 

This cannot be achieved in such a straightforward fashion for the
\typetwo\ maps that have been smoothed to 15\arcm, resulting in
correlated noise. Given the size of the maps, a fully dense matrix
description of the noise is not feasible. However, the set of 1,000
realistic noise simulations has been processed through the pipeline
(first smoothed and then linearly combined), and used to compute a 
$\sigma$-map, which is a good pixel noise approximation. The
\typethree\ standard deviation map is obtained in a similar fashion.

The $\sigma-$maps of all \Planck\ CO maps are plotted in the
right columns of Figs.~\ref{fig:fullskyCO10}, \ref{fig:fullskyCO21}
and \ref{fig:fullskyCO32}. We remind here that for the \typeone\ and
\typethree\ \cooz\ maps, the $\sigma-$maps correspond to the maps at their native resolutions,
while the CO maps have been smoothed to 15\arcm\ for display
purpose. These standard deviation maps have all been validated by checking
that their mean at high latitude was in agreement with standard
deviation measured directly in the CO map at these locations (where no
signal is expected, see lower panel of Fig.~\ref{fig:Planck_fluxes_on_cfa_nodetections}). 

Using a common resolution of 15\arcm\ to compare the
high-latitude noise level in the maps, we find the standard deviation of the
\cooz\ maps to be typically
1.77~K$_{\rm RJ}$\,km\,s$^{-1}$ for the
\typeone\ map, 0.45~K$_{\rm RJ}$\,km\,s$^{-1}$ for the \typetwo\ map and 0.16~K$_{\rm RJ}$\,km\,s$^{-1}$ for the
\typethree\ map. These uncertainties can be compared with typical uncertainties on
ground-based surveys for \cooz\, for example 0.6~K$_{\rm RJ}$\,km\,s$^{-1}$ for the 
and 1.2~K$_{\rm RJ}$\,km\,s$^{-1}$ for the \citet{2001ApJ...547..792D} and NANTEN \citep{2004ASPC..317...59M} surveys.
At the same resolution, the \typeone\ and \typetwo\ \coto\
maps have standard deviations of 0.74~K$_{\rm RJ}$\,km\,s$^{-1}$ and 0.12~K$_{\rm
  RJ}$\,km\,s$^{-1}$, respectively, while it is
0.73~K$_{\rm RJ}$\,km\,s$^{-1}$ \typeone\ \cott\ map.

\begin{figure}[!h]
\centering
\includegraphics[width=\columnwidth]{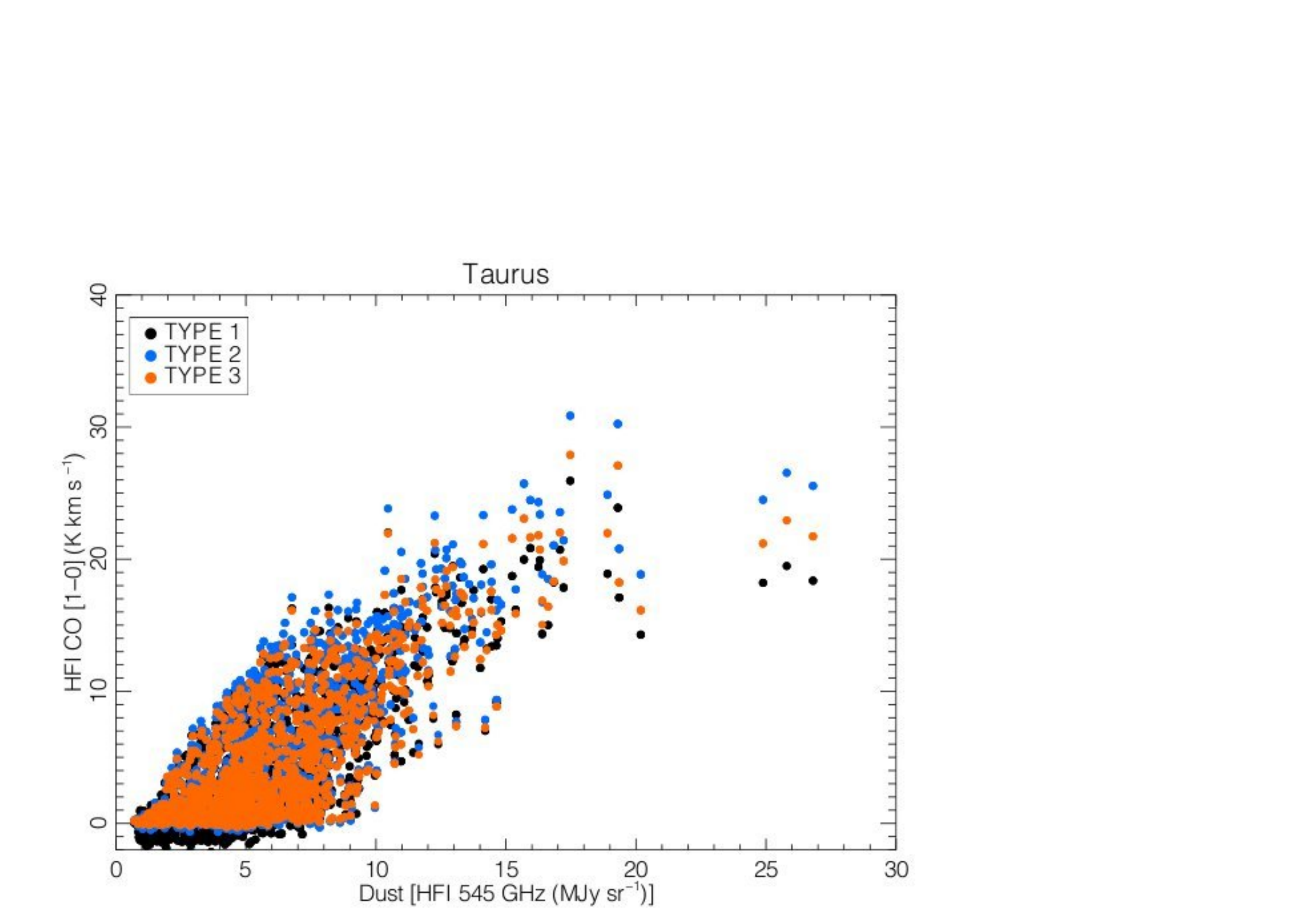}
\includegraphics[width=\columnwidth]{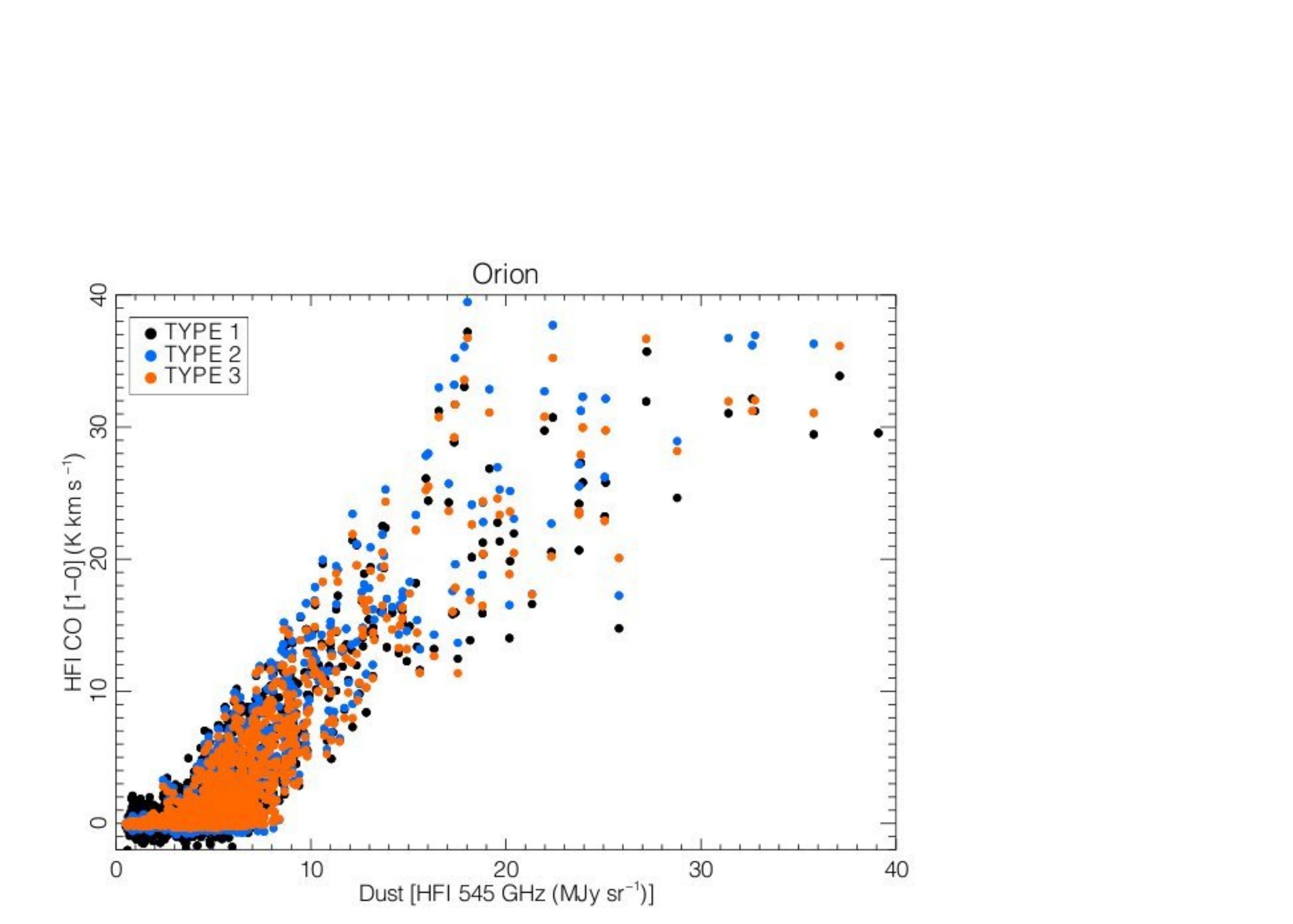}
\includegraphics[width=\columnwidth]{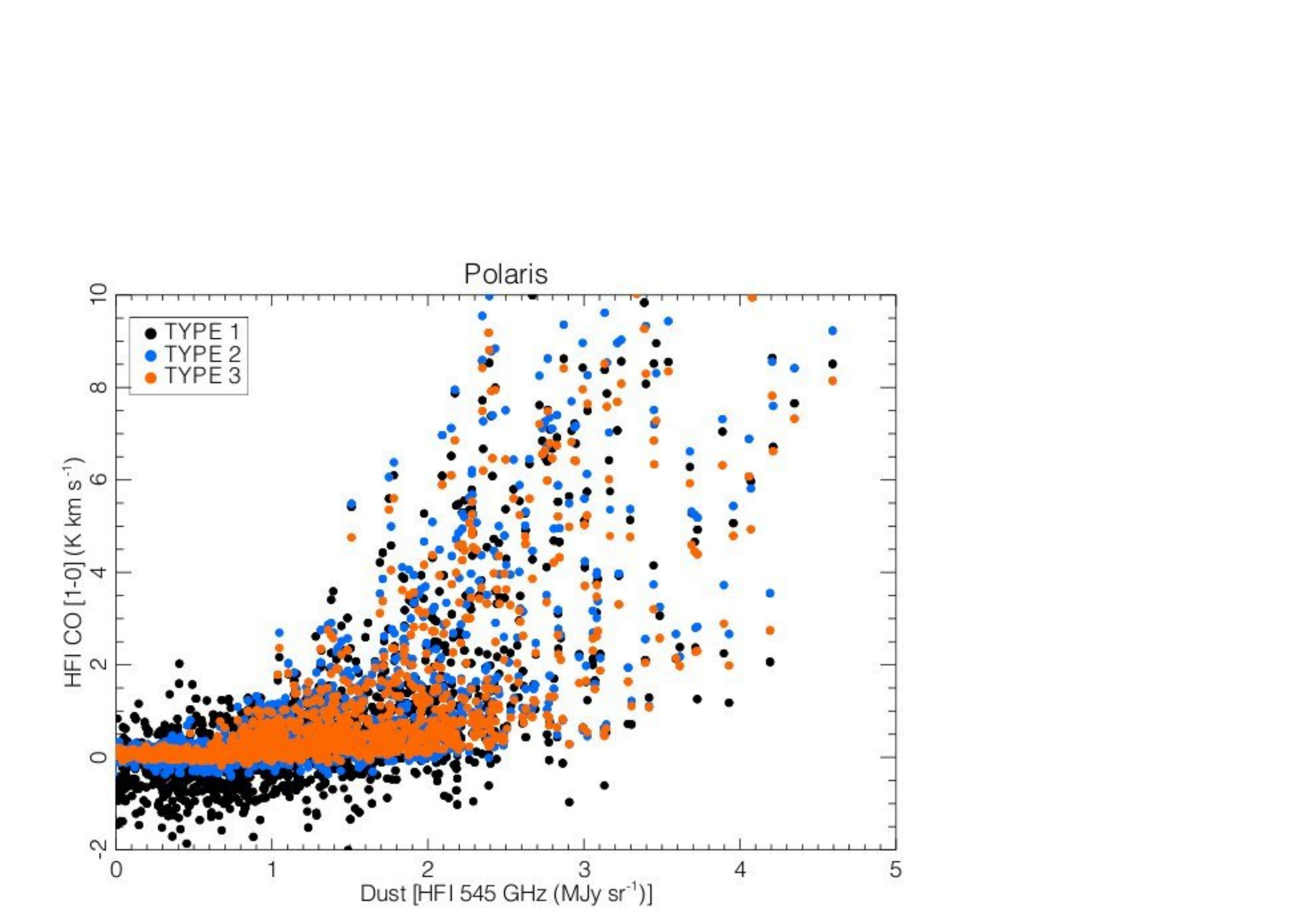}
\caption{Comparison of 3 types of CO product with a dust template map
  for the Taurus (top), Orion (middle) and
Polaris (bottom) molecular clouds.\label{fig:dust_correlation}}
\end{figure}

\subsection{Foregrounds}
\label{subsubsec:foregrounds}

All \Planck\ CO products suffer from systematic effects and foreground
contamination that need to be characterized; point sources or emission
from other CO lines affect all \Planck\ CO products while dust, free-free
emission and the Sunyaev Zeldovich effect (SZ) are important
for \typetwo\ and \typethree\ maps only (the \typeone\ product is the most
immune to contamination as it relies on single channel
information). Some of these foreground emissions (e.g., CMB, dust, free-free), need to be dealt with at the
stage of the component separation. Others, like point sources and the thermal SZ effect in clusters
of galaxies, may simply be masked afterwards. We discuss each of these foregrounds in detail below apart
from CMB emission for which the electromagnetic spectrum is well-known and can
be explicitly nullified within calibration errors in the \Planck\ frequency maps, as discussed in Sect.~\ref{subsec:compsep}.

\subsubsection{Dust}
\label{subsubsec:dust}

Dust emission is the main foreground as far as CO is concerned and is often
strong in the same regions, e.g. star forming regions. The
multi-channel approach (i.e. \typetwo\ products) is particularly sensitive to
the choice of the dust spectrum. One assumption we make in these
maps is that the dust spectrum is constant over the sky  and
described by a grey-body with $T_{\rm dust}=17$~K and $\beta_{\rm
  dust}=1.6$. This assumption is not correct since both the spectral index and temperature
of the dust are known to vary across the sky \citep[see for example][]{planck2011-7.0,2013AdAst2013E...3F}. These
values are, however, a good representation of the dust found in
CO-rich regions as can be seen on the $\beta_{\rm dust}$ and 
$T_{\rm  dust}$ maps of \citet{planck2013-p06}. 

First, let us remark that as far
as the dust transmission in HFI is concerned, $\beta_{\rm dust}$ and
$T_{\rm dust}$ are degenerate quantities. Testing different values of the
$\beta_{\rm dust}$ index\footnote{The temperature has little impact on the conversion
coefficients of the dust.}, we find this dust model to be the best
compromise for the multi-channel CO extraction. At $T_{\rm dust}=17$~K,
choosing $\beta_{\rm dust}=1.5$ results in
too much dust removal and a negative residual in the Galactic plane. Conversely,
$\beta_{\rm dust}=1.7$ generates a much larger remaining dust emission. In
the outer regions of the \typetwo\ maps ($l>90^\circ$ and
$l<270^\circ$), changing $\beta_{\rm dust}$ by $\pm 0.1$ yields an increase/decrease of $\lesssim 1$K$_{\rm RJ}$~km~s$^{-1}$ at 115~GHz and $\lesssim 2$
K$_{\rm RJ}$~km~s$^{-1}$ at 230~GHz. Still, these are larger than statistical errors. The inner regions of the Galactic plane
are far more sensitive to a change of the dust spectrum, with a shift of about 10~K~km~s$^{-1}$ at 115~GHz and 15~K~km~s$^{-1}$
at 230~GHz, but the signal is much larger.

\begin{figure*}
\centering
\includegraphics[width=0.60\columnwidth , trim=0cm 0cm 4cm 11cm,clip]{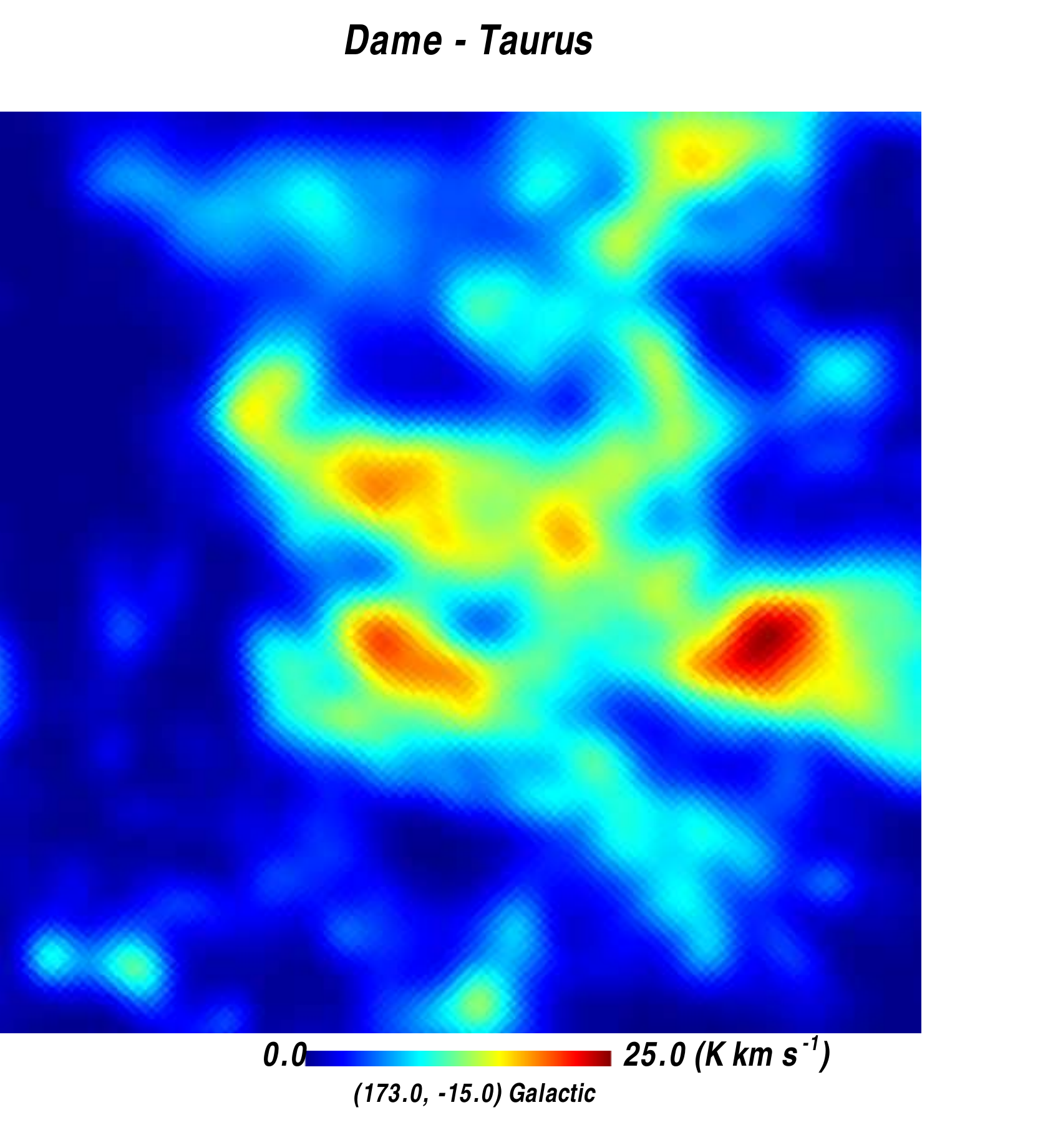}
\includegraphics[width=0.60\columnwidth ,trim=0cm 0cm 4cm 11cm,clip]{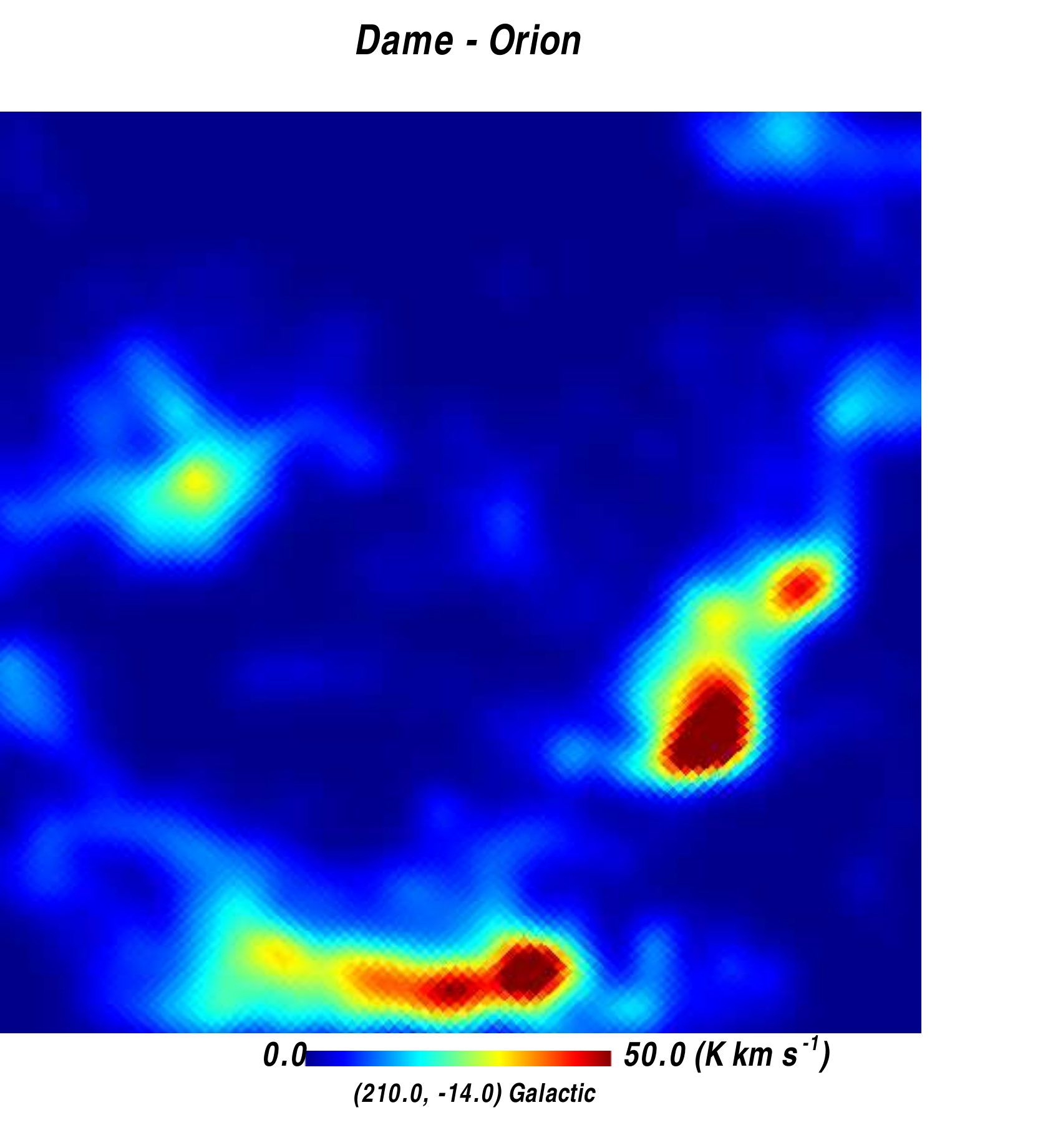}
\includegraphics[width=0.60\columnwidth ,trim=0cm 0cm 4cm 11cm,clip]{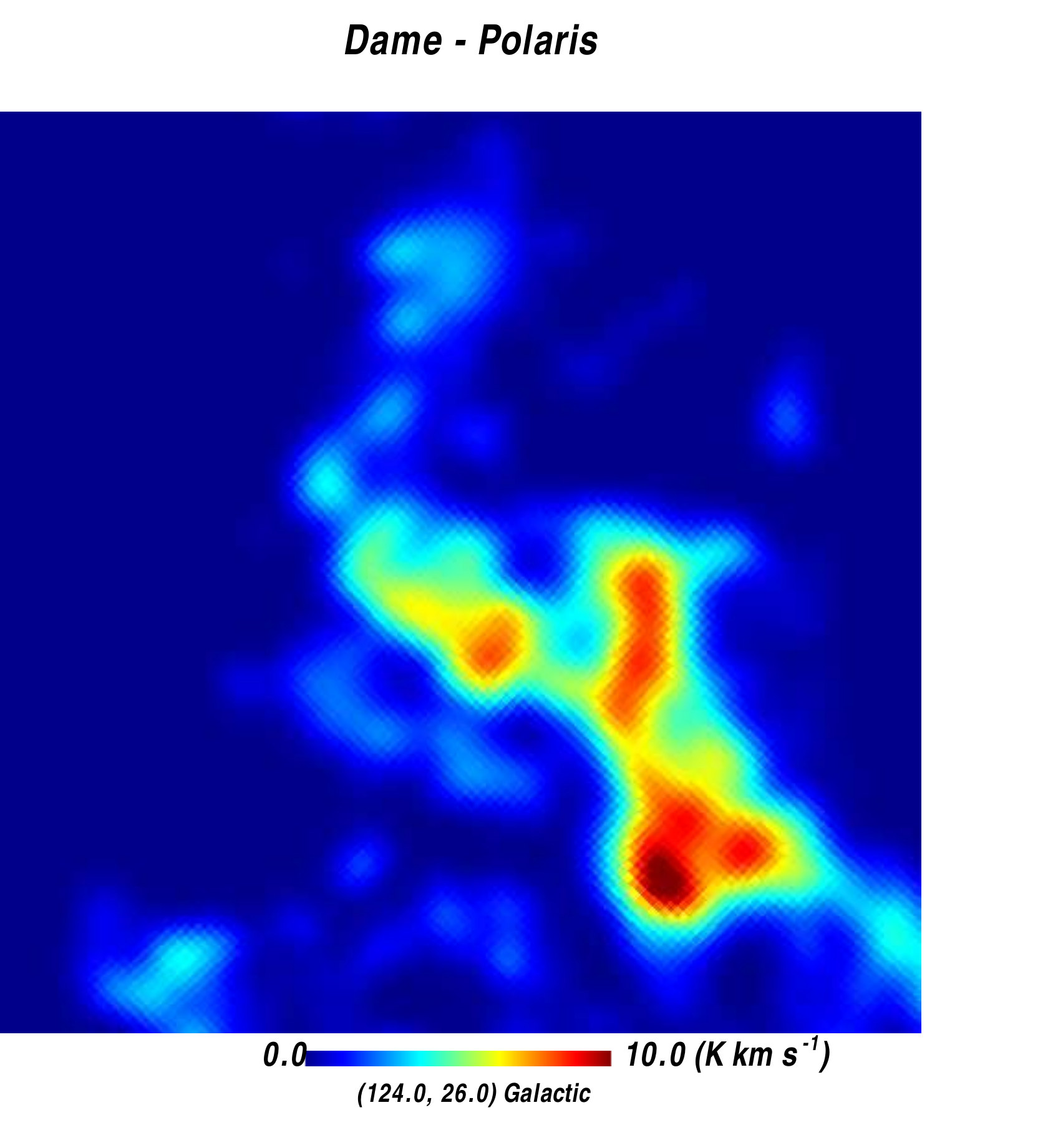}
\includegraphics[width=0.60\columnwidth ,trim=0cm 0cm 4cm 11cm,clip]{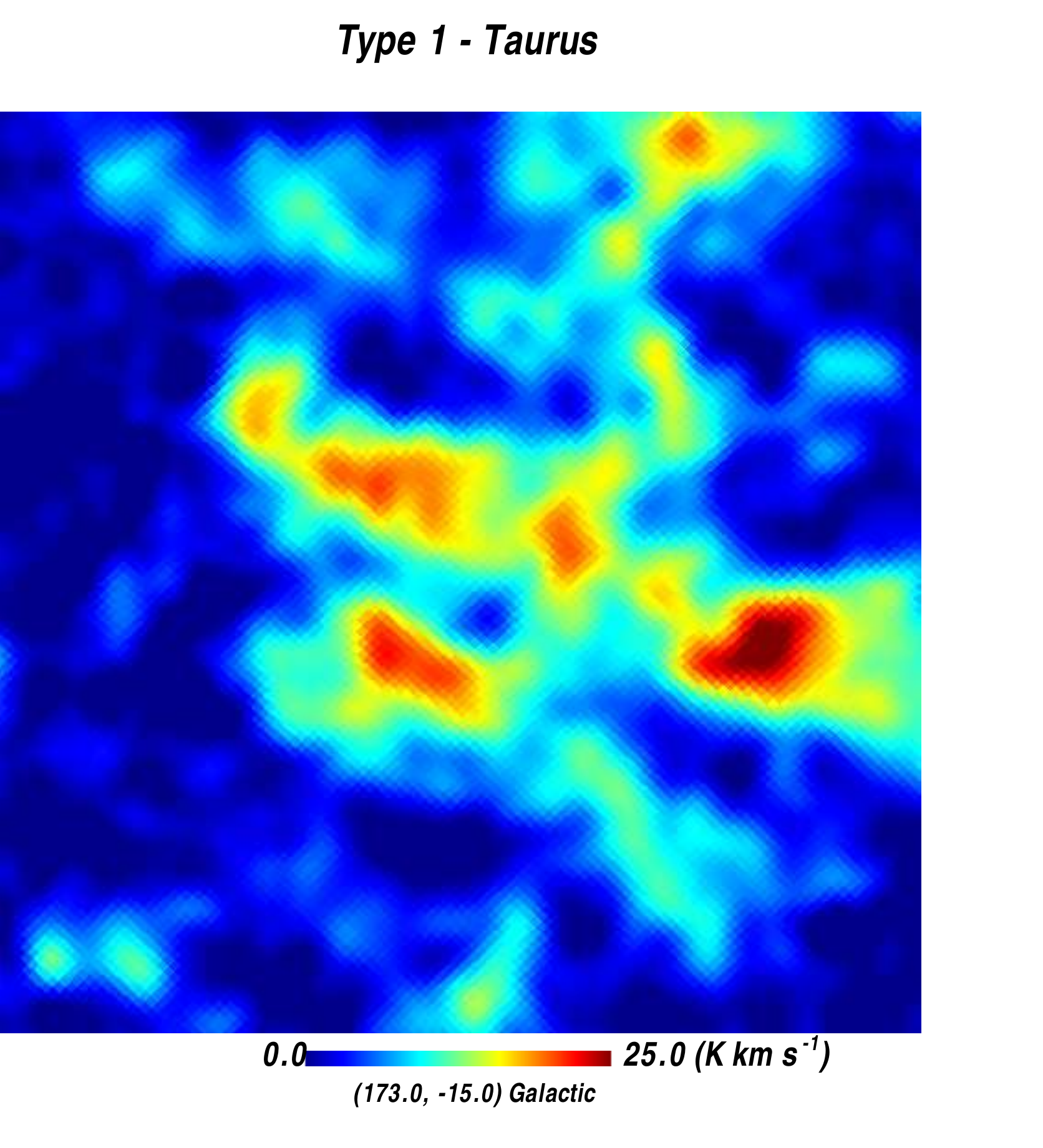}
\includegraphics[width=0.60\columnwidth ,trim=0cm 0cm 4cm 11cm,clip]{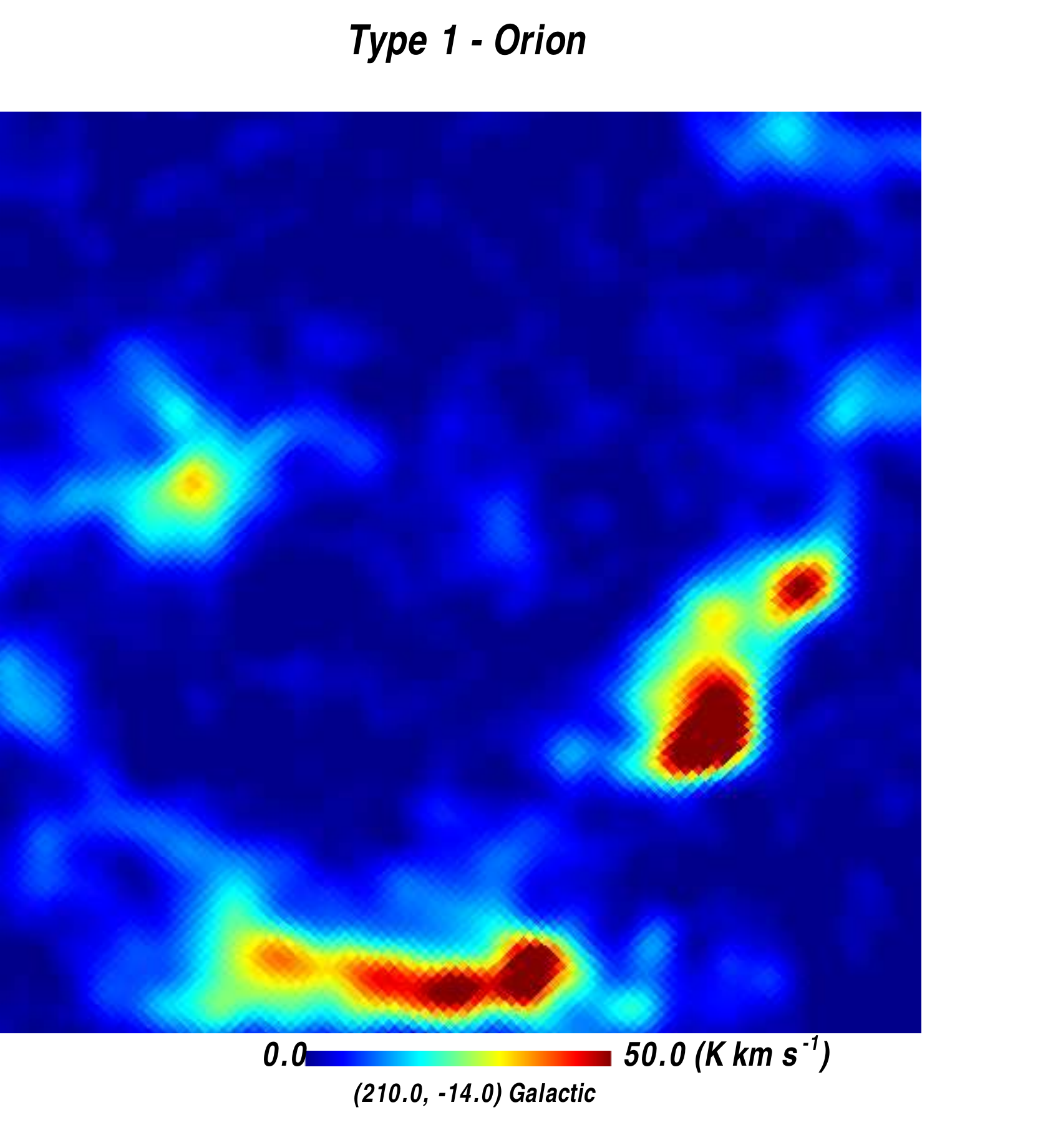}
\includegraphics[width=0.60\columnwidth ,trim=0cm 0cm 4cm 11cm,clip]{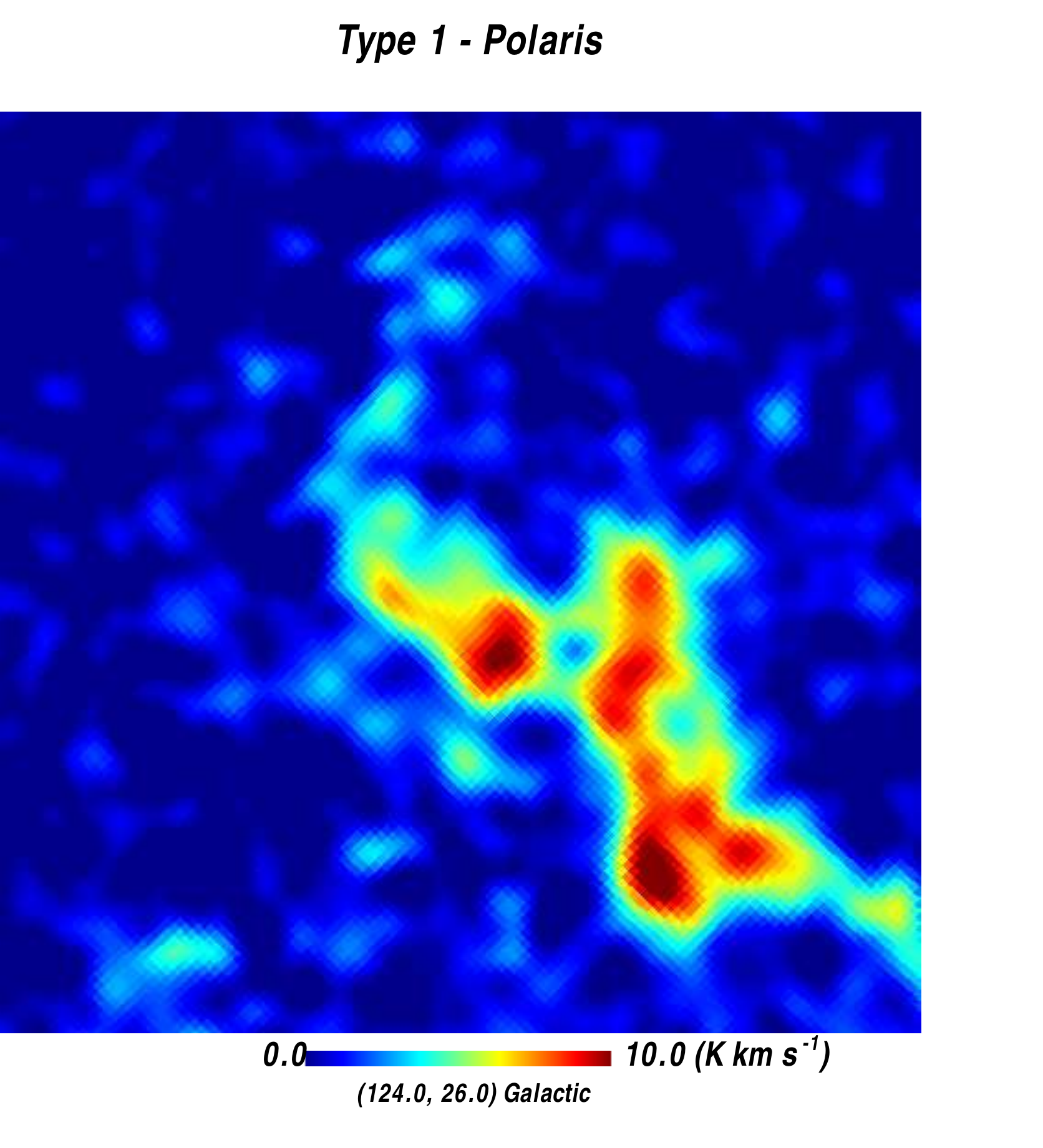}
\includegraphics[width=0.60\columnwidth ,trim=0cm 0cm 4cm 11cm,clip]{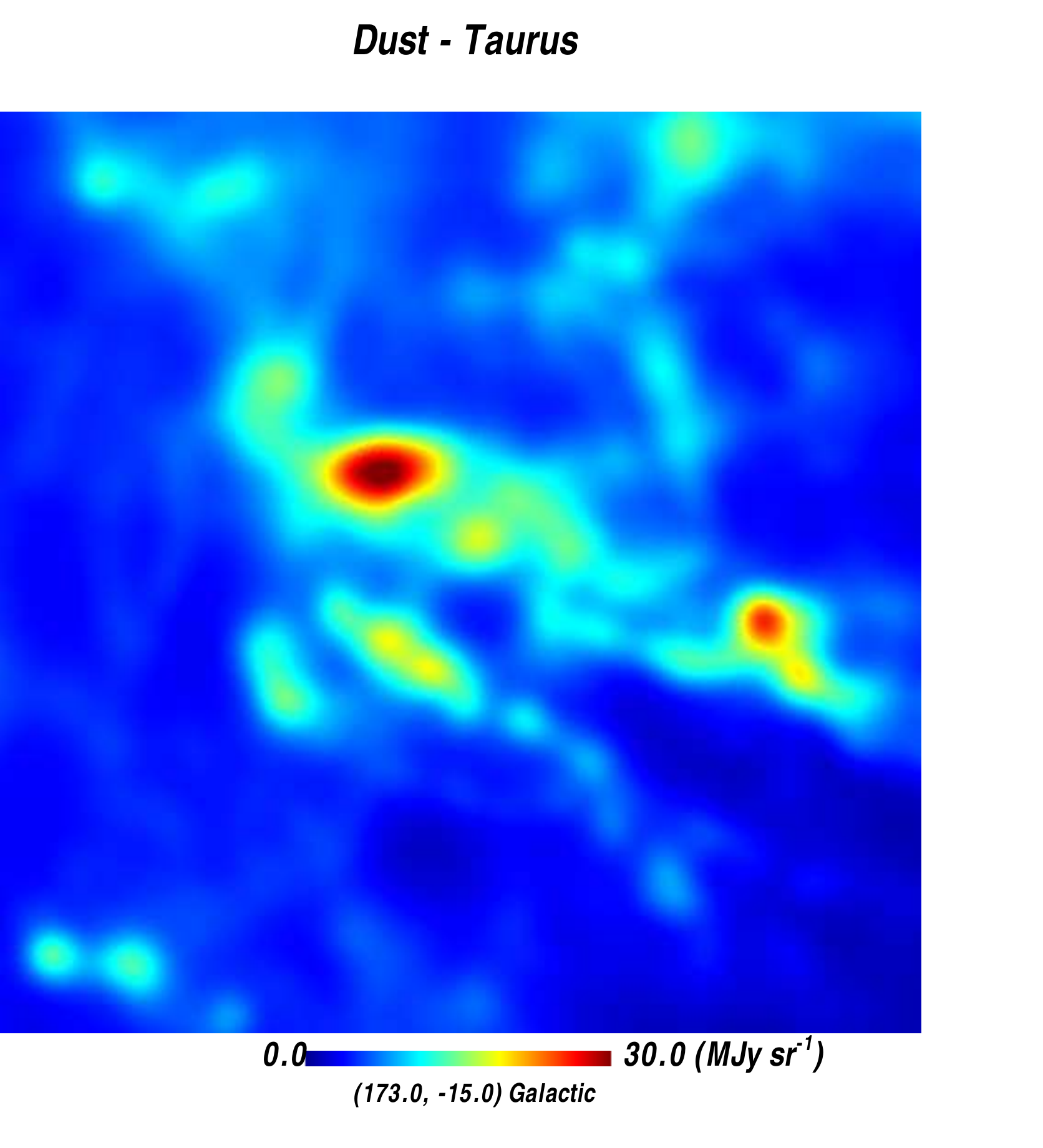}
\includegraphics[width=0.60\columnwidth ,trim=0cm 0cm 4cm 11cm,clip]{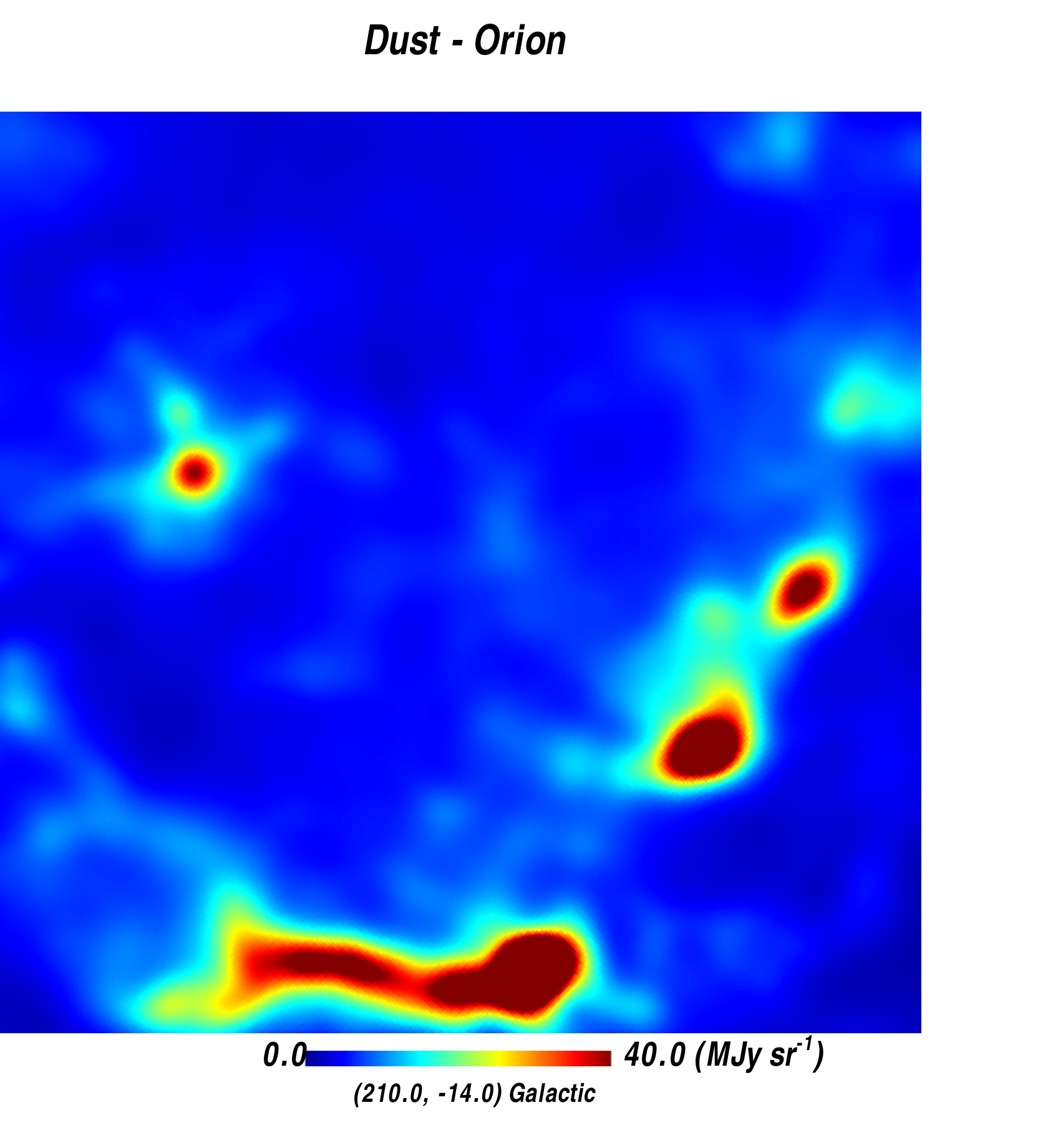}
\includegraphics[width=0.60\columnwidth, trim=0cm 0cm 4cm 11cm,clip]{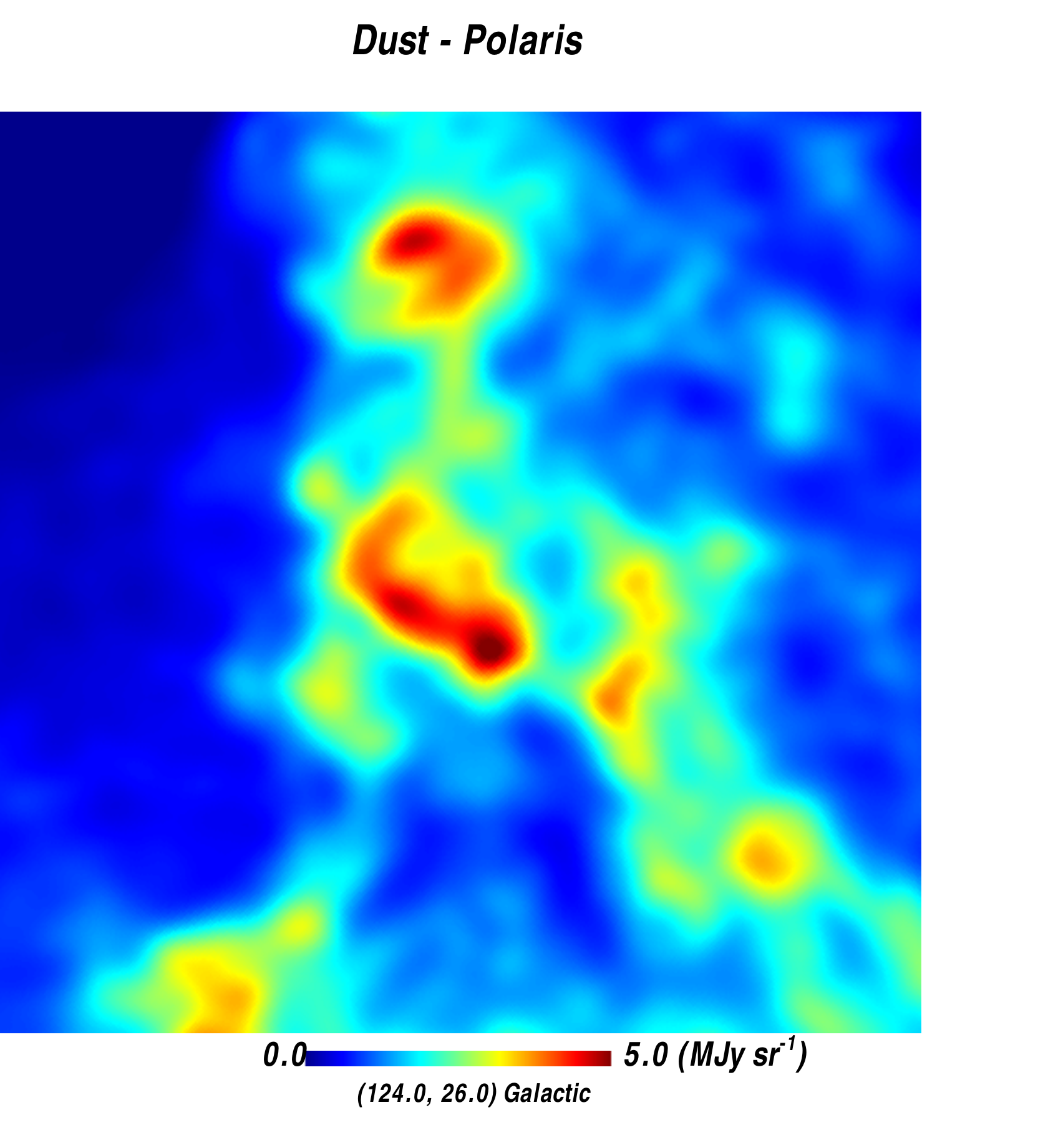}
\includegraphics[width=0.60\columnwidth ,trim=0cm 0cm 4cm 11cm,clip]{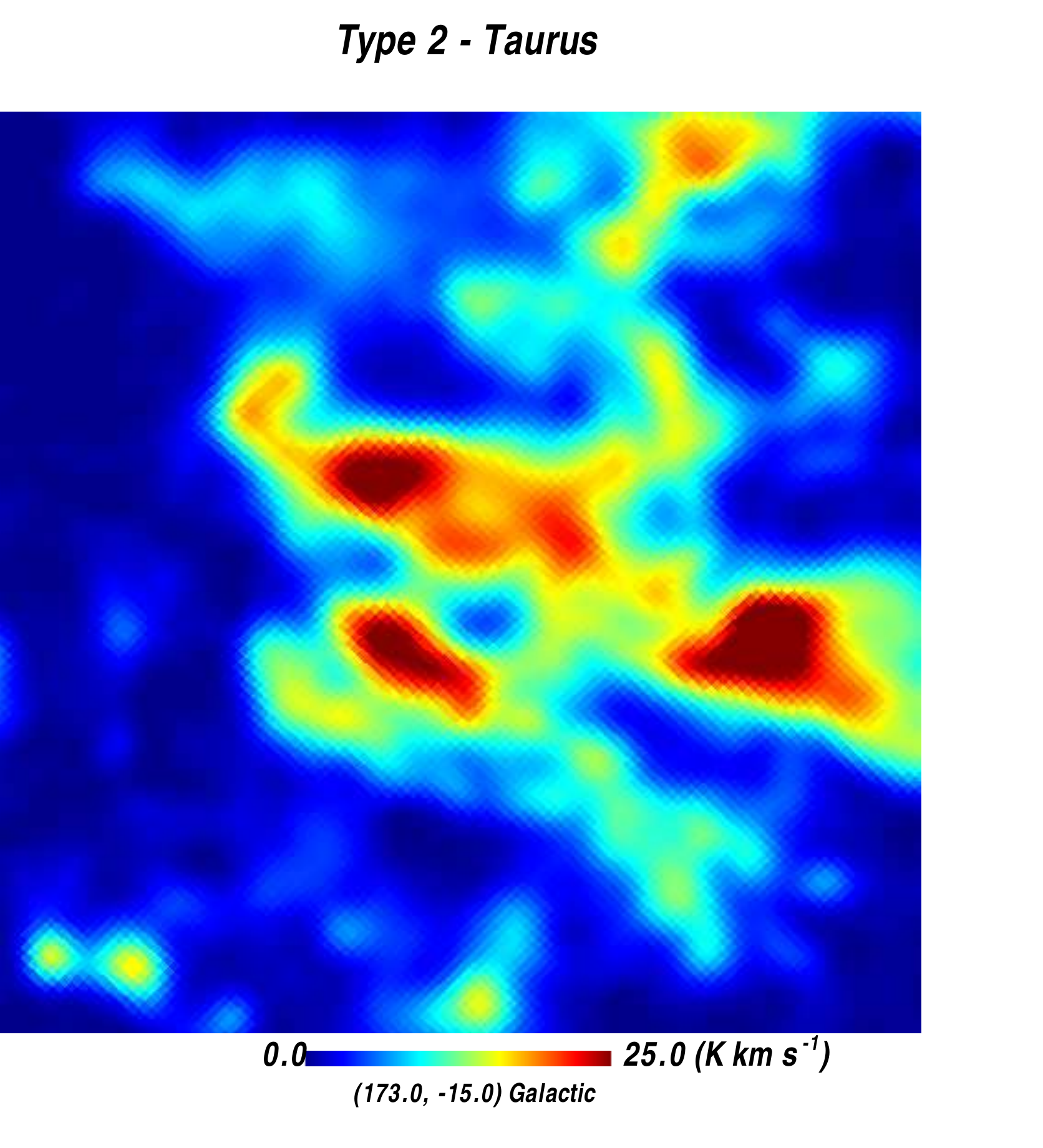}
\includegraphics[width=0.60\columnwidth ,trim=0cm 0cm 4cm  11cm,clip]{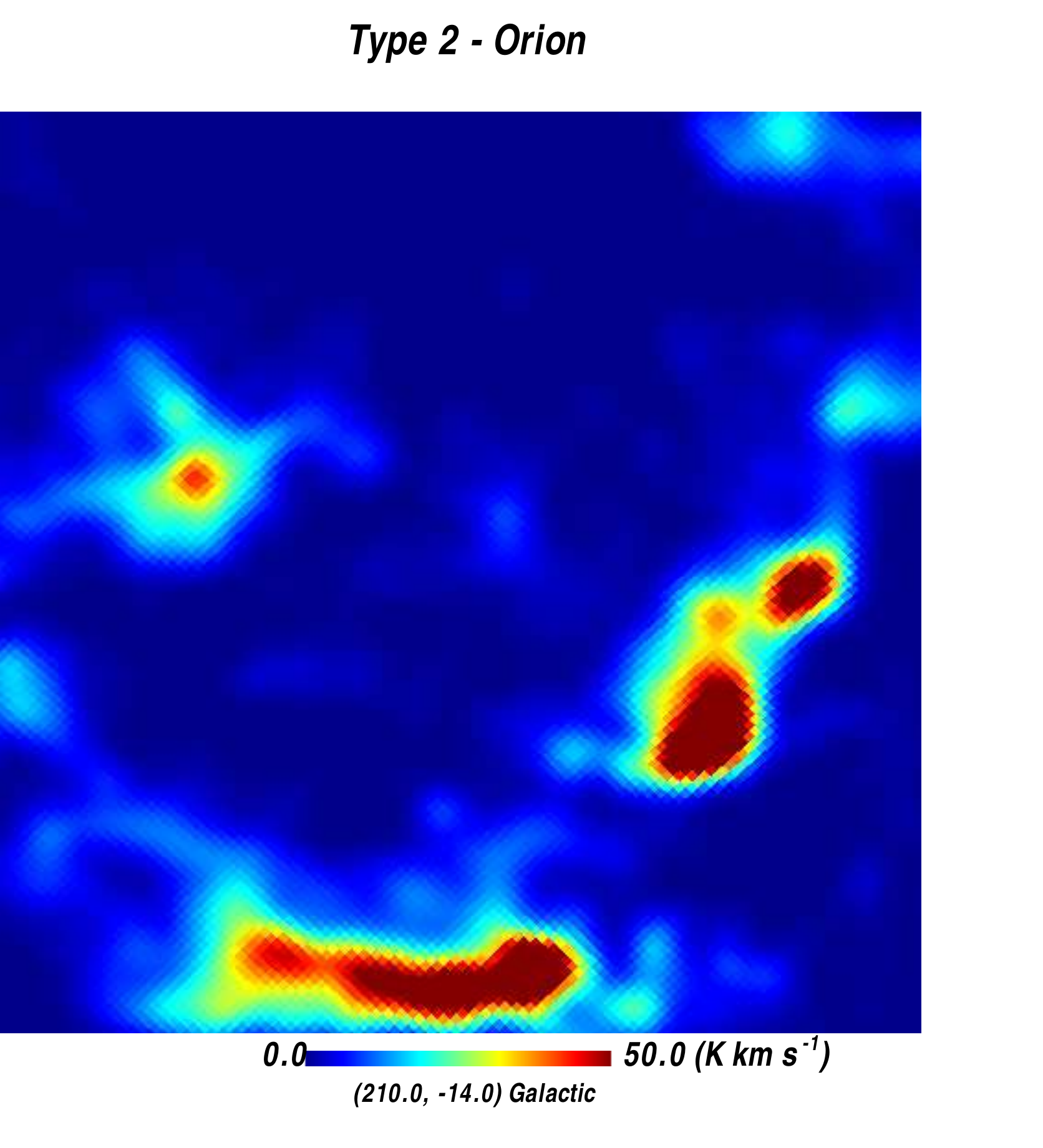}
\includegraphics[width=0.60\columnwidth ,trim=0cm 0cm 4cm 11cm,clip]{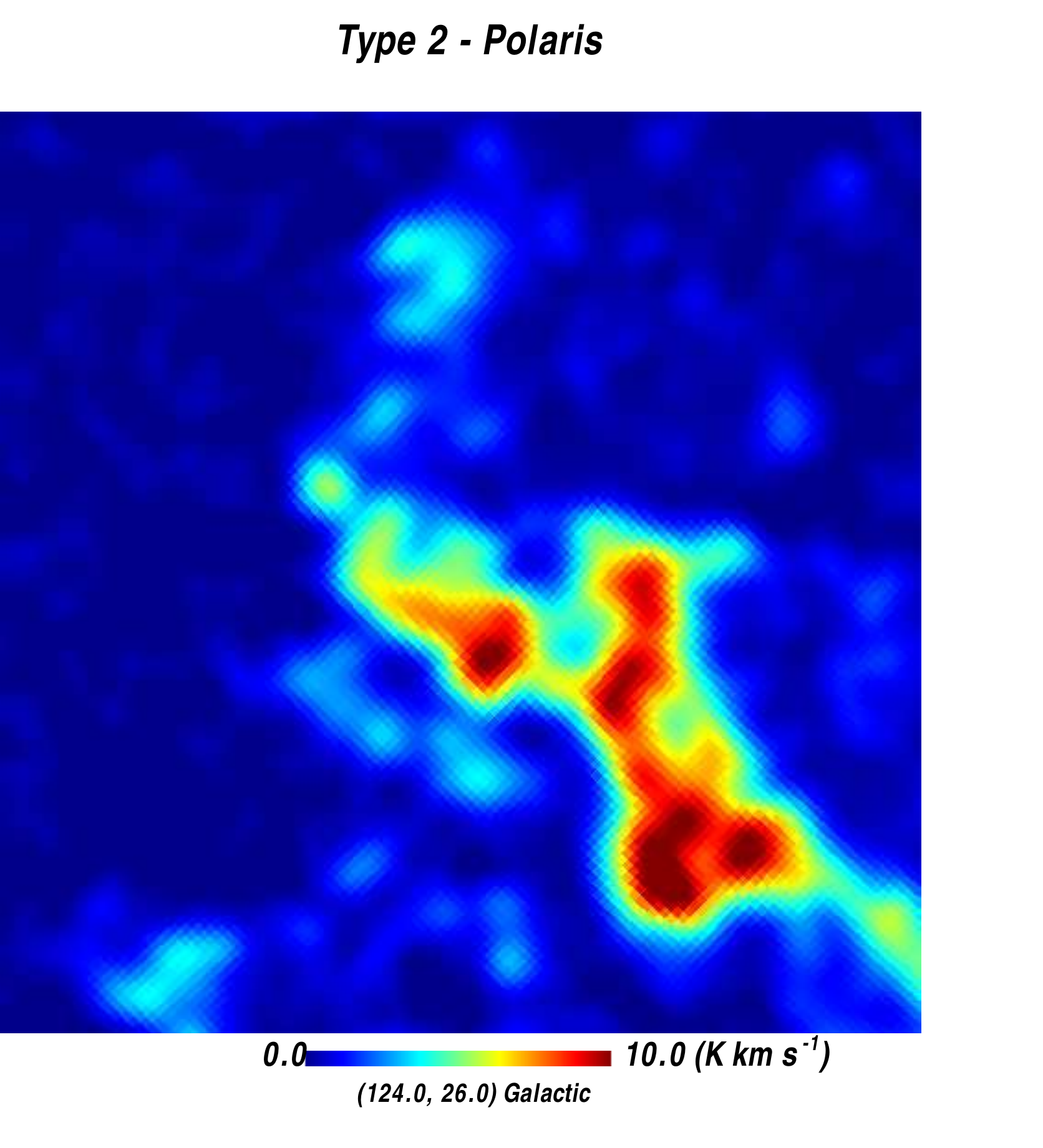}
\caption{Images ($12\deg.5 \times 12\deg.5$) of the Taurus (left column), Orion (middle column) and
  Polaris (right column) regions considered throughout the paper. The
  first row corresponds to the \citet{2001ApJ...547..792D} survey. The two
  middle rows show the \typeone\ and \typetwo\ $J$=1$\rightarrow$0 \Planck\ CO products. The \Planck-HFI 545~GHz channel map, used as a dust template, is shown in the last row. All maps are smoothed to 30\arcm.\label{fig:snapshots_regions}}
\end{figure*}

Given our chosen dust model, Fig.~\ref{fig:dust_correlation} shows the correlation between the
types of $J$=1$\rightarrow$0 CO maps discussed above and \Planck's
545~GHz map used here as dust template. All maps were smoothed to
30\arcm. The comparison  has been done in three
molecular clouds hosting different environments: i) Orion, a very
active massive star forming region; ii) Taurus, that hosts low-mass star
formation \citep{2008hsf1.book..405K}; and iii) Polaris, a high-Galactic latitude
translucent cloud with little to no star formation and presenting
both atomic and molecular gas \citep{2010A&A...518L.104M}. 
Figure~\ref{fig:snapshots_regions} presents smoothed images of the three
regions for the \typeone\ and \typetwo\ \Planck\ CO maps and for \Planck's
545~GHz channel. Snapshots of the \citet{2001ApJ...547..792D} $J$=1$\rightarrow$0 
 data are also shown in Fig.~\ref{fig:snapshots_regions} but will be discussed in Sect.~\ref{subsec:dame}.

Dust emission is quite intense in both Orion and Taurus
(Fig~\ref{fig:snapshots_regions}, right column) and, as can be
seen in Fig.~\ref{fig:dust_correlation}, some correlation is indeed found
between the \Planck\ CO maps and the dust template in these
locations not unexpectedly. This is
particularly true in Orion where the images show very similar
pattern of CO and dust. In the Polaris
region the correlation loosens a lot; dust is far less intense and
shows a different distribution than CO. For the weakest dust emission, probably in atomic
gas, there is no CO signal. These correlation plots against dust should be compared
with the ones performed against ground-based CO measurements, where the correlation is this time much
tighter (this is discussed in Sect.~\ref{subsec:dame}, see
Fig.~\ref{fig:dame_vs_all_10}). Looking at several other molecular
clouds (e.g. Ophiucus, Chameleon), we find that whatever the method, the
correlation with CO is always much tighter than the one with
dust, making us confident that dust is not a major issue for CO-rich
regions.

Nonetheless, \typetwo\ and \typethree\ maps do suffer some level of dust
contamination. This can be seen by eye in the \cooz\ images of the Taurus
regions in Fig.~\ref{fig:snapshots_regions} (top) where the \typetwo\ 
map shows more flux than the \typeone\ CO at the locations where the
dust is the brightest. However, it is in the Galactic plane of the \coto\
\typetwo\ map that contamination is the most important. This is
particularly visible in Fig.~\ref{fig:fullskyCO21}.

\subsubsection{Other CO lines}
\label{subsubsec:13co}

For all of the \Planck\ reconstructed CO maps we expect contamination
from other CO isotopologue transition lines. Lines from the $^{13}$CO isotopologue are the
main contaminants in the \typeone\ maps, while $^{13}$CO and
the $^{12}$\cott\ both contribute to the \typetwo\ maps.  The
contribution of each contaminant line can be estimated to first order
as followed:
\paragraph{\typeone:} Forgetting
about other contaminants, the CO content of the \typeone\ \Planck\ map can be written as
\begin{equation}
{\rm CO}_{\rm type1}^X  =  {^{12}{\rm CO}}^X  \sum_i^{N_{\rm bolos}} w_i
F_{12}^{\rm i} + {^{13}{\rm CO}}^X\sum_i^{N_{\rm bolos}} w_i
F_{13}^{\rm i} 
\label{eq:13co_type1}
\end{equation}
where $X$ represents the \joz, \jto\ or \jtt\ transition, 
$w_i$ are the weights of the linear combination and $F_{12,13}^i$ is the
 $^{12,13}{\rm CO}$ transmission in bolometer $i$. Using the
 weights of the \typeone\ linear combination and using the bandpass
 coefficients for  $^{13}{\rm CO}$, we find the quantity $\sum_i^{N_{\rm bolos}} w_i
F_{13}^{\rm i}$ to be equal to 0.53 at 115~GHz, 0.01 at
230~GHz and 0.36 at 353~GHz. Assuming a $^{13}{\rm CO}$ to  $^{12}{\rm CO}$ ratio of 0.2,
this translates in ${\rm CO}_{\rm type1}^{1-0} \approx 1.11 \times {^{12}{\rm
  CO^{1-0}}}$ and ${\rm CO}_{\rm type1}^{3-2} \approx 1.07 \times {^{12}{\rm
  CO^{3-2}}}$, i.e. an overestimation of about 10\% of the \typeone\
\joz\ and \jtt\ maps. The effect on the \jto\  transition is negligible. 

\paragraph{\typetwo:} The \typetwo\ maps are constructed as ${\rm CO}_{\rm type2}^{X} =
  \sum_{\nu} w^{X}_{\nu} M_{\nu}$, where $X$ stands for \joz\ or
  \jto, $M_{\nu}$ is the \Planck\ map at the channel frequency $\nu$ and
  $w_{\nu}$ is a weighting factor. Using the average CO transmissions
  in each channels, the CO content of the \typetwo\ maps formally reads
\begin{eqnarray}
\nonumber
{\rm CO}_{\rm type2}^X &=& {^{12}{\rm CO}}^X+ \frac{\langle F_{13}\rangle}{\langle F_{12}\rangle }\; {^{13}{\rm
    CO}^X}
 +w_{353} \langle F_{12}\rangle_{353}{^{12}{\rm
     CO}}^{[3-2]}\\\nonumber
 & = & {^{12}{\rm CO}}^X\left(1+w_{353}\langle F_{12}\rangle_{353}\frac{{^{12}{\rm CO}}^{[3-2]}}{{^{12}{\rm
       CO}}^{X}}\right) \\
&+& \frac{\langle F_{13}\rangle}{\langle F_{12}\rangle }\; {^{13}{\rm
     CO}^X}
\label{eq:13co_type2}
\end{eqnarray} 
where $\langle F_{12,13}\rangle$ are the average
conversion coefficients of ${^{12}{\rm CO}}$ and ${^{13}{\rm CO}}$ at 100 or 217~GHz, depending on the
transition under scrutiny and $\langle F_{12}\rangle_{353}$
is that of  ${^{12}{\rm CO}}$ in the 353~GHz channel. In the following, we assume the
the line ratios to be $^{12}$\cott\,/\,$^{12}$\cooz\,=\,0.2
and $^{12}$\cott\,/\,$^{12}$\coto\,=\,0.4 as obtained from the median of the line ratio distributions 
computed from the \typeone\ maps, \citep[see][]{planck2013-coscience}. 
The widths of the distributions are large and thus we consider uncertainties of $\pm 0.1$
in the two ratios. Notice that within these uncertainties the values considered here are
consistent with those derived from the Ruler analysis in Sect.~\ref{subsec:combined}.
Using these values we find ${\rm CO}_{\rm
  type2}^{[1-0]} \approx 1.02 \times {^{12}{\rm CO}}^{[1-0]}$
and ${\rm CO}_{\rm type2}^{[2-1]} \approx 0.75 \times {^{12}{\rm
    CO}}^{[2-1]}$. From this, we see that contamination by the $J$=3$\rightarrow$2
line can be neglected in the $J$=1$\rightarrow$0 \typetwo\ map, but reduces by
$\sim 25\%$ the ${^{12}{\rm CO}}$ signal in the $J$=2$\rightarrow$1 map. For this
reason, we decided to correct the \typetwo\ map from this effect by
dividing the output \typetwo\ ($J$=$2\rightarrow 1$) map by 0.75. With this correction,
the overall calibration of the \typeone\ and \typetwo\ ($J$=$2\rightarrow 1$) maps are
in agreement (see Sect.~\ref{sec:co_char}).

The second term of the contamination due to ${^{13}\rm CO}$ can now be
estimated. Using the bandpass estimate of the
coefficients, the quantity $\langle F_{^{13}{\rm CO}}\rangle / \langle
F_{^{12}{\rm CO}}\rangle$ is equal to 1.14 at 115~GHz and 0.82 at
230~GHz, which translate into ${\rm CO}_{\rm type2}^{[1-0]} \approx 1.2 \times {^{12}{\rm
  CO^{[1-0]}}}$ and ${\rm CO}_{\rm type2}^{[2-1]} \approx 1.2 \times {^{12}{\rm
  CO^{[2-1]}}}$ when assuming a ratio of 0.2 between
${^{13}{\rm CO}}$ and ${^{12}{\rm CO}}$ and the 0.75 correction factor
for the $J$=2$\rightarrow$1 line.

The values above are given as a rough estimate of the ${^{13}{\rm CO}}$
contamination. The latter depends on the isotopic ratio and relative optical depth, and will, in
practice, vary across the sky.\\

\subsubsection{Masking SZ clusters and point sources}

\Planck\ CO maps are also contaminated by other localized non-CMB signals
that have not been included in the component separation. This is the
case of radio point sources, which are present in all three types of maps.
Also present in the maps is the Sunyaev-Zeldovich (SZ) effect is a secondary anisotropy of the
CMB coming from the interaction of CMB photons with the hot electron
population of galaxy clusters. The \Planck\ mission has shown the
potential of galaxy cluster detection via the SZ effect with the
publication of a catalogue of a few thousands clusters
\citep{planck2013-p05a}. As far as CO extraction is concerned, SZ is yet another contaminant
foreground of the \typetwo\ and \typethree\ maps, as these rely on
multi-channel information. On the other hand \typeone\ maps are not affected.

For this first release of the \Planck\ CO maps, a point source mask is
provided with the \typeone\ and a point source + SZ mask for
the \typetwo\ and \typethree maps. The point source mask corresponds to
\Planck-HFI official 100~GHz mask (we chose the 4-$\sigma$, $S/N=10$
mask) where we unmasked any pixels located within $|b|<1.5^{\circ}$. This
modification was necessary given that many point sources in the mask 
are molecular cold cores located in the Galactic disk, so that using
the original mask hides most of the Galactic CO.
For the SZ mask we start from a simulated Compton parameter
map of the \Planck\ cluster sample
\citep{planck2013-p05a,planck2013-p05b} where a
universal generalized pressure profile \citep{arn10} was assumed. 
The mask is then generated by imposing a threshold of $4\times
10^{-6}$ in Compton parameter units to this map.

\subsection{Absolute calibration uncertainties}

Three main contributors to the absolute calibration uncertainties of the \Planck\ CO maps have
been identified: i) the \Planck-HFI calibration uncertainties of the
temperature maps; ii) the uncertainty on the CO bandpass conversion
coefficients; and iii) the uncertainty on the $^{13}$CO
contribution. We do not include the effect of dust in this absolute
calibration given that, conversely to $^{13}$CO, dust is not always
spatially correlated to CO and is therefore considered as a
systematic effect (see Sect.~\ref{subsubsec:dust}).

As described in \citet{planck2013-p03b}, the absolute
calibration of the \Planck\ maps is about 0.55\% from 100~GHz to 217~GHz,
1.25\% at 353~GHz, and 10\% at 545 and 857~GHz. 
Uncertainties on the averaged bandpass CO conversion coefficients
are at the 5\% level at 100~GHz and 1\% at 217 and 353~GHz
\citep{planck2013-p03d}. We described how $^{13}$CO affected the
\Planck\ CO maps in Sect.~\ref{subsubsec:13co}; using the results
obtained with $^{13}$CO/${^{12}{\rm CO}}=0.2$ \citep{1979ApJ...232L..89S}, and assuming a conservative
uncertainty of $\pm 0.1$ on this ratio, we can estimate the $^{13}$CO
contribution to the absolute uncertainty in each CO map. 

Final calibration uncertainties are summarized in Table~\ref{tab:abscal}. 
A conservative estimate of the total absolute calibration uncertainties of each map is given in the
last column of the table. For most maps, the calibration uncertainties are of the order of 10 \% and
are dominated by the $^{13}$CO contribution. Notice that for \typeone\ \coto\ and \cott\ the uncertainties
are significantly smaller as they are less contaminated by $^{13}$CO.

\begin{table}[tmb]
\begingroup
\newdimen\tblskip \tblskip=5pt
\caption{Absolute calibration uncertainties of \Planck\ CO maps.\label{tab:abscal}}                         
\nointerlineskip
\vskip -6mm
\footnotesize
\setbox\tablebox=\vbox{
   \newdimen\digitwidth 
   \setbox0=\hbox{\rm 0} 
   \digitwidth=\wd0 
   \catcode`*=\active 
   \def*{\kern\digitwidth}
   \newdimen\dpwidth 
   \setbox0=\hbox{.} 
   \dpwidth=\wd0 
   \catcode`!=\active 
   \def!{\kern\dpwidth}
\halign{\hbox to 1.5cm{\hfil#\hfil}\tabskip 2em&
     \hfil#\hfil \tabskip 1em&
     \hfil#\hfil \tabskip 1em&
     \hfil#\hfil \tabskip 1em&
     \hfil#\hfil \tabskip 1em&
     \hfil#\hfil \tabskip 0em\cr 
\noalign{\doubleline}
 Type & Line & HFI calib. & $\langle F_{\rm CO}^{\rm BP}\rangle$ & $^{13}$CO & Total \cr
     &     &      [\%]     &   [\%] &  [\%]   &  [\%] \cr
\noalign{\vskip 3pt\hrule\vskip 5pt}
\omit \typeone &\cooz & 0.55 & 5 & 5 &  10\cr
\omit \typeone &\coto & 0.55 & 1 & \dots & 2  \cr
\omit \typeone &\cott & 1.25 & 1   & 3 & 5 \cr
\noalign{\vskip 4pt} 
\omit \typetwo &\cooz & 0.70 & 5 & 10 & 15\cr
\omit \typetwo &\coto & 0.70 & 1 & 10 & 11\cr
\noalign{\vskip 4pt}
\omit \typethree & \dots & 3 & 5& \dots & 11\cr
\noalign{\vskip 3pt\hrule\vskip 5pt}
}
}
\endPlancktable 
\endgroup
\end{table}

\begin{figure}[!h]
   \centering
\setlength{\unitlength}{\columnwidth}
\begin{picture}(1,2.1)
\put(0,1.4){\includegraphics[width=\columnwidth]{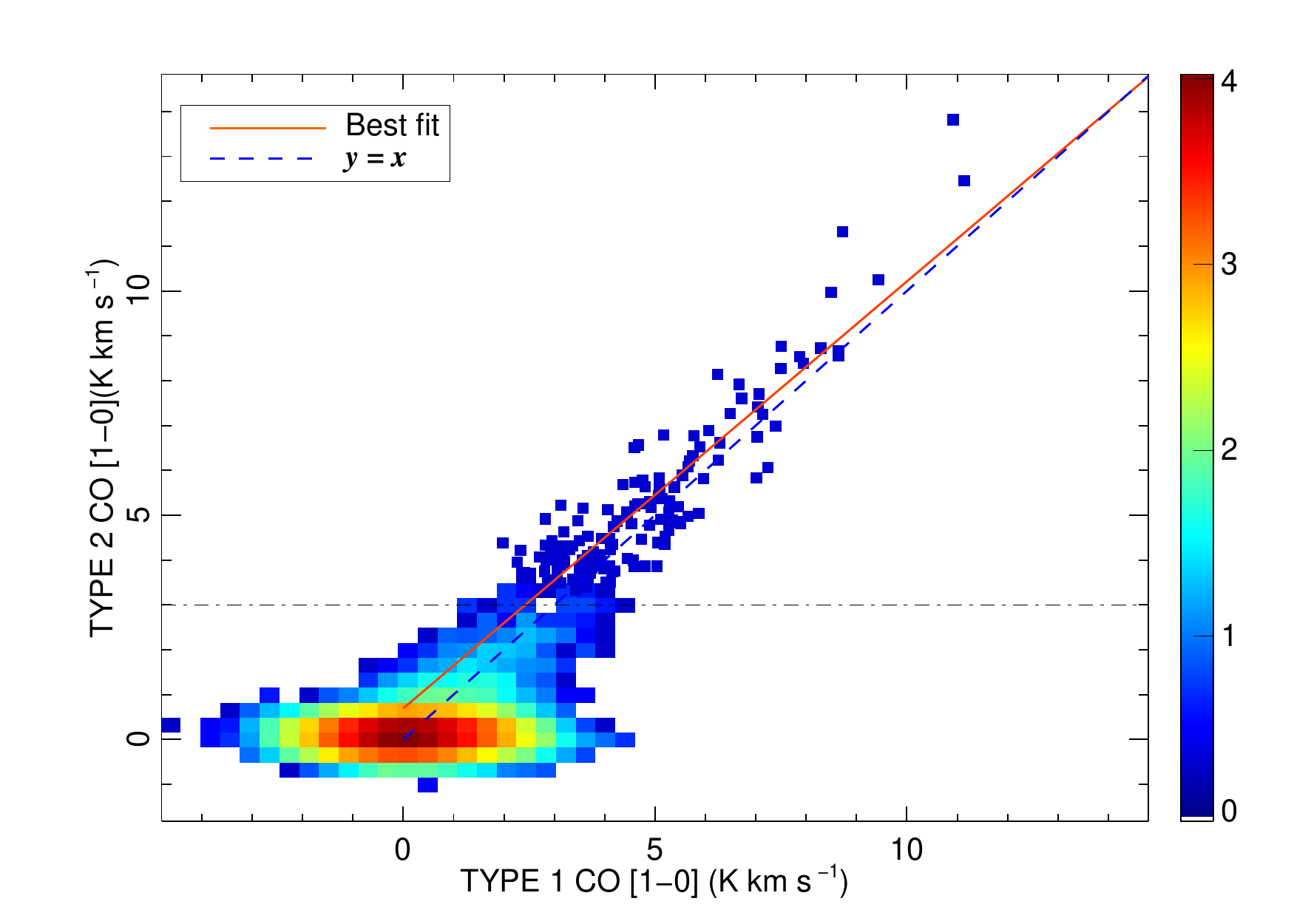}}
\put(0,0.7){\includegraphics[width=\columnwidth]{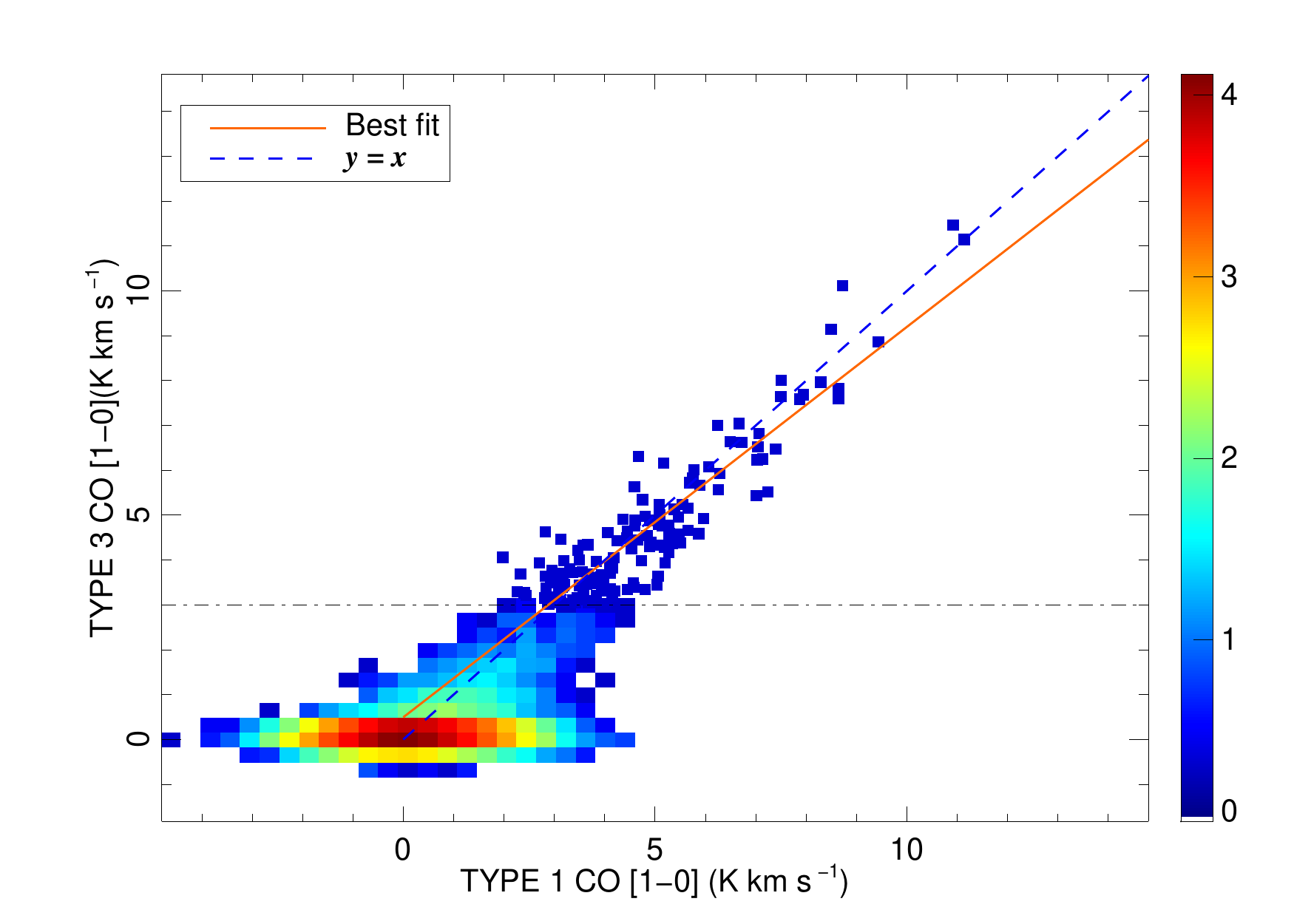}}
\put(0,0){\includegraphics[width=\columnwidth]{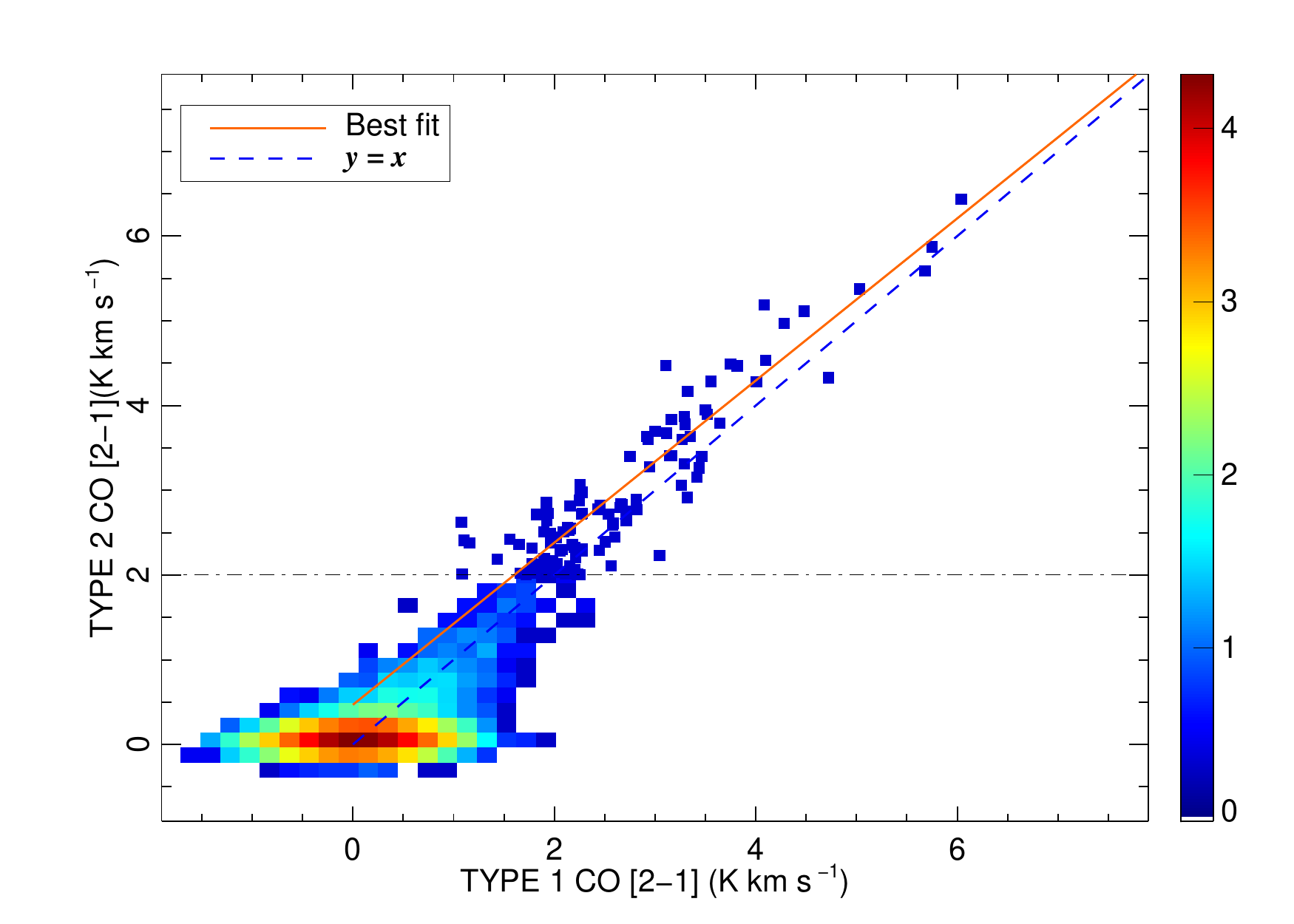}}
\put(0.05,2.05){(a)}
\put(0.05,1.35){(b)}
\put(0.05,0.65){(c)}
\end{picture}

 \caption{Correlation between the \typeone\ and \typetwo\ (top) and \typeone\
   and \typethree\ (middle) $J$=1$\rightarrow$0 CO maps at high Galactic latitudes $25^\circ  <
   |b| < 90^\circ$. Correlation between the \typeone\ and
   \typetwo\ $J$=2$\rightarrow$1 CO maps is shown on the bottom plot.The solid line corresponds to $y=x$. Below the
   dashed horizontal line the log colour scale gives the number of pixels in a
   given bin. The best linear fits (solid red line) are computed from the individual pixels
   plotted above the dashed line. The best-fit slopes and uncertainties are: (a) $(0.93 \pm 0.17)$; (b) $(0.87 \pm 0.20)$;
   and (c) $(0.96 \pm 0.12)$. The intercept is compatible with zero within the error bars.
   \label{fig:comparison10_highlat}}
 \end{figure}
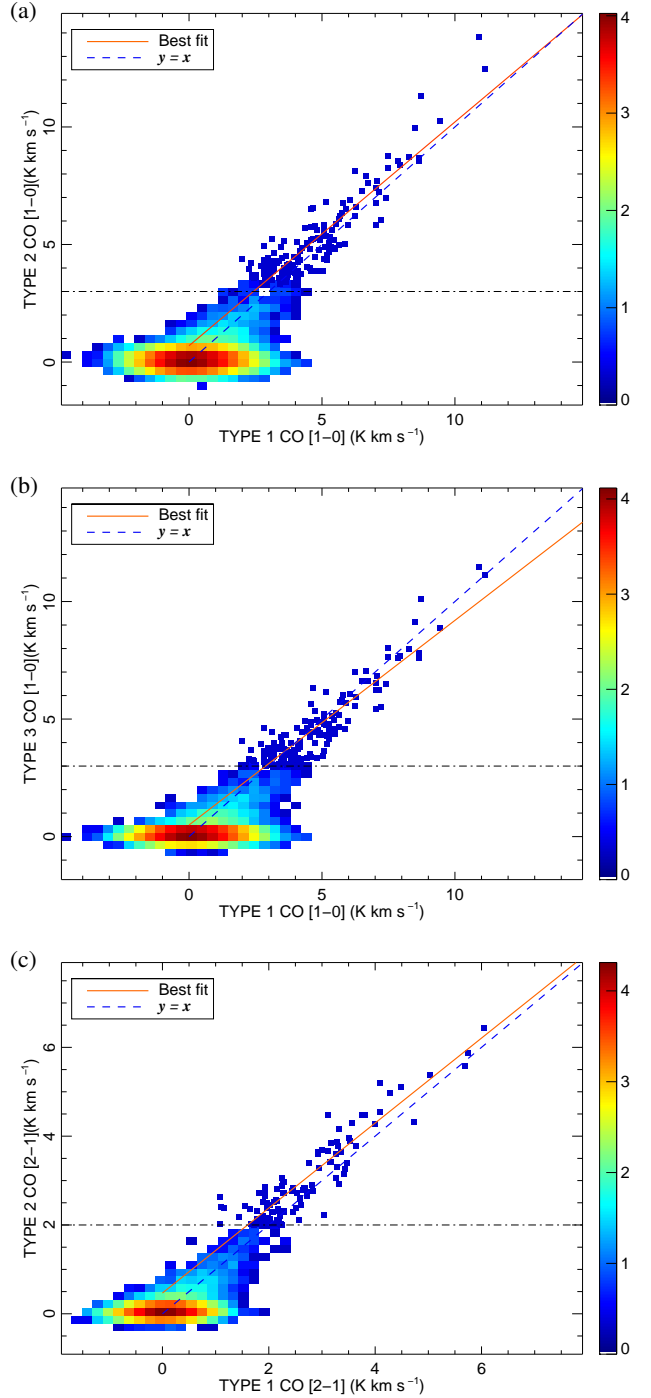


\section{Internal validation of the CO maps}
\label{sec:co_char}

Before validating these maps on external data, it is possible to perform some
internal checks by comparing the CO maps of different types.
As mentioned previously, the \typeone\ and \typetwo\ CO products can differ
in the Galactic plane because of the higher level of contamination
affecting the \typetwo\ products. However, at high Galactic latitudes,
where no significant free-free or dust emission is expected,
the products may be compared in order to assess the overall
inter-calibration of the \typeone, 2 and 3 CO maps. 
To perform this comparison, the maps are first smoothed to a common
resolution of 30\arcm\ and degraded to $N_{\rm side}=128$ to avoid noise correlation
between samples. After
masking point sources, correlation plots are produced for all
remaining pixels located at Galactic latitudes $|b| \ge 25^\circ$.

Figure~\ref{fig:comparison10_highlat} compares both
\typeone$-$\typetwo\ and \typeone$-$\typethree\ maps. At high latitudes, CO emission is very sparse so that
most pixels have very low emission. Therefore, below an empirical
threshold, shown as the dashed line in each figure, the number of
points is such that a contour representation of binning suffices. 
Above the dashed line, individual pixels are plotted directly and used
to compute the best fit. 
The orange line gives the best-fit values and is compatible within
errors with $y=x$. This indicates that the different
types of product share the same overall calibration, be it for the
$J$=1$\rightarrow$0 maps (Fig.~\ref{fig:comparison10_highlat}, left and middle panel) or $J$=2$\rightarrow$1 maps 
(Fig.~\ref{fig:comparison10_highlat}, right panel).

\begin{figure*}
   \centering
\setlength{\unitlength}{\columnwidth}
 \begin{picture}(2,2.1)
\put(0,1.4){ \includegraphics[width=\columnwidth]{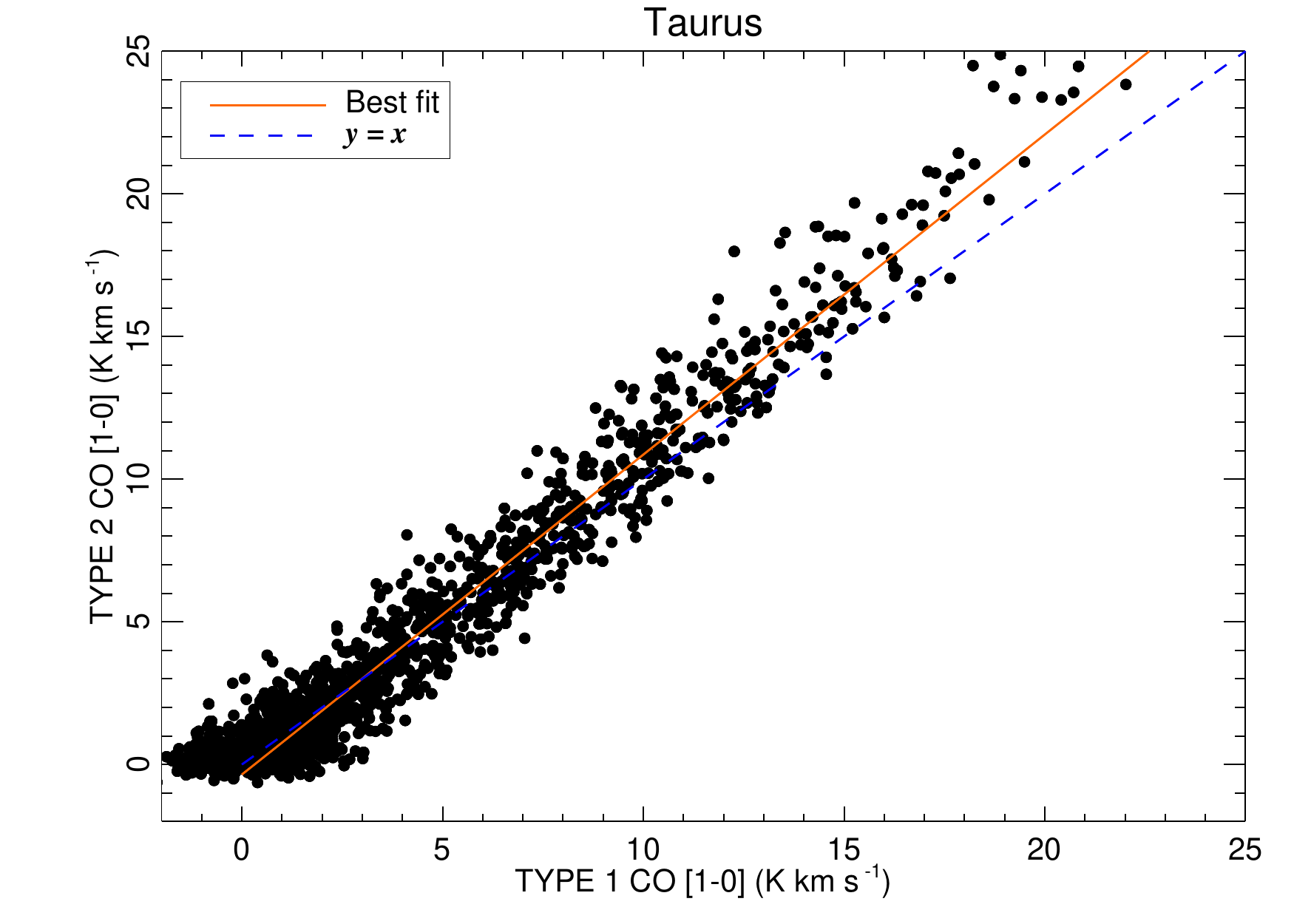}}
\put(1,1.4){ \includegraphics[width=\columnwidth]{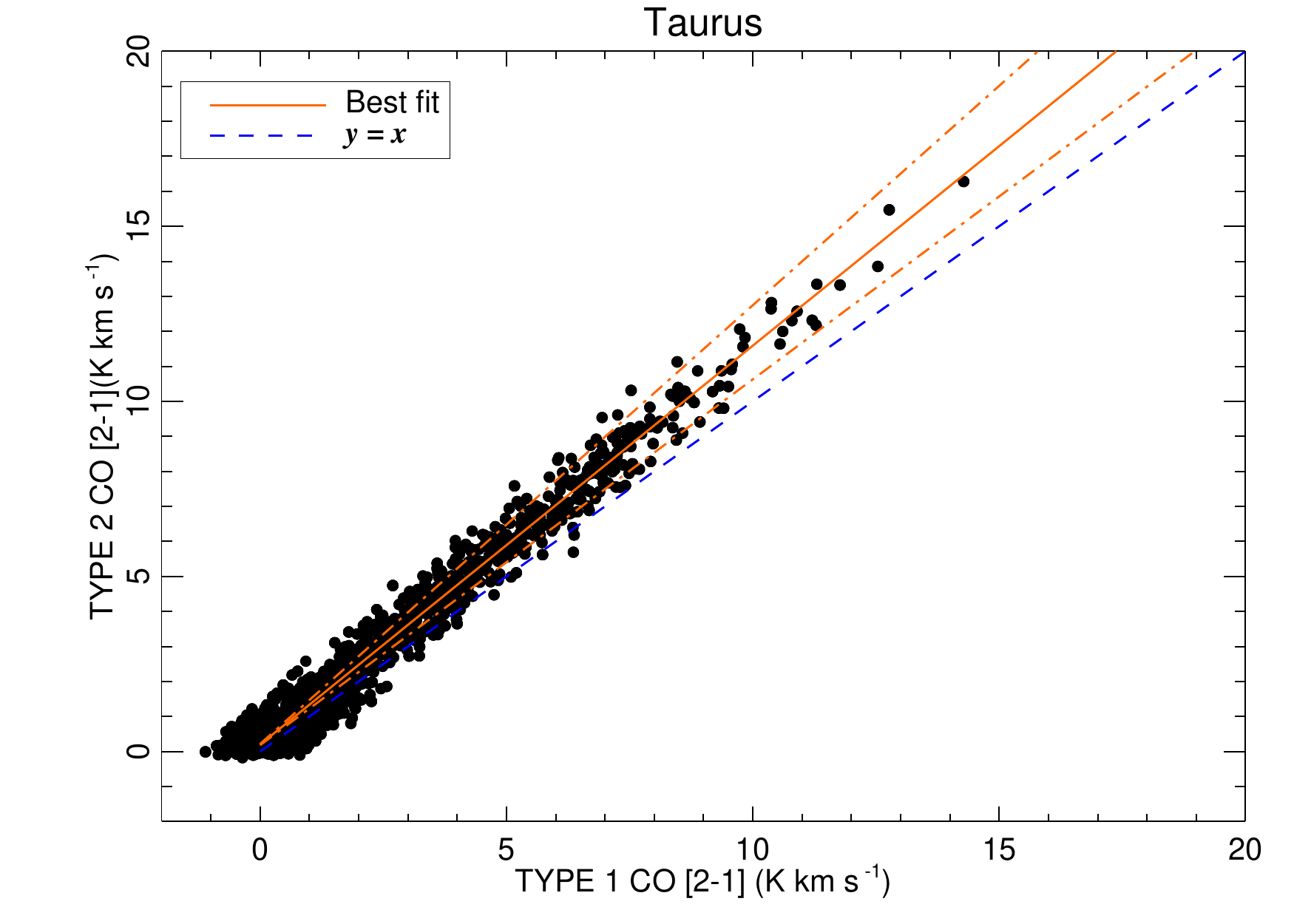}}
\put(0,0.7){\includegraphics[width=\columnwidth]{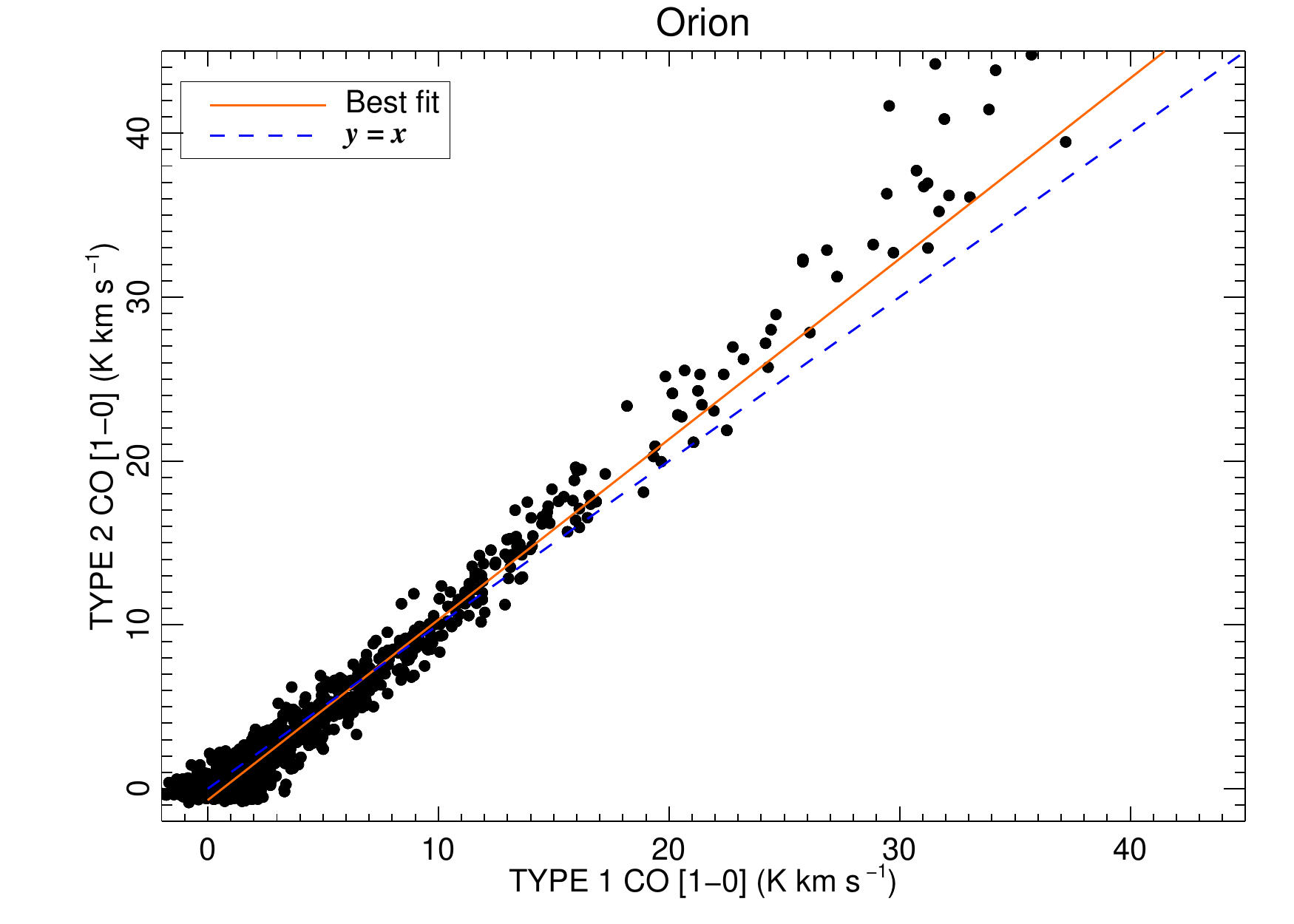}}
\put(1,0.7){ \includegraphics[width=\columnwidth]{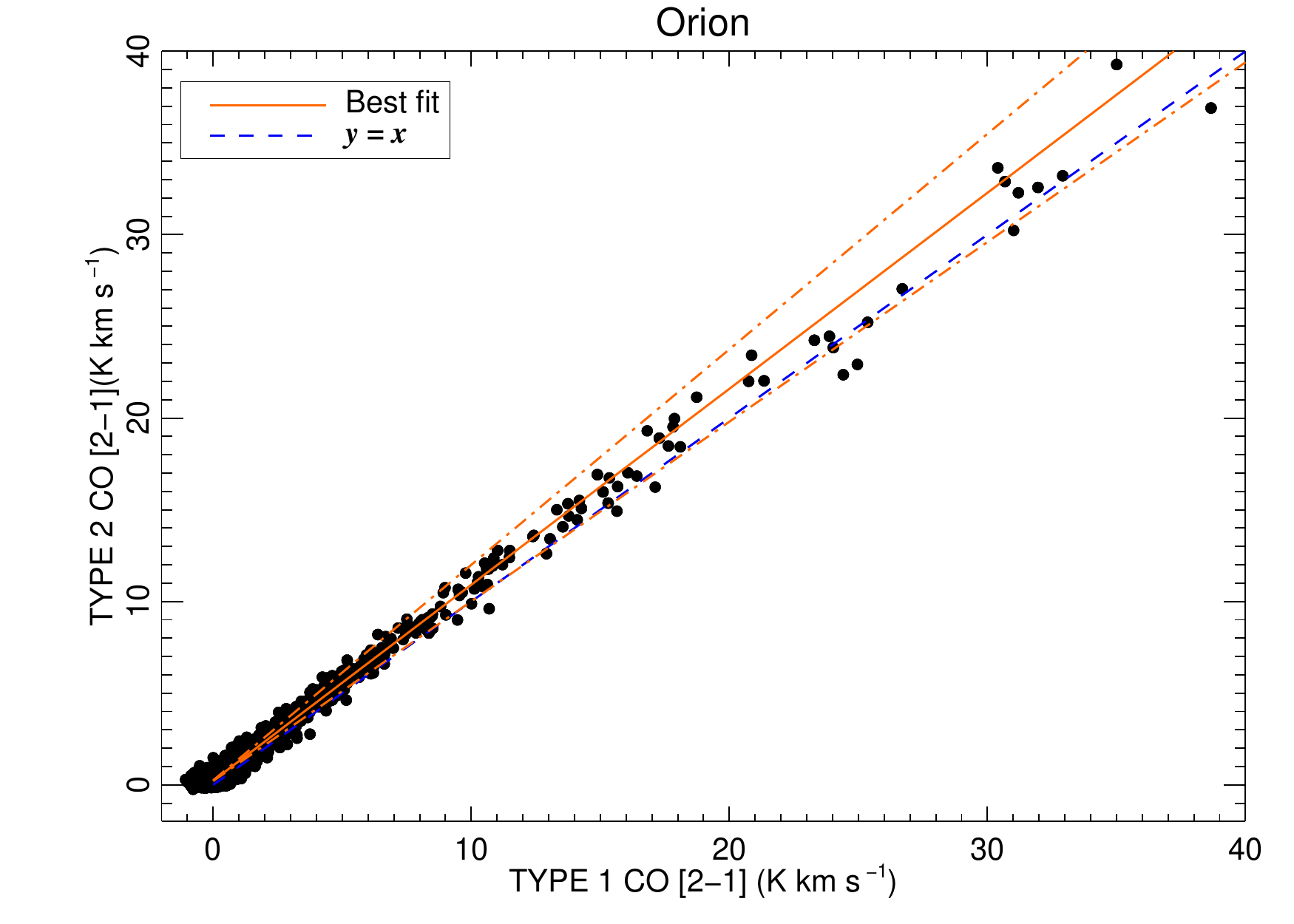}}
\put(0,0){ \includegraphics[width=\columnwidth]{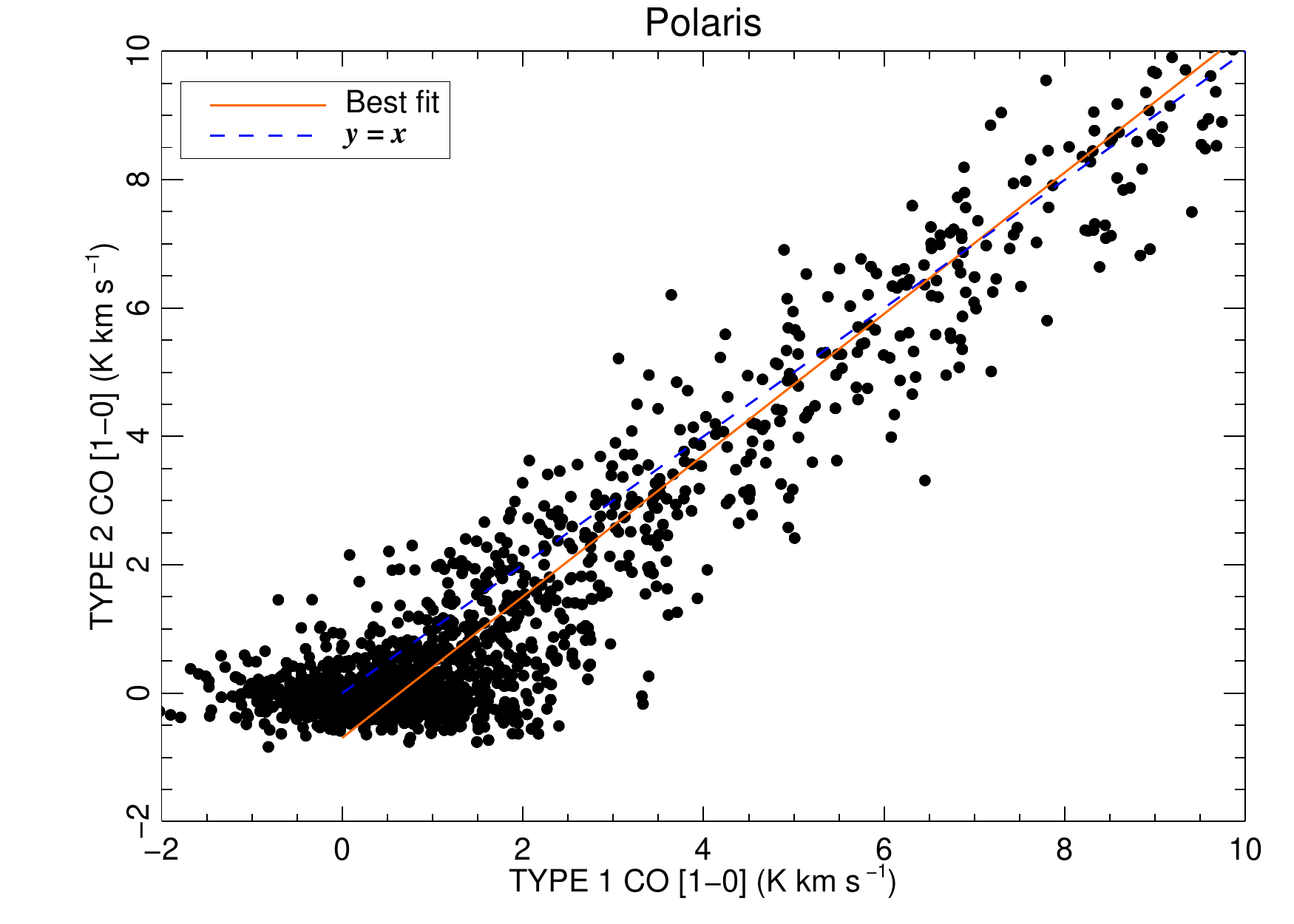}}
\put(1,0){\includegraphics[width=\columnwidth]{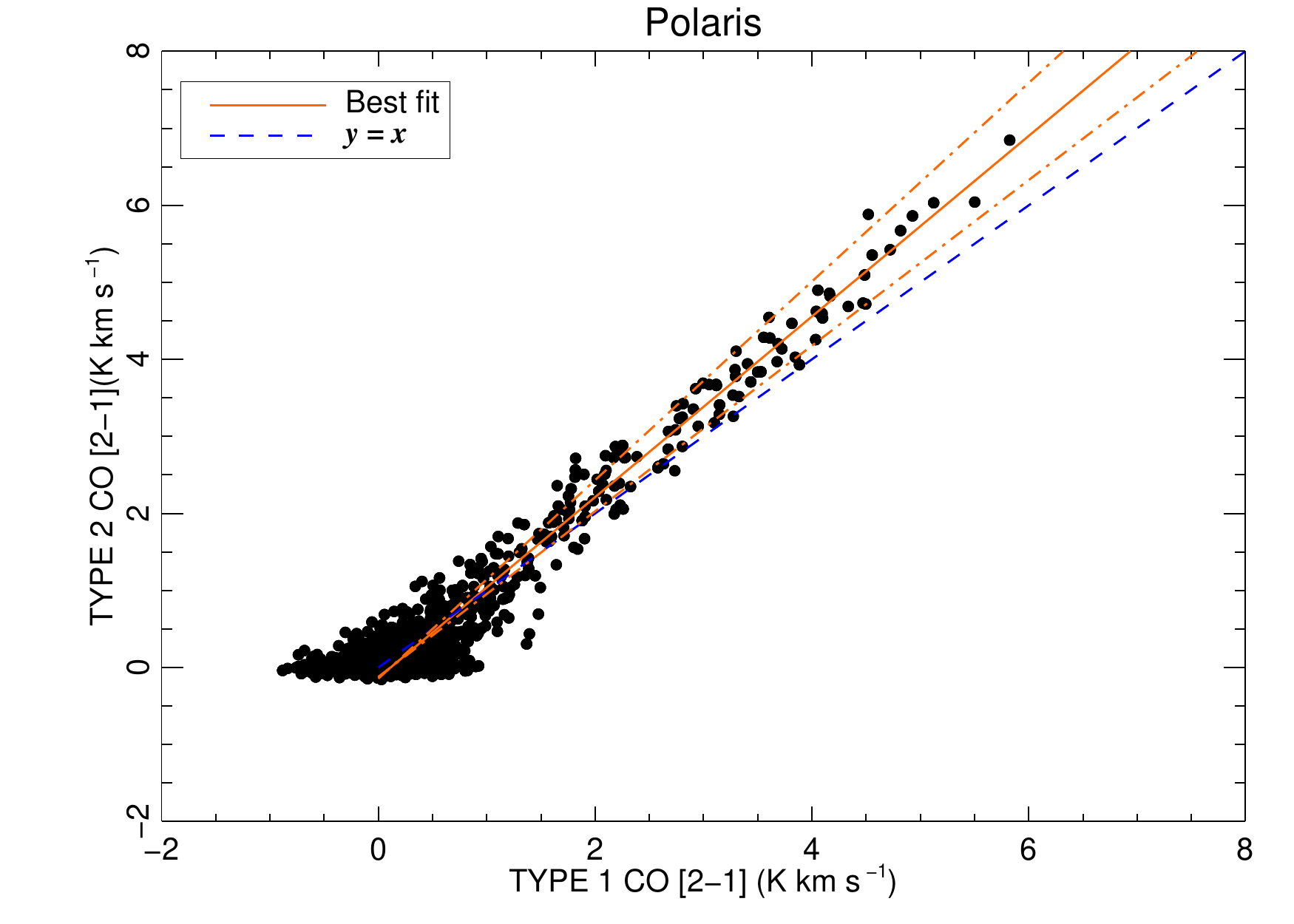}}
\put(0.05,2.05){(a)}
\put(1.05,2.05){(b)}
\put(0.05,1.35){(c)}
\put(1.05,1.35){(d)}
\put(0.05,0.65){(e)}
\put(1.05,0.65){(f)}
\end{picture}
   \caption{Correlation between \typeone\ and \typetwo\ CO maps at
     115~GHz (left column) and 230~GHz (right column) in the Taurus (top), Orion
     (middle) and Polaris (bottom) molecular clouds. The best linear fits  (orange solid line) have been computed between 2
     and 15 K\,km\,s$^{-1}$ at 115~GHz and between 1 and 15
     K\,km\,s$^{-1}$ at 230~GHz, where bulk of the data lie and
     avoiding the noise around zero. The best-fit slopes and uncertainties are: 
     (a) $(1.12 \pm 0.03)$; (b) $(1.14 \pm 0.01)$; (c) $(1.10 \pm 0.04)$; (d) $(1.07 \pm 0.01)$; (e) $(1.07 \pm 0.06)$; and (f) $(1.17 \pm 0.04)$.
     The intercepts are compatible with zero within error bars.
     The orange dot-dashed lines in the bottom panels correspond to the recalibration uncertainty due to the contribution of the $J$=3$\rightarrow$2   line in the
  \typetwo\ $J$=2$\rightarrow$1 product. See text for detail. 
 \label{fig:mapsintercomparison10}}
 \end{figure*}
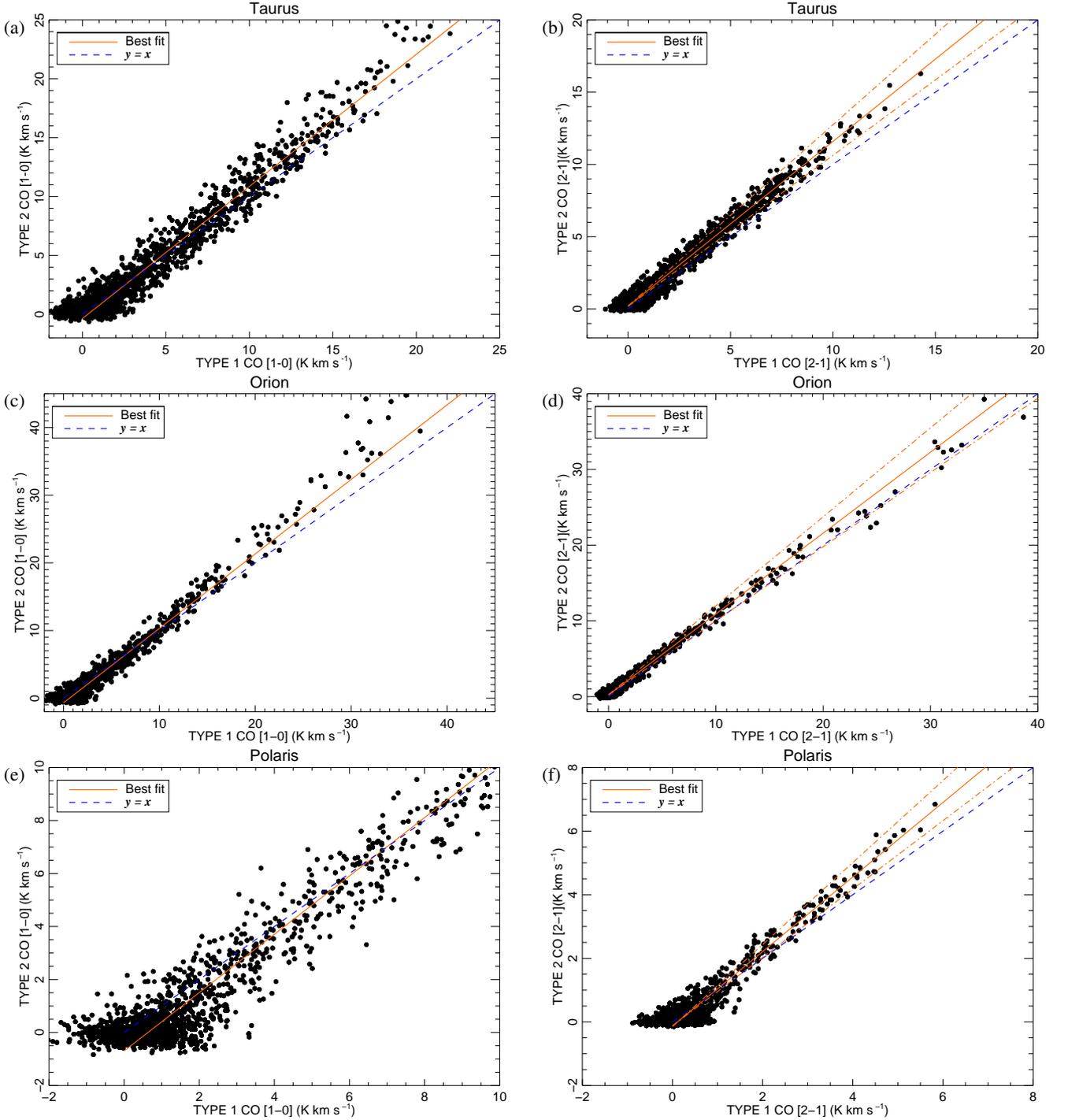

The different types of product may also be compared in specific
molecular clouds, as shown in the top and bottom panels of Fig.~\ref{fig:mapsintercomparison10}
for the $J$=1$\rightarrow$0 (top) and $J$=2$\rightarrow$1 (bottom) lines,
respectively. As for the comparison to dust, the Taurus (left), Orion (middle) and Polaris (right)
molecular clouds have been chosen because of the three very
different molecular environments they host. Figure~\ref{fig:mapsintercomparison10} shows
tight correlations between the \typeone\ and \typetwo\ \cooz\ products in these
three molecular clouds. The best fits are computed for all points in the
range $[2-15]$~K\,km\,s$^{-1}$, where the bulk of the emission
lies. The fit uncertainties are dominated by the errors bars (not shown
in the figure) of the \typeone\ points. The \typetwo\ \cooz\ map
shows more flux than the \typeone\, with best-fit slopes of about $1.1$ for Taurus, Orion and Polaris. Such
behavior is nonetheless expected, given that the \typetwo\ \cooz\
map suffers from more $^{13}$CO and dust contamination than the 
\typeone\ \cooz\ map as described in Sects.~\ref{subsubsec:13co} and \ref{subsubsec:dust}.
Similar trends are observed for the \coto\ maps
(bottom panel), but here we also give as dashed lines the best-fits
that would have been obtained for a change of $\pm 0.1$ in the
$^{12}$\cott\,/\,$^{12}$\coto\ ratio assumed to correct the
\typetwo\ map from the $^{12}$\cott\ contamination (see
Sect.~\ref{subsubsec:13co}). Given the absolute calibration
and statistical uncertainties quoted above for each of the maps, 
the residuals between the \typeone\ and \typetwo\ maps are consistent
within the errors.

\section{Comparison with external data}
\label{sec:ext_comp}

In this section, we validate the Planck CO products described in
Sect.~\ref{sec:COmaps} using existing ground-based CO data of the
first three transition lines. 

\subsection{Comparison with the \citet{2001ApJ...547..792D} CO $J$=1$\rightarrow$0 survey}
\label{subsec:dame}
The publicly available data of
\citet{2001ApJ...547..792D} represent
the most complete survey of Galactic $^{12}$CO $J$=1$\rightarrow$0 emission to date. The original data consist of a composite map
constructed from a set of 37 independent surveys taken by the 1.2
meter Millimetre-Wave Telescope at the CfA\footnote{The complete
  data set can be retrieved from
  {\url{http://www.cfa.harvard.edu/rtdc/CO/}}}. In order to compare the
\Planck\ CO maps to the \citet{2001ApJ...547..792D} data, 
we use the velocity-integrated {\tt {\tt HEALPix}} rendition of the survey which is available on the
Lambda website\footnote{\url{http://lambda.gsfc.nasa.gov/}}.

\subsubsection{Molecular clouds}
\begin{figure}[h]
\centering
\setlength{\unitlength}{\columnwidth}
\begin{picture}(1,2.1)
\put(0,1.4){\includegraphics[width=\columnwidth]{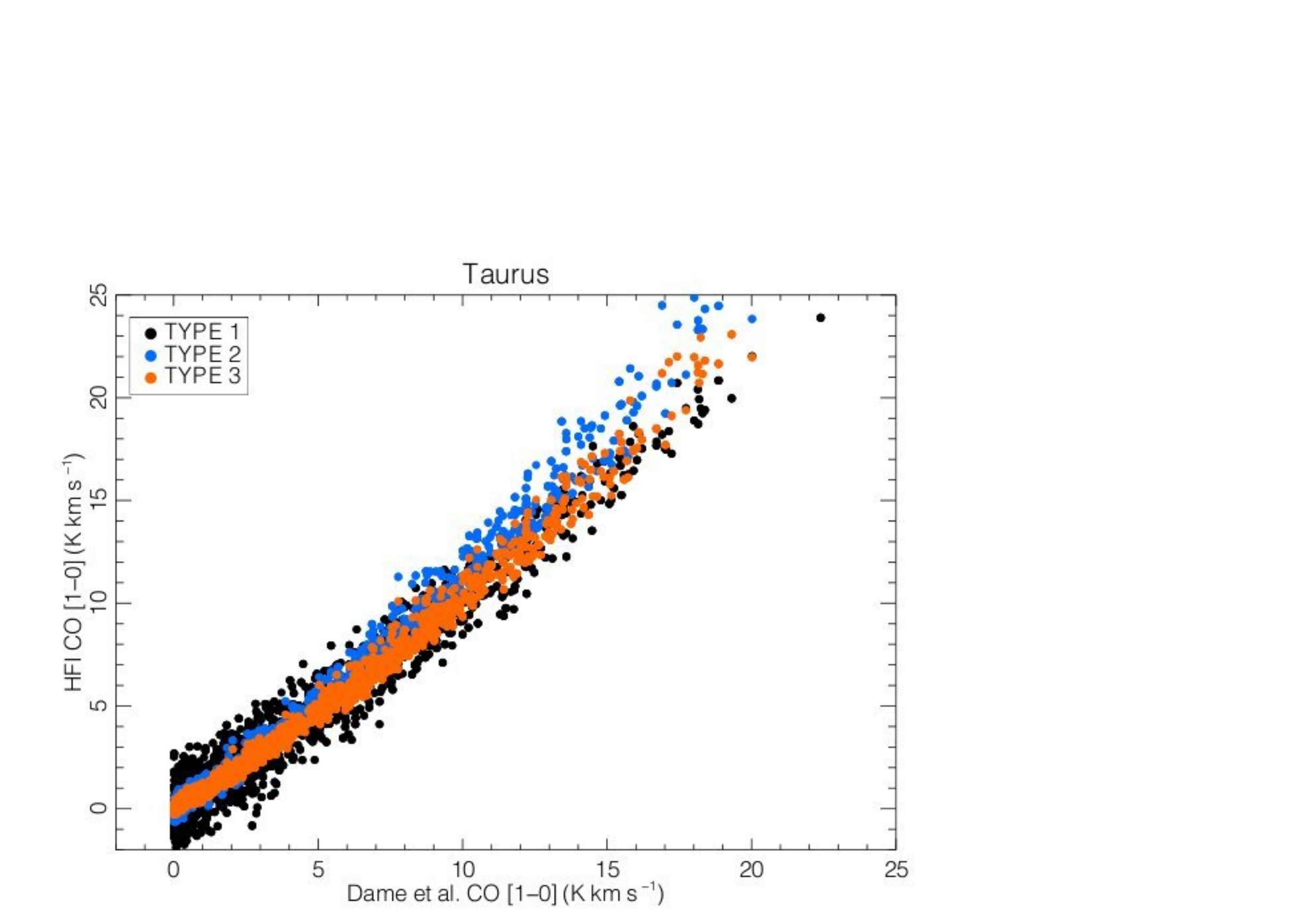}}
\put(0,0.7){\includegraphics[width=\columnwidth]{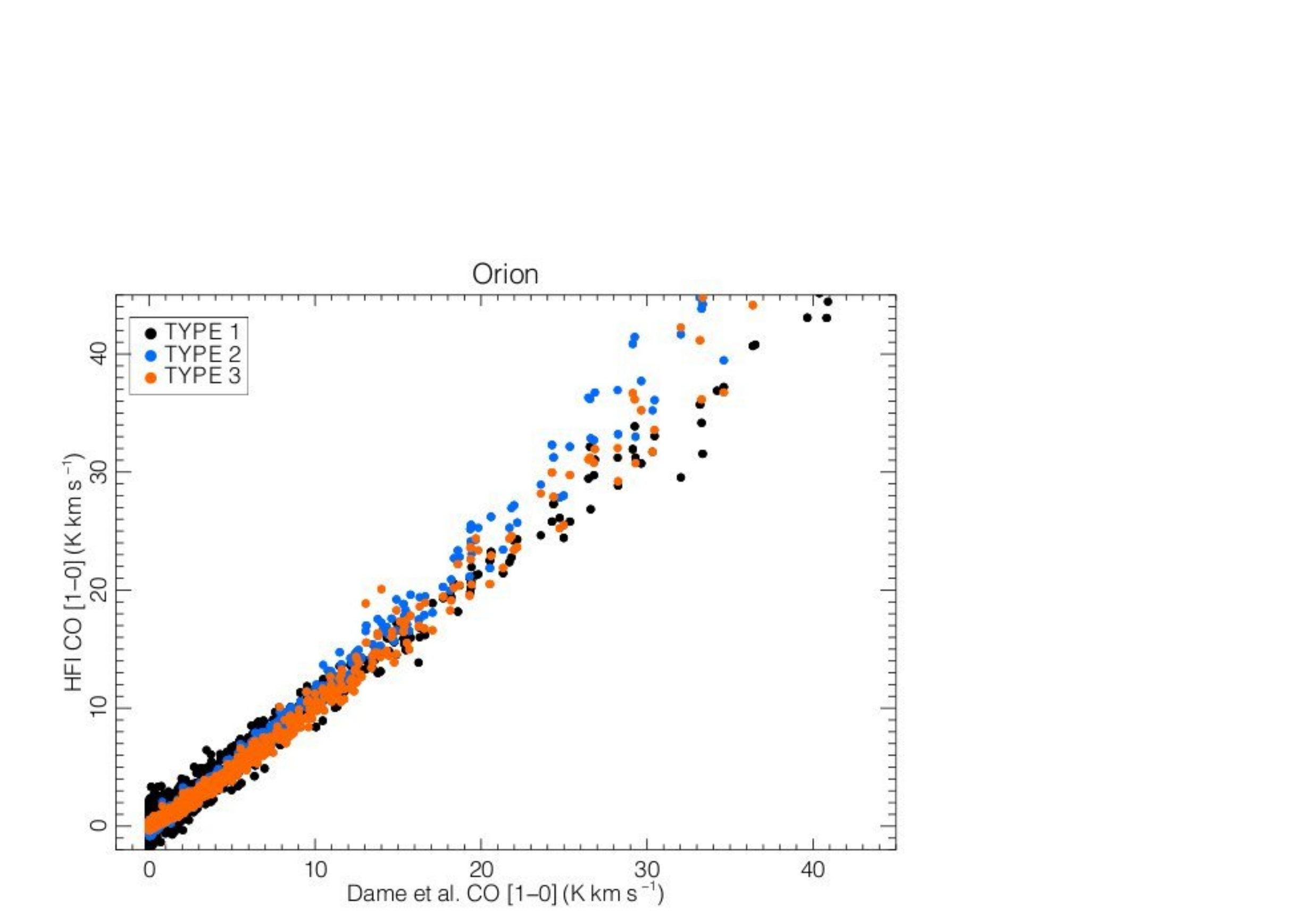}}
\put(0,0){\includegraphics[width=\columnwidth]{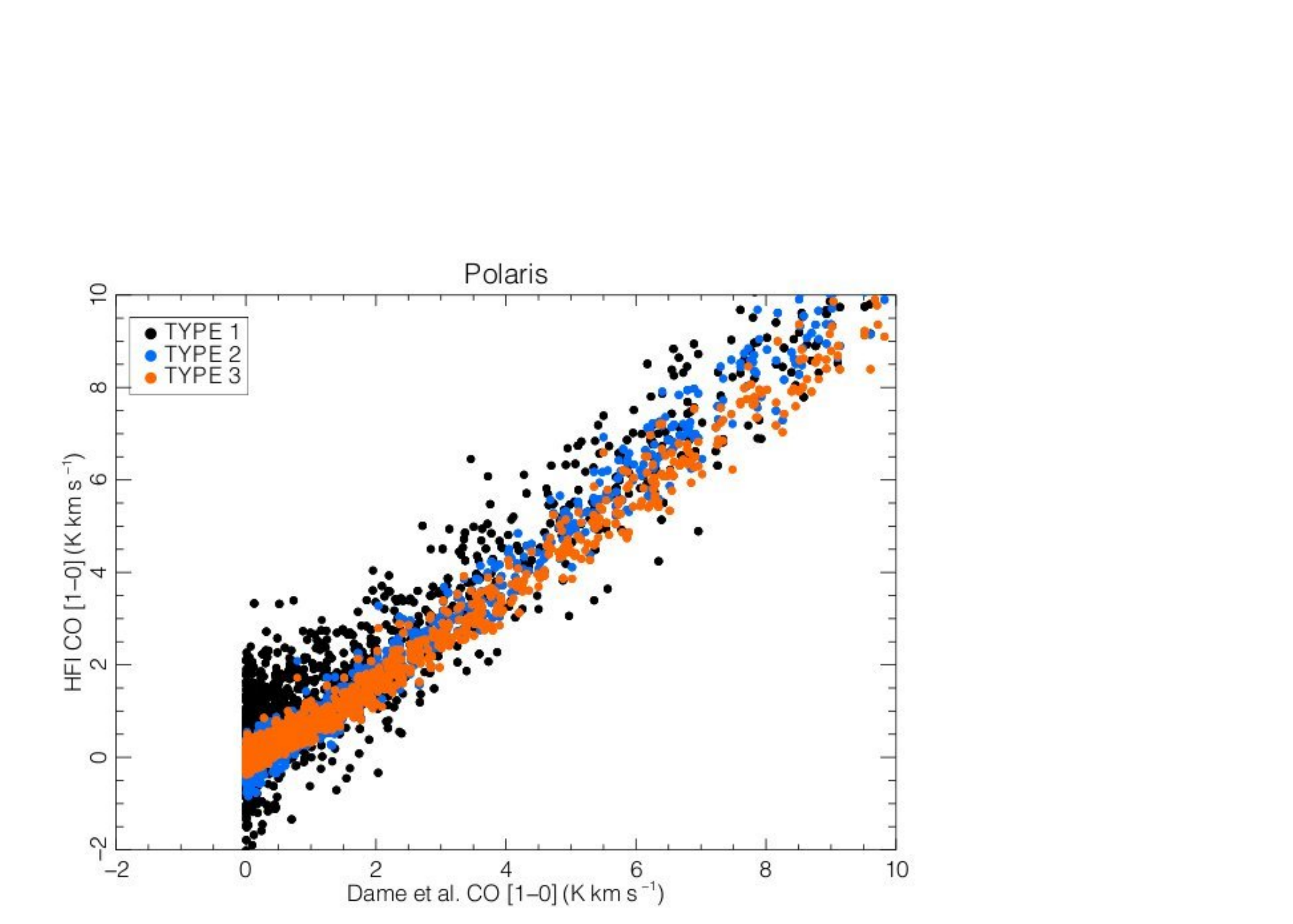}}
\put(0.05,2.05){(a)}
\put(0.05,1.35){(b)}
\put(0.05,0.65){(c)}
\end{picture}
\caption{Correlation between the three types of CO products with the \citet{2001ApJ...547..792D} data in the Taurus (top), Orion (middle) and Polaris (bottom) molecular clouds. The best-fit slope and intercept are given in Table~\ref{tab:fig10}. Notice that the intercepts are in all cases below 0.7~K~km~s$^{-1}$.\label{fig:dame_vs_all_10}}
\end{figure}
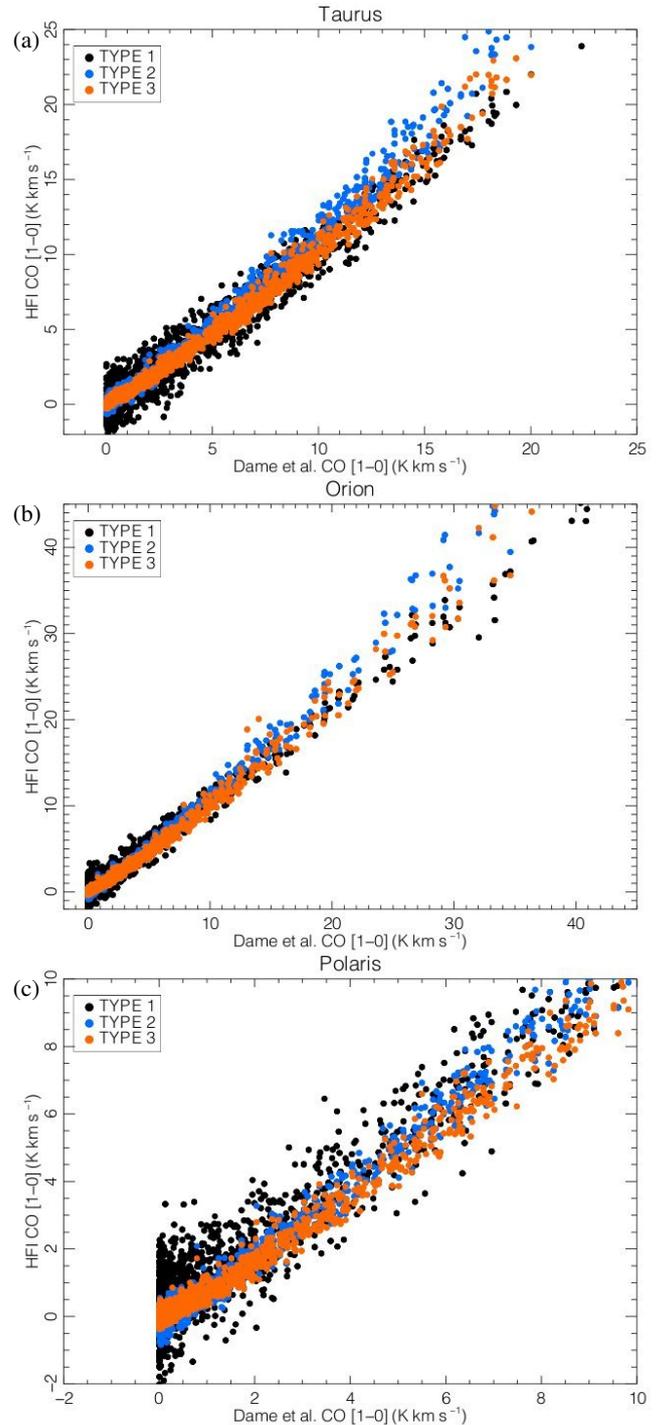

We focus again on the three molecular clouds
Taurus, Orion and Polaris. Images of the
\citet{2001ApJ...547..792D} and \Planck\ data in these fields are
shown in Fig.~\ref{fig:snapshots_regions}. The maps were smoothed
to a common resolution of 30\arcm\ and degraded to $N_{\rm
  side}=128$ (to avoid noise correlation between samples) before plotting the correlations shown in
Fig.~\ref{fig:dame_vs_all_10}. In each panel of this figure, the
three types of \Planck\ CO [1-0] maps (shown in different colours) are correlated with the
\citet{2001ApJ...547..792D} data. The slopes of the best-fit linear
regressions for the three products are given in the legend of each
panel (they are found with a 1-$\sigma$ statistical error smaller than
1\%). The spread in the data points is reminiscent of the level of
noise in the maps, showing once again that the \typeone\ map is the
noisiest while the \typethree\ map has the best signal-to-noise ratio. 

\begin{table*}[tmb]
\begingroup
\newdimen\tblskip \tblskip=5pt
\caption{Best linear fit parameters, slope and intercept, for the correlation plots shown in Fig.~\ref{fig:dame_vs_all_10}\label{tab:fig10}}                         
\nointerlineskip
\vskip -6mm
\footnotesize
\setbox\tablebox=\vbox{
   \newdimen\digitwidth 
   \setbox0=\hbox{\rm 0} 
   \digitwidth=\wd0 
   \catcode`*=\active 
   \def*{\kern\digitwidth}
   \newdimen\dpwidth 
   \setbox0=\hbox{.} 
   \dpwidth=\wd0 
   \catcode`!=\active 
   \def!{\kern\dpwidth}
\halign{\hbox to 1.5cm{\hfil#\hfil}\tabskip 1em&
     \hfil#\hfil \tabskip 1em&
     \hfil#\hfil \tabskip 1em&
     \hfil#\hfil \tabskip 0em\cr 
\noalign{\vskip 10pt\hrule\vskip 5pt}
\omit Panel & \typeone\  & \typetwo\ & \typethree\ \cr
\noalign{\doubleline}
\noalign{\vskip 5pt}
\omit(a)      &   $(1.054 \pm 0.016)$;  $(-0.231 \pm 0.097)$   &   $(1.195 \pm 0.001)$;  $(*0.409 \pm 0.070)$    & $(1.081 \pm 0.001)$;  $(*0.334 \pm 0.003)$     \cr
\omit(b)      &   $(1.056 \pm 0.009)$;  $(*0.141 \pm 0.069)$   &   $(1.290 \pm 0.001)$;  $(-0.654 \pm 0.005)$   &  $(1.163 \pm 0.001)$;  $(-0.488 \pm 0.002)$     \cr
\omit(c)      &   $(1.104 \pm 0.029)$;  $(-0.070 \pm 0.052)$   &   $(1.156 \pm 0.002)$;  $(*0.102 \pm 0.004)$   &   $(1.012 \pm 0.001)$;  $(*0.112 \pm 0.002)$    \cr
\noalign{\vskip 3pt\hrule\vskip 5pt}
}
}
\endPlancktable 
\endgroup
\end{table*}

The \typeone\ CO is in good agreement with the \citet{2001ApJ...547..792D} data, with a slope of about $1.05$ in both Taurus and Orion. 
The excess with respect to the 1-to-1 correlation is at the level expected from $^{13}$CO contamination (see Sect.~\ref{subsubsec:13co}). 
For Polaris the slope is $\sim 1.1$, but in such faint regions, the level of noise of the \typeone\ map
makes the comparison more difficult. For all clouds, the \typetwo\ CO shows roughly a 20\% excess compared to 
the  \citet{2001ApJ...547..792D} data that is due to a combination of both dust and
$^{13}$CO contamination. It performs particularly poorly in Orion
where the dust is significantly correlated with CO emission. Finally, the
\typethree\ CO does as well as the \typeone\ product in Taurus, 
but also suffers from some contamination in Orion with a correlation
of 1.16. This test was conducted in other molecular regions (not shown
here) and similar trends were found. The level of the residuals
is always compatible with the calibration uncertainty of the maps.

From these, we conclude that the
\typeone map is the most robust $J$=1$\rightarrow$0 map in terms of CO
extraction, but its low S/N ratio makes it unsuitable for the study of faint
CO regions. The \typetwo\ and \typethree\ maps both suffer from some level
of dust contamination but allow us to probe fainter regions and to perform discovery
studies.

\subsubsection{Velocity effect in the \typeone\ $J$=1$\rightarrow$0 map}
\begin{figure}[h!]
\centering
\includegraphics[width=\columnwidth,trim=0cm 3.4cm 0cm 3.4cm, clip ]{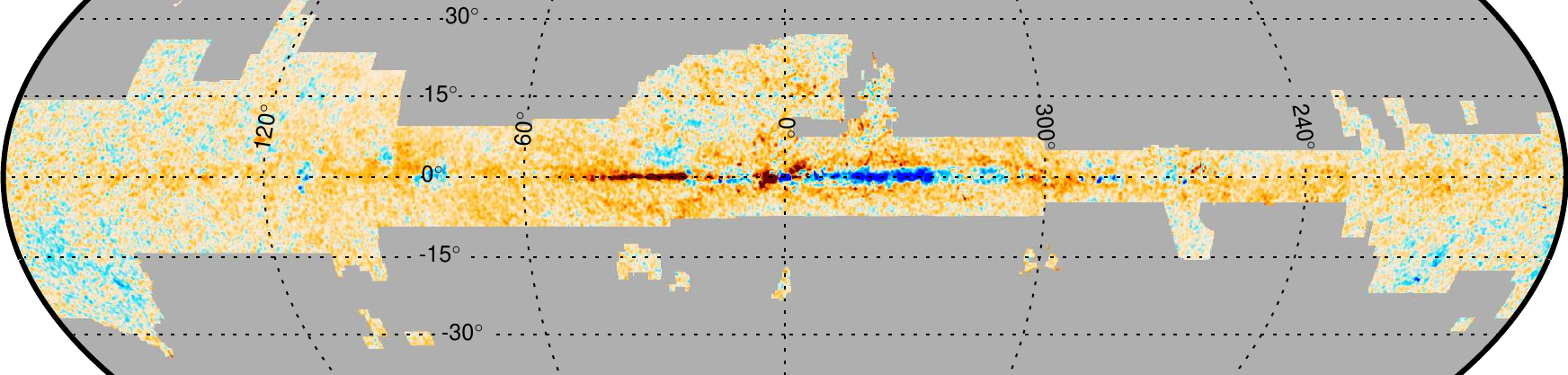}
\includegraphics[width=\columnwidth,trim=0cm 0cm 0cm 10cm, clip ]{./figs/type1_dame_residual.pdf}
\caption{Residual map with respect to the \cite{2001ApJ...547..792D}
  map for the \typeone\ $J$=1$\rightarrow$0 \Planck\ CO map. The Planck
  maps have been recalibrated to Dame before performing the subtraction.\label{fig:damediffmaps}}
\end{figure}

After checking the behaviour of our maps in these specific locations, we
focus here on the Galactic plane as a whole.  Still
working at 30\arcm, the \Planck\ map has been degraded to $N_{\rm
  side}=512$ in order to match the {\tt HEALPix} resolution of the \citet{2001ApJ...547..792D}.
The \Planck\ map has been recalibrated to the Dame et al. data by
performing a linear regression on all pixels (${\rm CO}_{\rm type1}
\approx 1.16 \times {^{12}{\rm CO}_{\rm Dame}}$), after which operation
the difference was taken. Figure~\ref{fig:damediffmaps}
shows the residual between the \typeone\ \cooz\ map and
\citet{2001ApJ...547..792D} Galactic plane composite survey.
We observe a clear modulation pattern in the Galactic disk, producing
a positive residual at longitude $l_{gal}  < 90^{\circ}$ and a
negative residuals at $l_{gal} > 270^{\circ}$. This residual can reach
5\% of the total intensity of CO and it is interpreted as the
signature of the rotation of the Galactic disk. Because the CO line frequency
is shifted due to the Doppler effect, the emission along
different lines-of-sight with different velocities produce a different
CO response in a given bolometer. The effect is then averaged over all
bolometers in the final reconstructed CO map. Using the mean velocity
of CO emission in the \citet{2001ApJ...547..792D} survey, we can fit
the residuals as a linear function of the velocity. We find that this
effect at the \joz\ transition is well represented by ${\rm CO}_{\rm
  type1}^{[1-0]} = \left( 1 + \mathrm{v}/800\;\rm km\;s^{-1} \right) \ {\rm CO}_{\rm
  true}^{[1-0]}$, where v is the radial velocity of the gas. This formula
was also found consistent with the difference observed between survey
maps (for which the \Planck\ satellite velocity is modulated by the
satellite revolution around the Sun) and may be used to correct the
\Planck\ map for this velocity effect. Note that this effect is not
seen when performing the residuals in the same way for the
\typetwo\ and \typethree\ producta as it is i) somewhat averaged out by the use of
channel maps instead of bolometer maps; and ii) hidden by the higher
Galactic plane contamination of these maps.

\subsection{Comparison with NANTEN~II $^{12}$CO and
  $^{13}$CO $J$=1$\rightarrow$0 data}
\label{subsec:nanten}

Observations in both $^{12}$CO and $^{13}$CO have been
carried out with the NANTEN~II  millimetre-submillimetre
telescope at Atacama, between February 2010 and
October 2012. The 4-m dish provides a half-power
beam width (HPBW) of 2\parcm6 in $^{12}$CO $J$=1$\rightarrow$0. The spectrometer is a digital
Fourier spectrometer with a frequency resolution of 61 kHz and a
bandwidth of 1 GHz. A Hamming window function is applied when the FFT is performed in the
spectrometer, which results in a final frequency resolution of $79.3$~kHz. 
The velocity coverage and resolution are $\sim$ 2600
km\,s$^{-1}$ and $0.21$~km\,s$^{-1}$ for $^{12}$CO, and $2720$
~km\,s$^{-1}$ and $0.22$ ~km\,s$^{-1}$ for $^{13}$CO respectively.
All observations have been carried out in On The Fly (OTF) mode,
where the telescope constantly drives across the aree, with an
output grid of 60\arcs.
The standard size of an OTF block is 1\deg\ by 1\deg\ and at least one scan of longitudinal direction and
lateral direction have been done for each OTF block. Each scanning
data set was combined by the basket weaving method
\citep{1988A&A...190..353E} to reduce scanning noise in each OTF direction.
The intensity calibration was made with the chopper-wheel method \citep{1981ApJ...250..341K}
and absolute intensity calibration performed on
$\rho$-Oph, 
Ori-KL, 
M17SW, 
and Perseus. 

\begin{figure}
\centering
\includegraphics[width=\columnwidth, trim=0cm 0cm 0cm 7.5cm, clip]{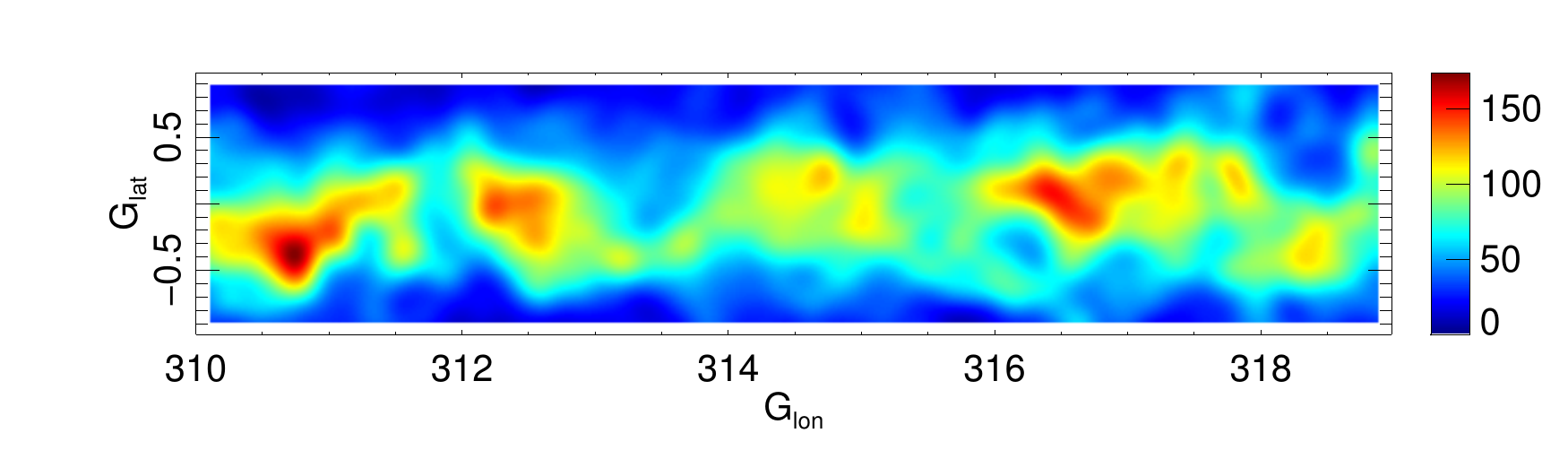}
\includegraphics[width=\columnwidth, trim=0cm 0cm 0cm 7.5cm, clip]{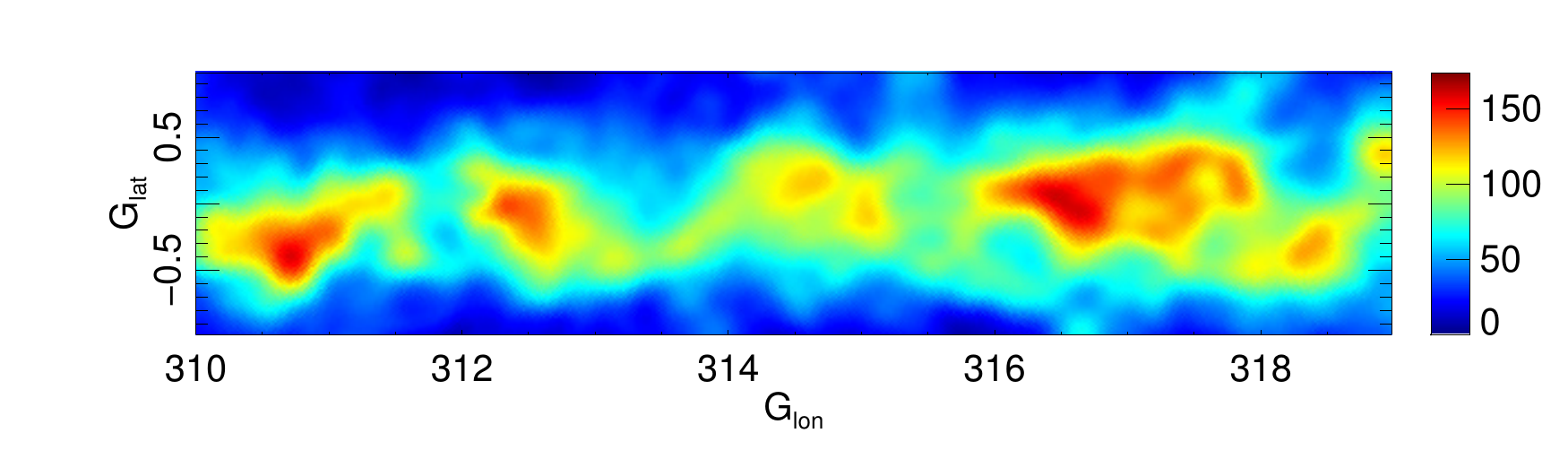}
\includegraphics[width=\columnwidth]{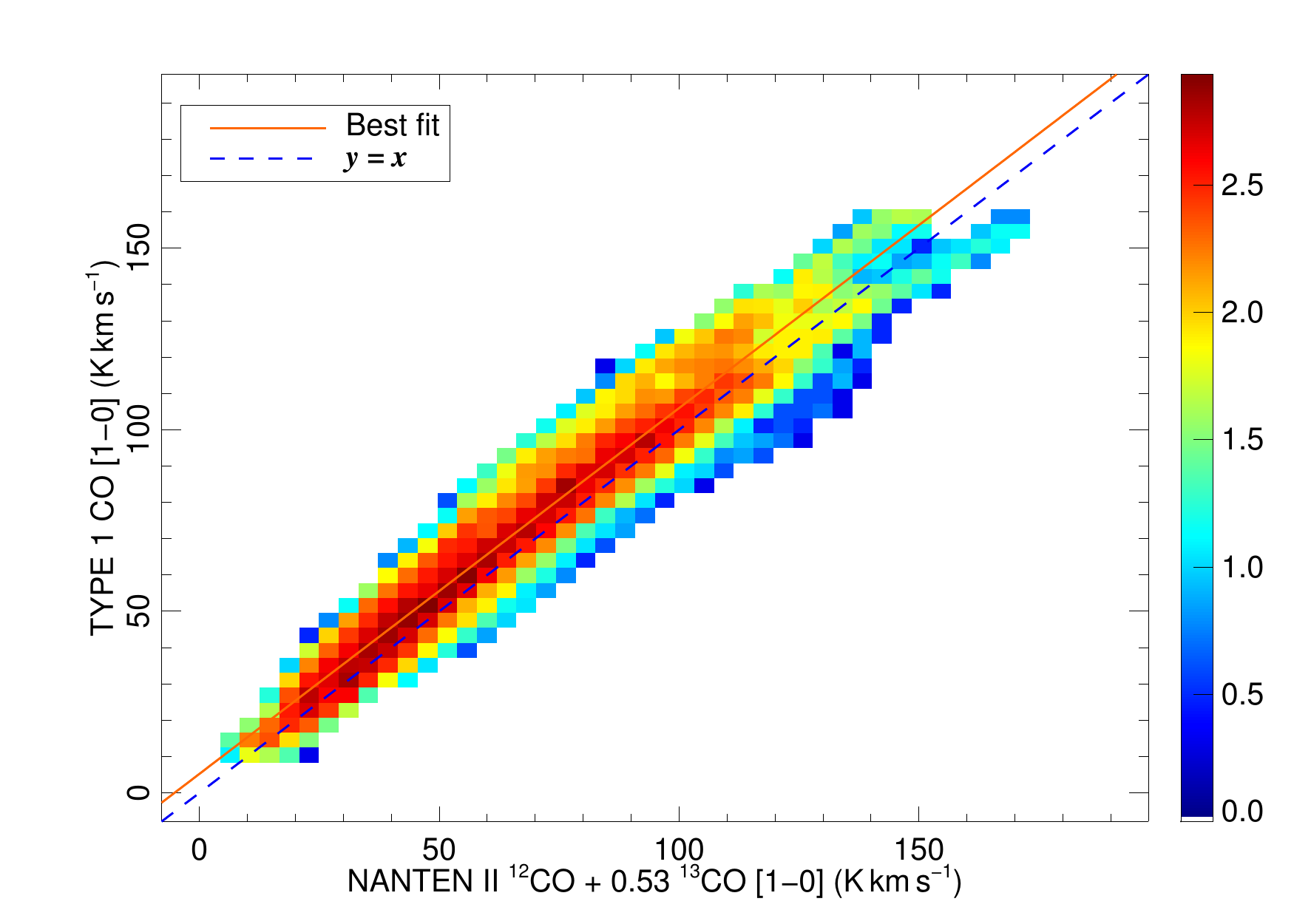}
\caption{Comparison of the  \Planck\ ($J=1\rightarrow 0$) \typeone\ map to the
  NANTEN~II survey in a $9^\circ \times 2^\circ$ section of the
  Galactic plane. \emph{Top}: Combined
  NANTENII map as ${^{12}{\rm CO}}+0.53\times {^{13}{\rm CO}}$, where the factor applied to
  $^{13}\rm CO$ corresponds to the bandpass estimate of the $^{13}\rm
  CO$ contamination in the \typeone\ map (see Sect.~\ref{subsubsec:13co}). \emph{Middle}:
  \Planck\ \typeone\ $J$=1$\rightarrow$0 map. 
  \emph{Bottom}: correlation plot between the two maps. The log
  colour scale represents the number of pixels in a given intensity
  bin. The solid and dashed lines represent $y=x$ and the best linear fit, respectively.
  The best-fit slope and intercept are $(1.007 \pm 0.001)$ and $(5.243 \pm 0.032)$, respectively.
  \label{fig:nantencomparison}}
\end{figure}

The soon-to-be-published data we use here consist of a $9^\circ \times 2^\circ$ Galactic plane patch
around Galactic coordinates (315\deg, 0\deg).
We smoothed the NANTEN and \Planck\ maps to a 15\arcm\ resolution
before carrying out the comparison. This resolution is chosen as it
corresponds to that of the \typetwo\ CO map and allows a significant
noise reduction in the \typeone\ map (which comes with a 9.65\arcm
native resolution). Remembering that we expect a $0.53\times
{^{13}\rm{CO}}$ contribution to the \typeone\ map (see
Sect.~\ref{subsubsec:13co}), we show in Fig.~\ref{fig:nantencomparison}
the combined NANTEN~II ${^{12}{\rm CO}}+0.53\times {^{13}{\rm CO}}$ on
the top panel, the \typeone\ map in the middle and the correlation
plot in the bottom panel. The agreement between the two sets of data
is good  with a correlation of 0.95. While not shown here, we
perform this test for the \typetwo\ CO map as well, using this time
$1.14\times {^{13}{\rm CO}}$ as is estimated in
Sect.~\ref{subsubsec:13co}. The agreement is once again satisfactory 
with a best-fit slope of 1.00. In both cases, the level of the residuals between
the best-fit and the combination of NANTEN data is found to be
$\lesssim 10\%$ 
and thus compatible with the calibration uncertainties.

This test is important, as it validates our understanding of the
${^{13}{\rm CO}}$ contamination in the \Planck\ \cooz\ maps. It also
highlights that the \typetwo\ \cooz\ map does not suffer from major dust
contamination in the Galactic plane since, had this been the case, the
correlation with NANTEN would have presented a slope greater than
one. We also performed an independent check
of the $^{13}$CO contamination based on FCRAO data, which corroborates
the results obtained with NANTEN. This is the purpose of the next section.

\subsection{Comparison with FCRAO $^{13}$CO $J$=1$\rightarrow$0 data}
\label{subsec:fcrao}
Along the same lines as the comparison performed with NANTEN~II data, it
is possible to gain further confidence in our estimation of the
$^{13}$CO contamination using FCRAO $^{13}$\cooz\ data along the
$^{12}$\cooz\ data of \citet{2001ApJ...547..792D}. In
this analysis, we use the publicly available Boston University FCRAO Galactic Ring
Survey\footnote{The data of the BU-FCRAO-GRS may be retrieved at
  \url{http://www.bu.edu/Galacticring/}} (FCRAO-GRS) that has been described in
\citet{2006ApJS..163..145J}. It consists of a 75.4~deg$^2$ survey of
the Galactic plane, between 18\deg$< l <$55\pdeg7 and $|b|<1\deg$ at
46\arcs\ resolution.

\begin{figure}
\centering
\includegraphics[width=\columnwidth]{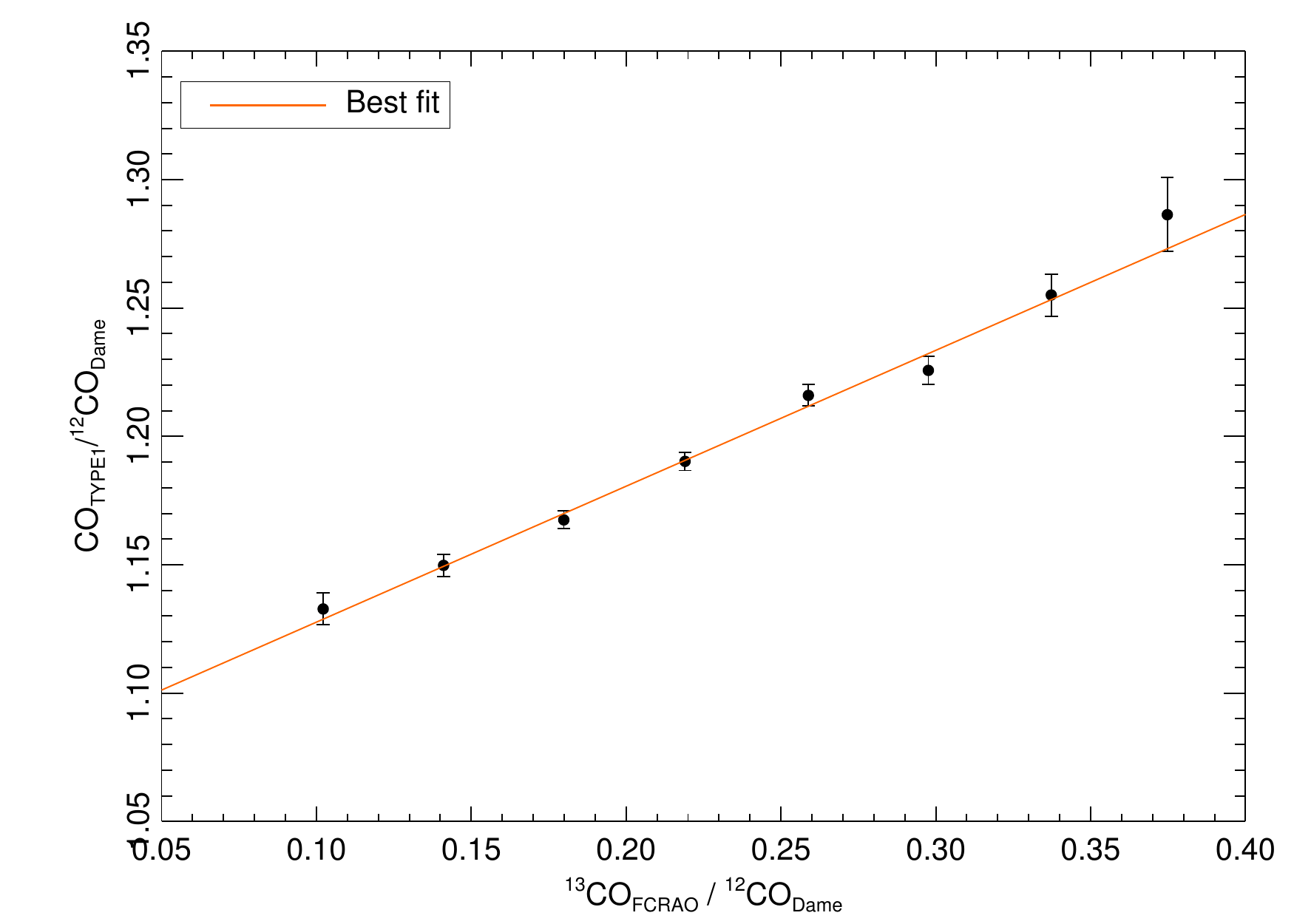}
\caption{Average ratio $\langle {\rm CO}_{\rm type1}\rangle/\langle{\rm
  {^{12}\rm CO}}_{\rm Dame}\rangle$ in bins of the ${^{13}{\rm CO_{\rm
        FCRAO}}}/{^{12}{\rm CO}}_{\rm Dame}$ ratio. The best linear fit is represented by
        the orange solid line. The best-fit slope and intercept are
        $(0.53 \pm 0.03)$ and $(-1.07 \pm 0.01)$, respectively.\label{fig:fcrao}}
\end{figure}

We first re-project the FCRAO-GRS data into the {\tt HEALPix} pixelization scheme
($N_{\rm side}=2048$) using a nearest grid-point approach and correct
for the 0.48 beam efficiency given in \citet{2006ApJS..163..145J}. Binning over the ratio ${^{13}{\rm CO}}_{\rm
  FCRAO} /{^{12}{\rm CO}}_{\rm
  Dame} $, we compute
the average ratio $\langle{\rm CO}_{\rm type1}\rangle/\langle{^{12}{\rm CO}}_{\rm
  Dame}\rangle$ in each bin and plot the correlation in
Fig.~\ref{fig:fcrao}. Error bars are obtained assuming white noise,
the amplitude of which is estimated at high Galactic latitude
($|b|>60^\circ$). The standard deviation in each bin is then simply
estimated using the number of pixels in the bin. We assume the noise in the
\citet{2001ApJ...547..792D} map to be negligible.

The correlation between the two ratios is good and from the
best-fit of the linear regression we see that the
\Planck\ \typeone\ map can be written as
\[
{\rm CO}_{\rm {\sc type1}}^{[1-0]} = 0.53 \times {^{13}{\rm CO_{\rm FCRAO}}} + 1.07 \times
{^{12}{\rm CO_{\rm Dame}}}\;,
\]
which is agreement with the $0.53\times ^{13}$CO
contamination estimate we made from the bandpass coefficients (see
Sect.~\ref{subsubsec:13co}). This test shows, in a completely independent
manner to that of the NANTEN comparison, that we have an excellent
grasp on the CO content of the \typeone\ map.

\subsection{Comparison with the AMANOGAWA-2SB CO $J$=2$\rightarrow$1 survey}
\label{subsec:amanog}

The AMANOGAWA-2SB survey \citep{2012ASPC..458..221H,2010PASJ...62.1277Y}
carried out simultaneous Galactic plane ($l=10^{\circ}-245^{\circ}$) observations of
the $^{12}$CO and $^{13}$\coto\ transitions with the Tokyo-NRO 60-cm
telescope, with a resolution and grid-spacing of 9 and
3.75\arcm\ respectively. 
Here, we use a $35^\circ \times 8^\circ$ section of the  AMANOGAWA-2SB $^{12}$CO
Galactic plane survey for comparison to the \Planck\ \coto\ maps. 

\begin{figure}
\centering
\includegraphics[width=\columnwidth,  trim=0cm 0cm 0cm 6cm, clip]{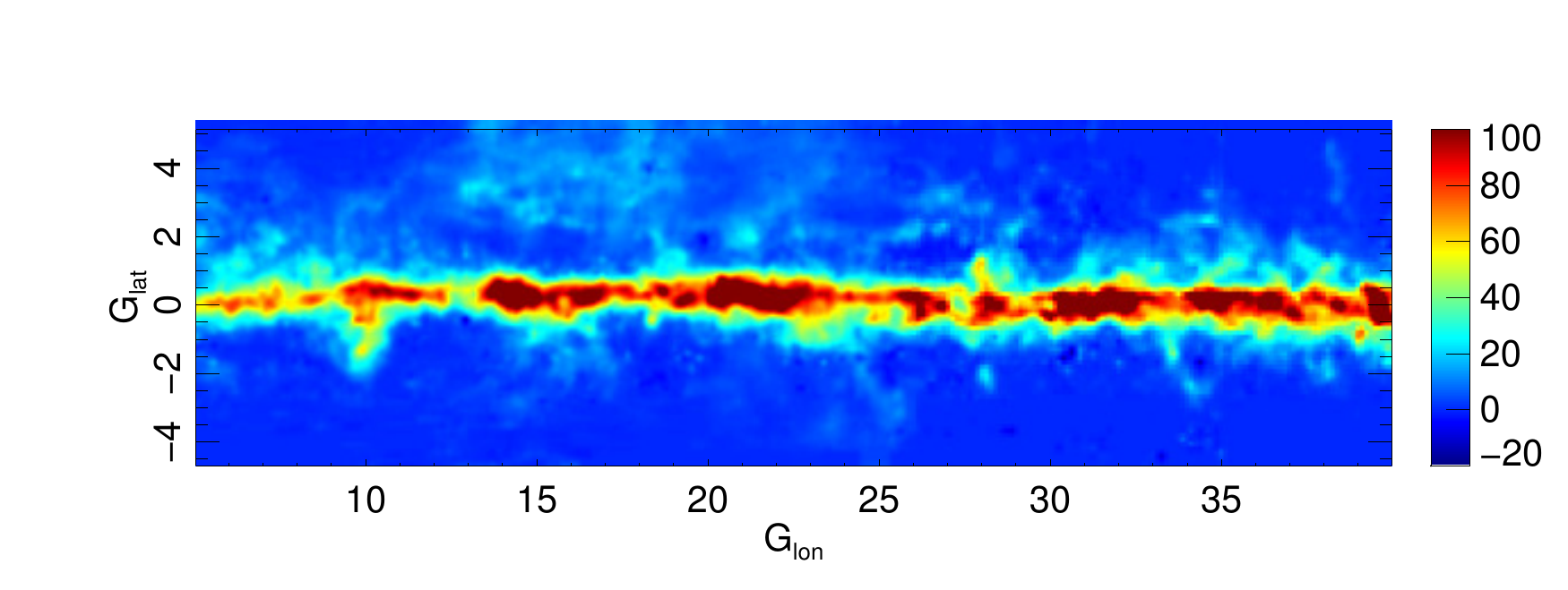}
\includegraphics[width=\columnwidth,  trim=0cm 0cm 0cm 6cm, clip]{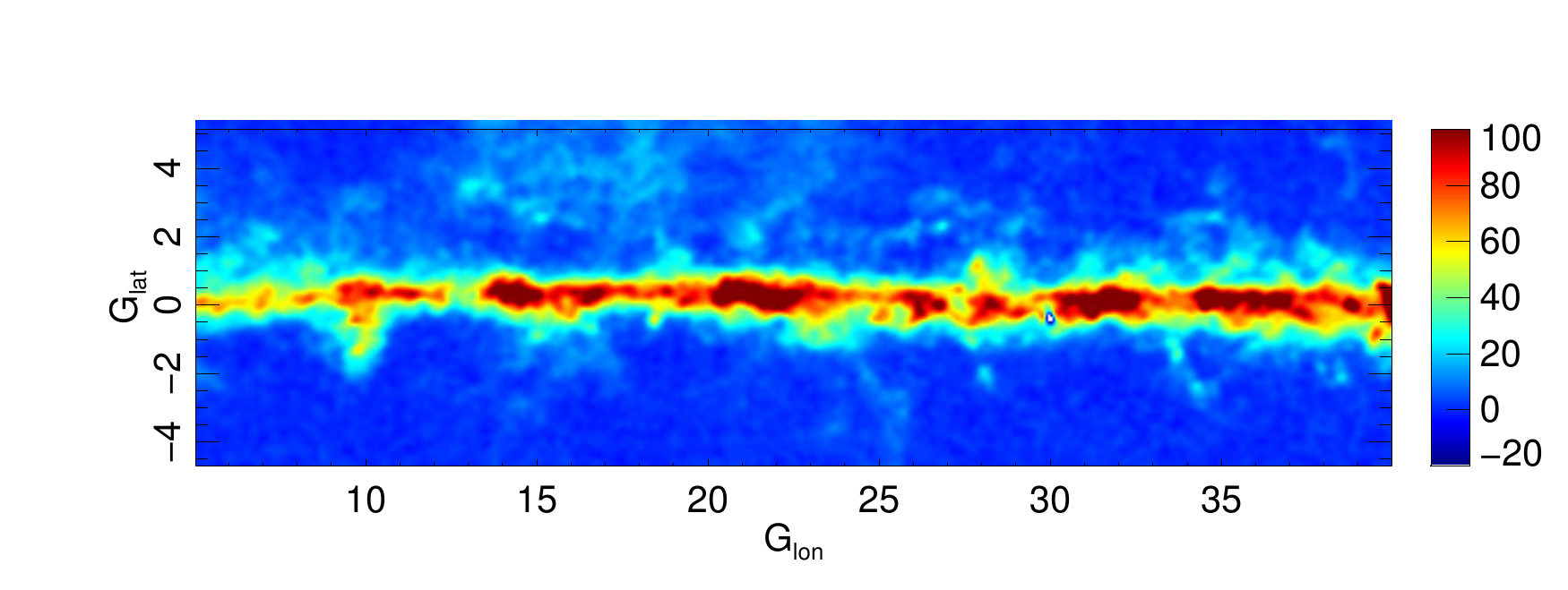}
\includegraphics[width=\columnwidth]{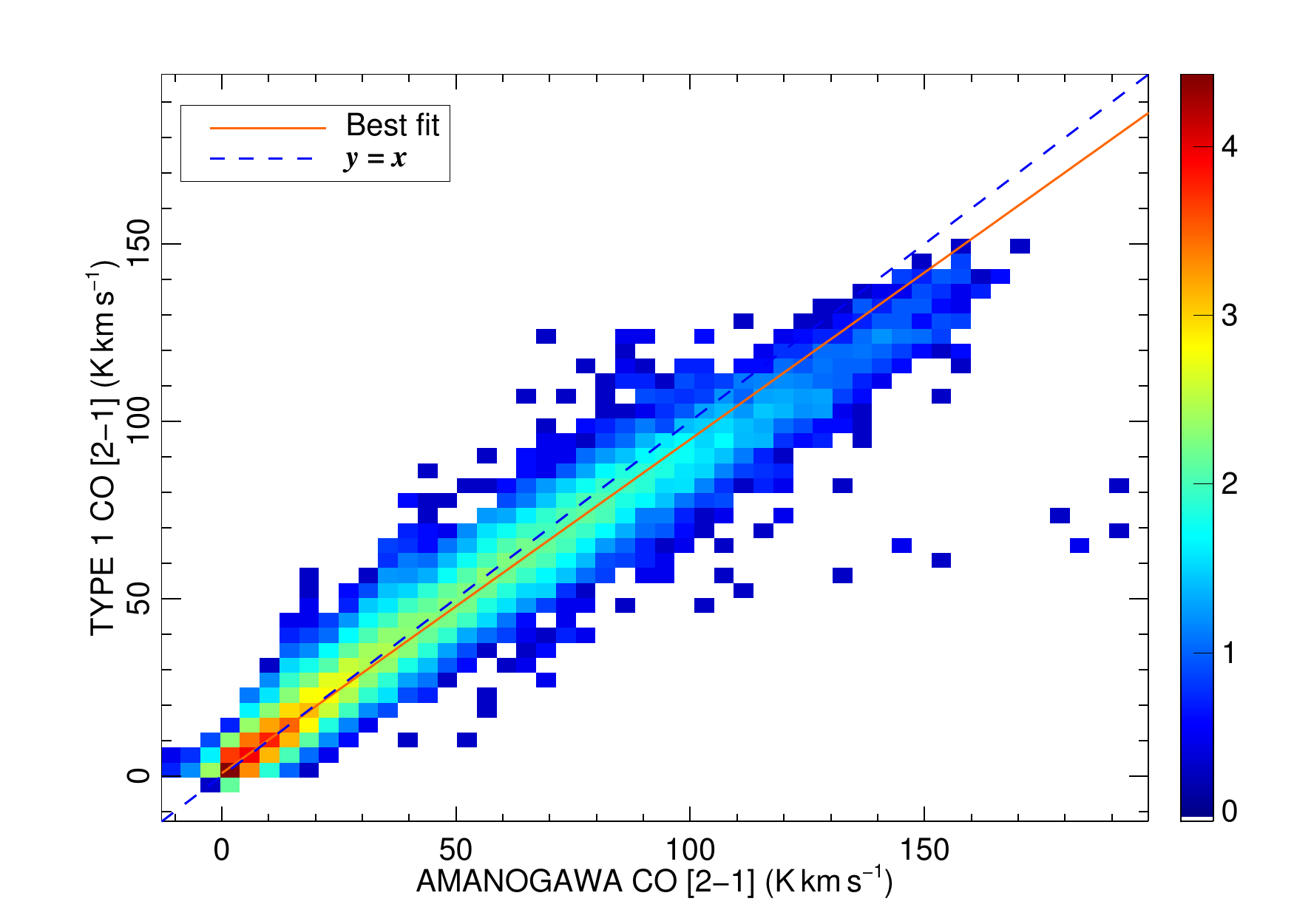}
\caption{Comparison of the \Planck\ J=2-1 \typeone\ CO map to the
  AMANOGAWA-2SB Galactic plane survey. From top to bottom we display
  the AMANOGAWA data, the \typeone data in the same region and the
  correlation between the two maps. The log
  colour scale represents the number of pixels in a given intensity
  bin. The solid orange and dashed blue lines represent $y=x$ and the best linar fit, respectively.
  The best-fit slope and intercept are $(0.941 \pm 0.001)$ and $(0.792 \pm 0.009)$, respectively.\label{fig:AMANOGAWA-2SB_maps}}
\end{figure}

The area under scrutiny is shown in
Fig.~\ref{fig:AMANOGAWA-2SB_maps} for the AMANOGAWA data set (top) and the
\Planck\ \typeone\ CO [2-1] map (middle) at the working resolution of
15\arcm. 
The correlation between the two data sets in given in the bottom
panel and shows remarkable agreement with a best-fit value for the
correlation of 0.95. Conversely to the \cooz\ map, no $^{13}$CO
contribution was required to match the two data sets. This is once
again in agreement to what was predicted in
Sect.~\ref{subsubsec:13co} where we found that the $^{13}$CO
contribution to the \typeone\ \joz\ line should be negligible, given
the weighted bandpass transmission of the isotopologue in the 217~GHz
channel. The residuals between
the best-fit and the AMANOGAWA data are found to be
about $5\%$, which is compatible with the combined calibration uncertainties
of \Planck\ and AMANOGAWA.

\subsection{Comparison with the FIRAS CO \joz, $J$=2$\rightarrow$1 and $J$=3$\rightarrow$2 surveys}
\label{subsec:firas}
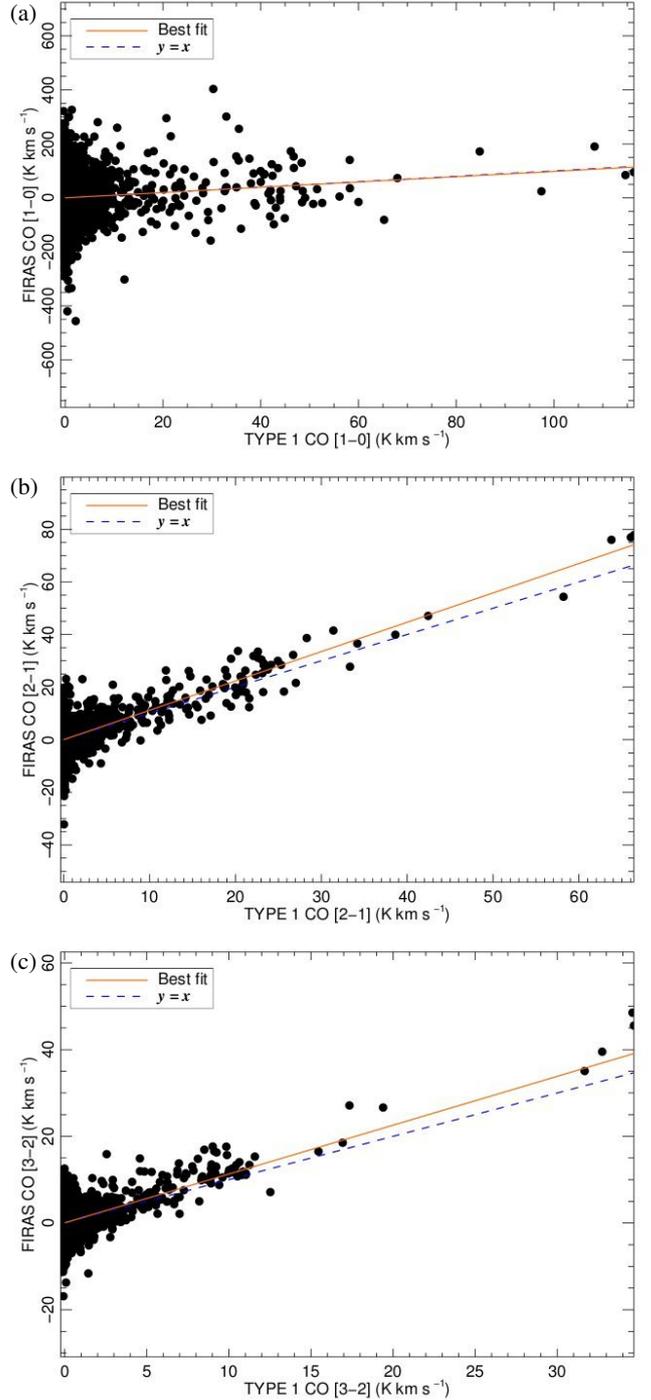
\begin{figure}[th]
\setlength{\unitlength}{\columnwidth}
\begin{picture}(1,2.1)
\put(0,1.4){\includegraphics[width=\columnwidth]{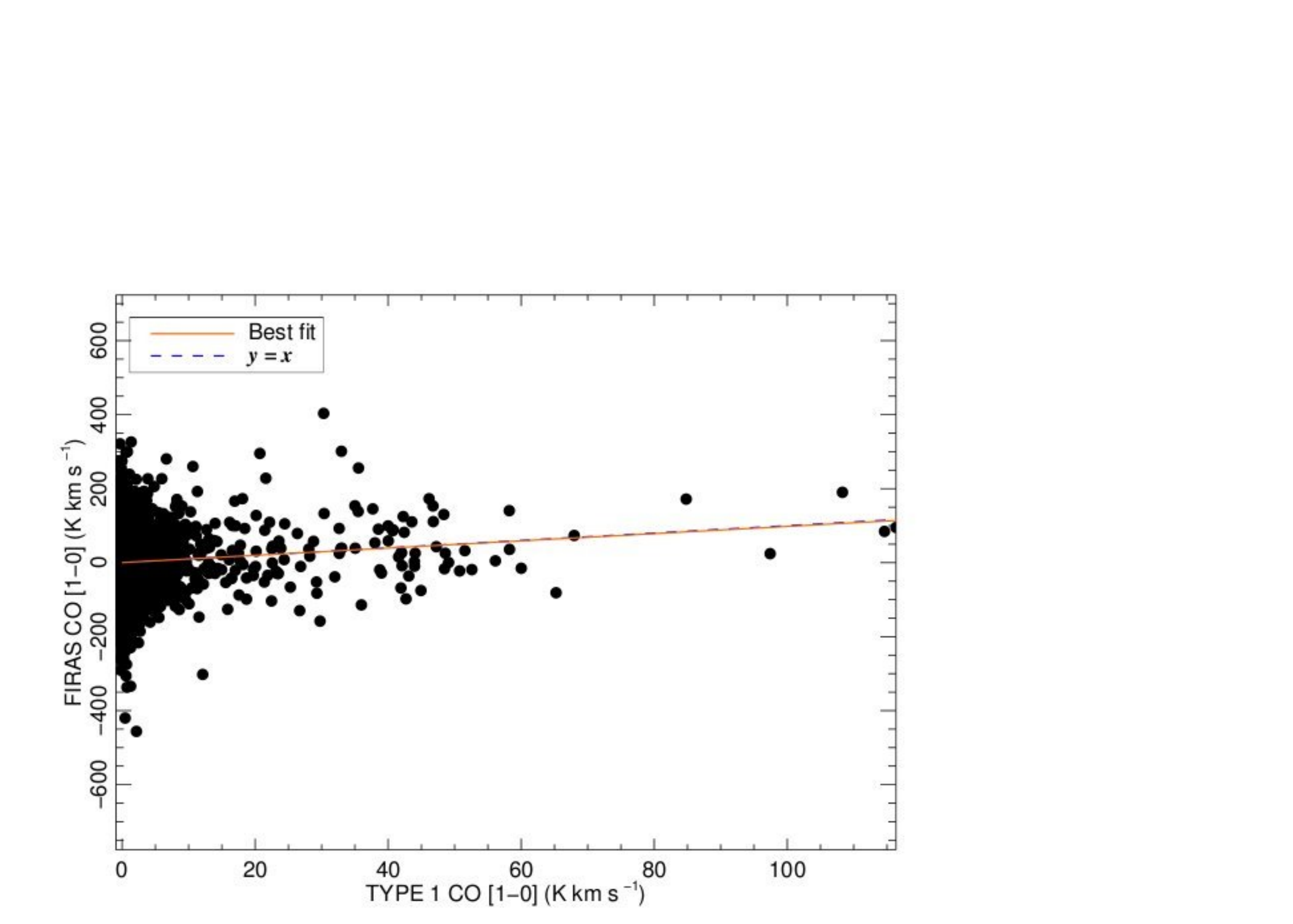}}
\put(0,0.7){\includegraphics[width=\columnwidth]{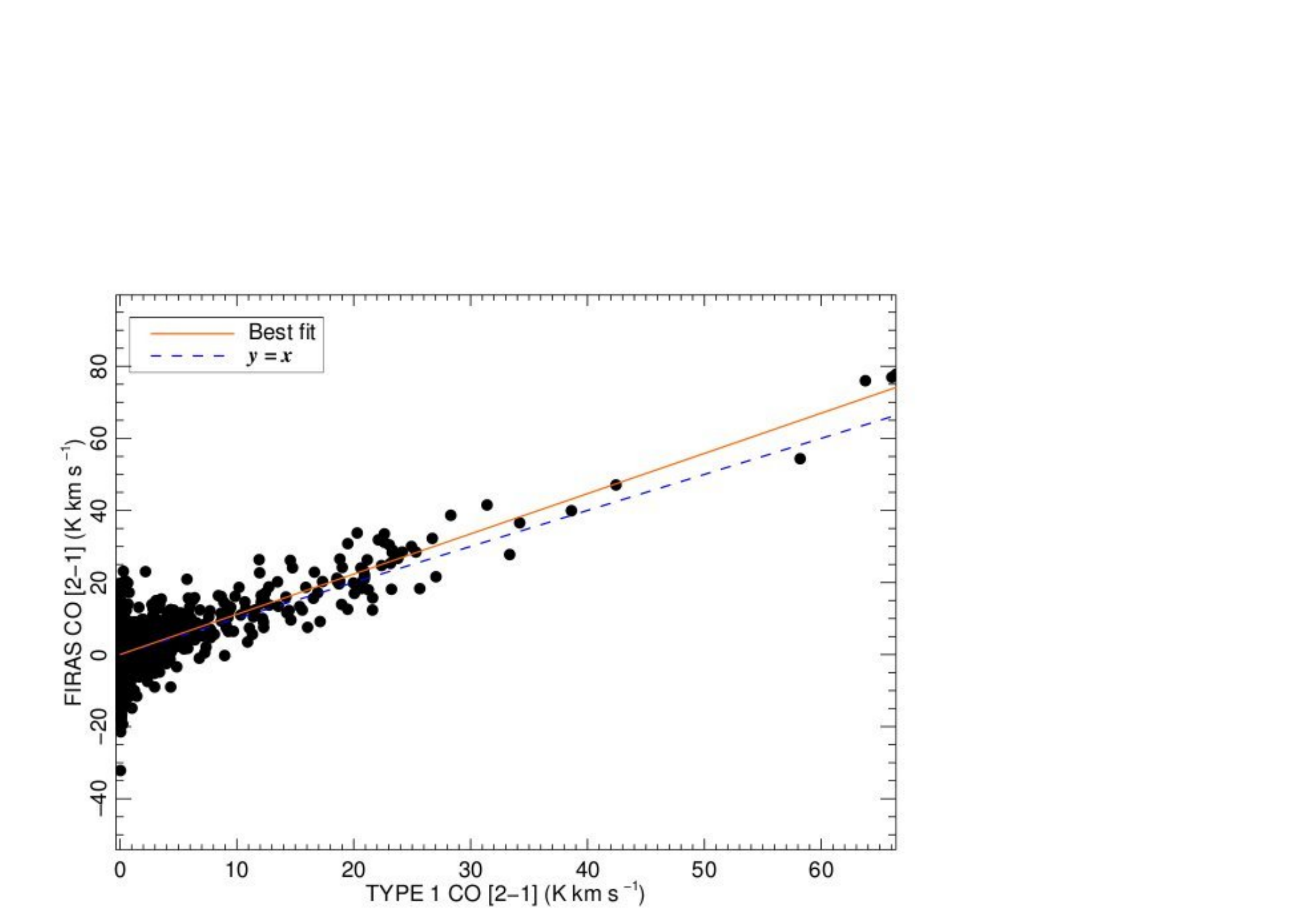}}
\put(0,0){\includegraphics[width=\columnwidth]{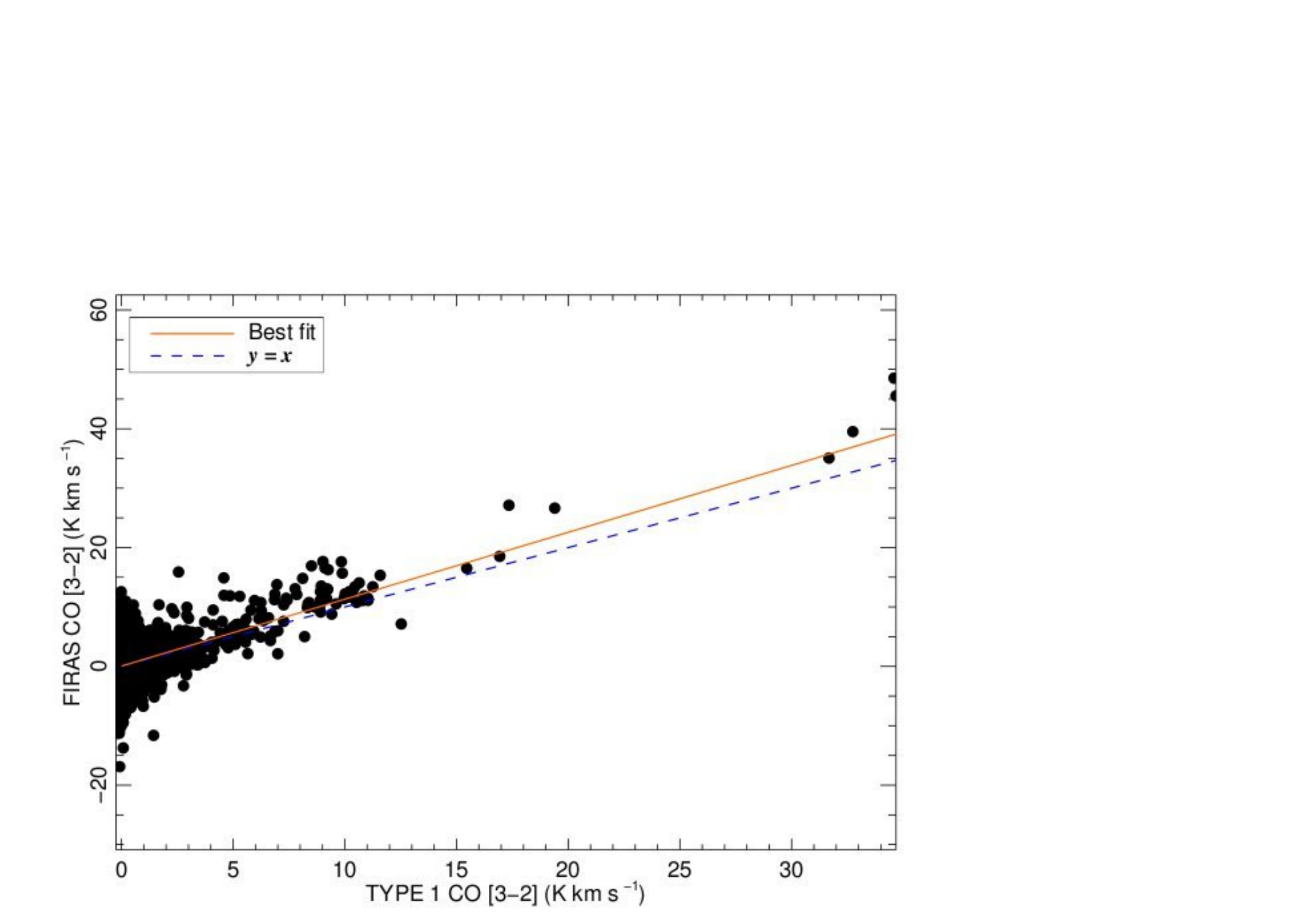}}
\put(0.05,2.05){(a)}
\put(0.05,1.35){(b)}
\put(0.05,0.65){(c)}
\end{picture}
\caption{Comparison of the \typeone\ \Planck\ CO maps with the FIRAS
\joz\ (top), \jto\ (middle) and \jtt\ (bottom) data.
 The solid orange and dashed blue lines represent $y=x$ and the best linear fit, respectively.
  The best-fit slopes and intercepts are: (a) $(0.98 \pm 0.17)$ and $(0.33 \pm 0.65)$; (b) $(1.12 \pm 0.02)$ and $(-0.01 \pm 0.04)$; and (c) $(1.13 \pm 0.03)$ and $(0.04 \pm 0.02)$.\label{fig:compare_firas}}
\end{figure}

\cite{1999ApJ...526..207F} have analyzed a by-product of the
Far Infrared Absolute Spectrophotometer (FIRAS on {\it COBE}
launched in 1989) sky maps, namely integrated velocity maps of well-known
Galactic lines. These include carbon, nitrogen and water lines, but,
more importantly for the present study, the complete $^{12}$CO
rotation ladder. These are absolutely calibrated all-sky measurements.
FIRAS has a better spectral resolution (13.5~GHz) than \Planck-HFI but
a much lower angular resolution. 
FIRAS integrated CO lines are retrieved from the Lambda
website\footnote{\url{http://lambda.gsfc.nasa.gov}}. 
The first three transitions
at 115.27, 230.54 and 345.80~GHz come from the low frequency emission
line maps. They are converted from $\mathrm{nW\, m^{-2}\, sr^{-1}}$ to 
$\mathrm{K\, km\,  s^{-1}}$ with the coefficient
$10^{-12}c^3/2k\nu_{\rm CO}^3$. Error maps are processed in the same way and we neglect gain
uncertainties.

\Planck\ CO maps are convolved with the FIRAS beam in the same way as
described in Appendix A of \cite{planck2013-p03b} and compared to FIRAS line maps
within the FIRAS pixelization scheme. This is simply done by taking the
\Planck\ value at the grid point nearest to the FIRAS pixel within 15\arcm.
Uncertainties are much larger for FIRAS than \Planck; the noisiest FIRAS values are not kept
in the comparison. The cut-off is chosen as 160, 11 and 7~$\mathrm{K\, km\, s^{-1}}$
for the \cooz, \coto, and \cott\ lines, respectively. The FIRAS Galactic centre
values may not be reliable because of the large velocity spread and we
therefore discard any pixels within 4\deg of the Galactic centre.
There is no possible tuning in this straightforward comparison. \\

Fig.~\ref{fig:compare_firas} shows the correlation
plot between the \typeone\ \Planck\ CO data and the FIRAS \joz\ (top), 
\jto\ (middle) and \jtt\ (bottom) data. In each panel, the dashed blue line
corresponds to $y=x$,  while the orange line gives
the best fit.  We find that FIRAS and \Planck\ all-sky CO data are consistent with a linear
relationship.  For the \cooz\ line, \cite{1999ApJ...526..207F} claim a detection
only in the Galactic centre. With \Planck\ we obtain a statistical detection
of \cooz\ with FIRAS outside the Galactic centre at the 7~$\sigma$ level.
For the two other lines, FIRAS measurements tend to overestimate the CO
emission with respect to \Planck\ by at most 10\%. This is well within the
absolute calibration error for \Planck\ and beam uncertainties for
FIRAS.

\subsection{Comparison with HARP/ACSIS CO $J$=3$\rightarrow$2 data}
\label{subsec:harp}

Surveys in of the \cott\ transition are not as numerous as those of
lower transitions, which makes assessing the quality of the
\typeone\ \jtt\ map more difficult. 
We used part of the JCMT HARP/ACSIS
$^{12}$\cott\ Galactic plane data (soon to be published). Details on
the HARP/ACSIS system and calibration can be found in
\citet{2009MNRAS.399.1026B,2012MNRAS.422.2992P}.

\begin{figure}
\centering
\includegraphics[width=\columnwidth]{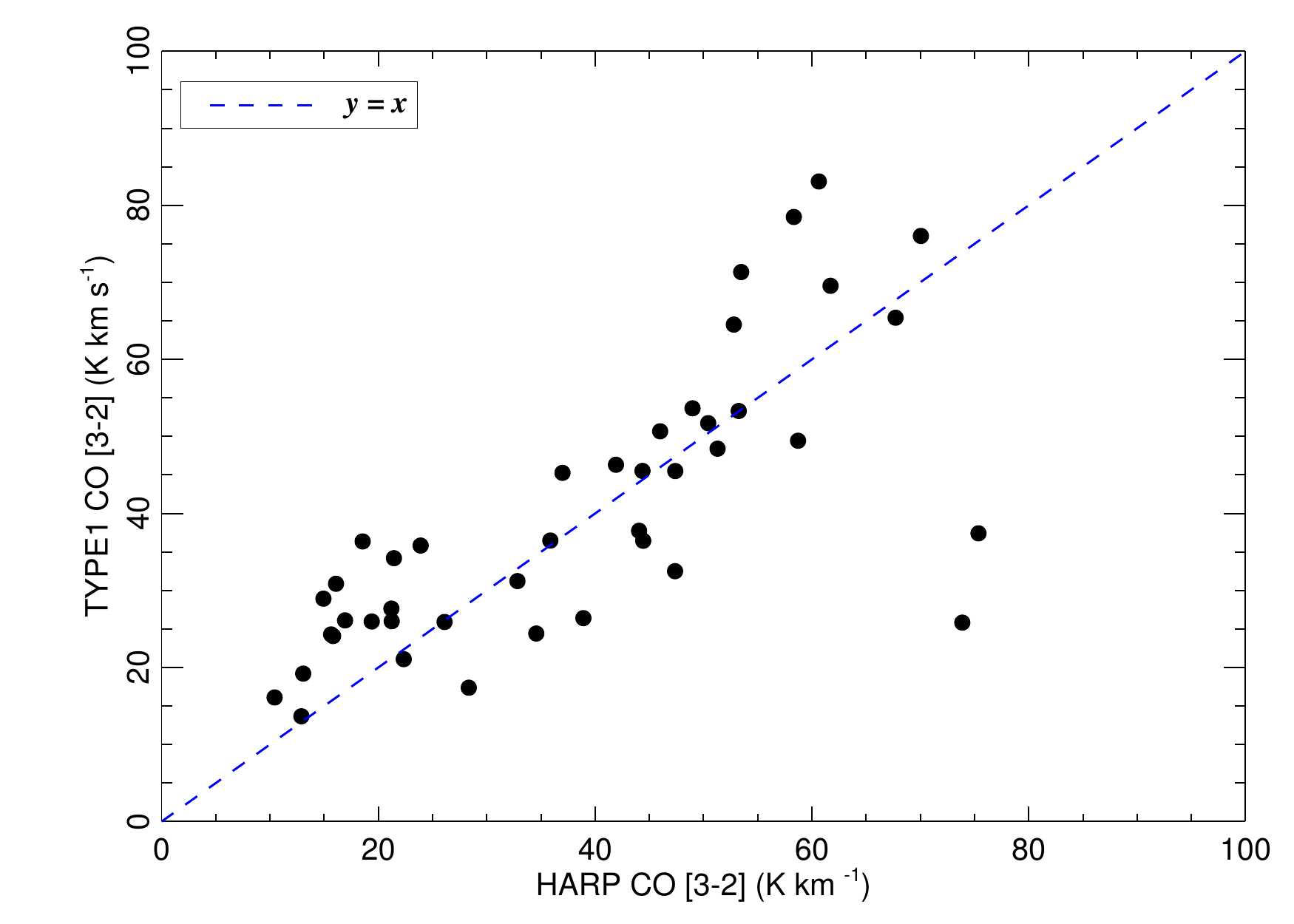}
\caption{Correlation plot between the latitude-collapsed longitudinal
  profiles (see text for details) of \typeone\ $J$=3$\rightarrow$2 HFI CO map and
  the HARP $J$=3$\rightarrow$2 survey.  The solid orange and dashed blue lines represent $y=x$.
 .\label{fig:HARP_3-2}}
\end{figure}

We use a 2$^\circ$ long section of the Galactic plane from $l=12.5$ to $l=14.5$ where the
survey has latitude range $\pm 0.5\deg$. The original HARP data were not clipped at some  n $\sigma$
threshold but collapsed over the whole spectral range, after performing a
heavy smooth on all regions with signal $<3\sigma$. This ensures there
is no significant emission missing, while minimizing the addition of
noise. The data were then smoothed to about the 4.5\arcm\ resolution, to mimic
the \Planck\ resolution but with pixels of around 5\arcs. The two maps
were then brought down to the same pixelization, the least constraining
rebinning resulting in a $43\times 23$ pixels patch, with about $2.7$\arcm\ pixels. Because of the noise level inherent to the
\typeone\ map, a large dispersion is seen when simply performing the
correlation between all pixels of these two patches. Therefore, we
perform  another rebinning, collapsing the latitude dimension to get a
43 point ``average'' longitudinal profile of the two maps. The
correlation of these two profiles is plotted in
Fig.~\ref{fig:HARP_3-2}.

Despite our rebinning, the correlation remains quite noisy, but the
figure shows that the $y=x$ line is a fair representation of the data
when ignoring the two outliers. More data would be required to
perform a thorough characterisation of the \typeone\ \jtt\ map.

\subsection{High Galactic latitude structures}
\label{subsec:mhhst}

\begin{figure}
\centering
\includegraphics[width=\columnwidth]{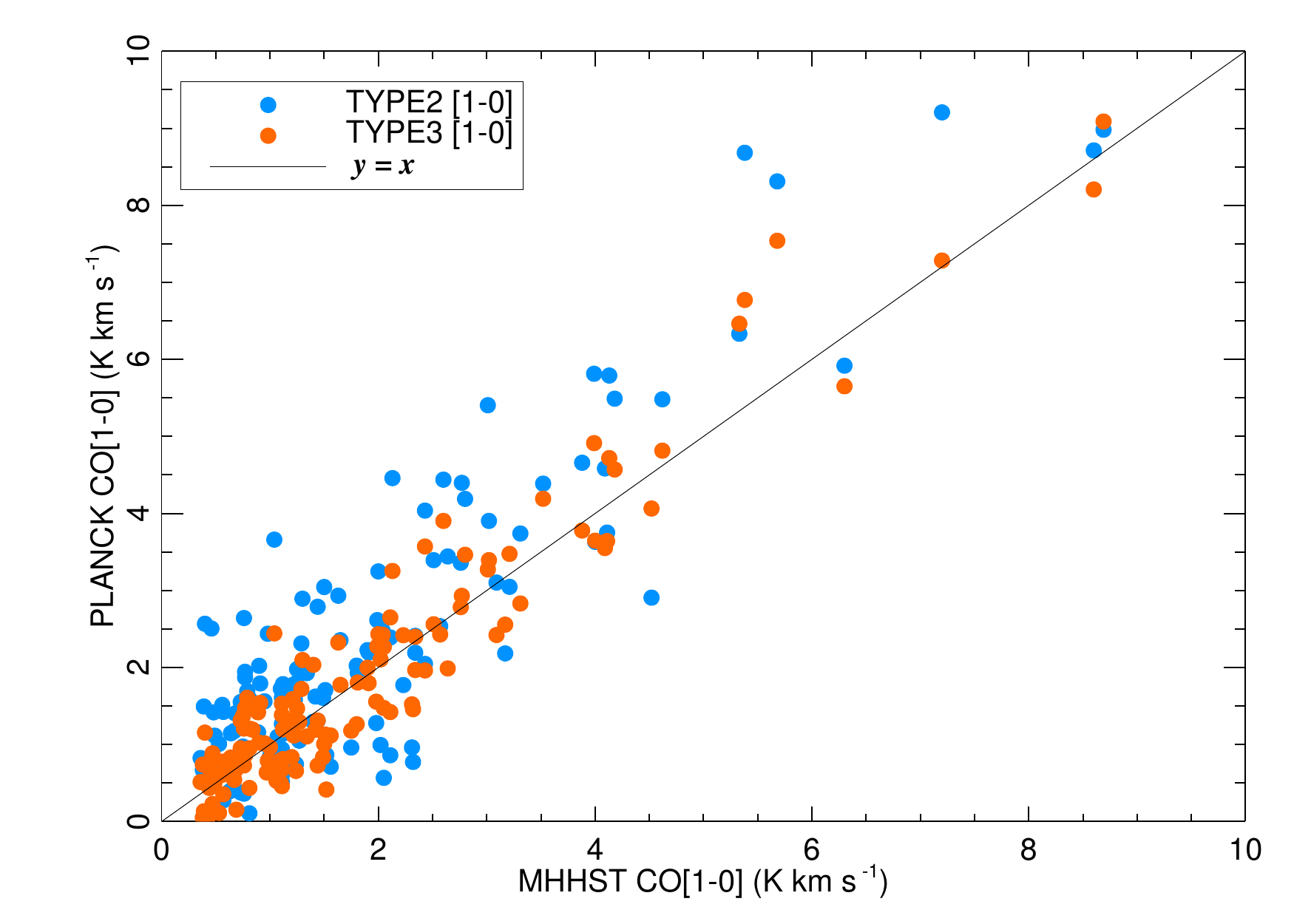}
\includegraphics[width=\columnwidth]{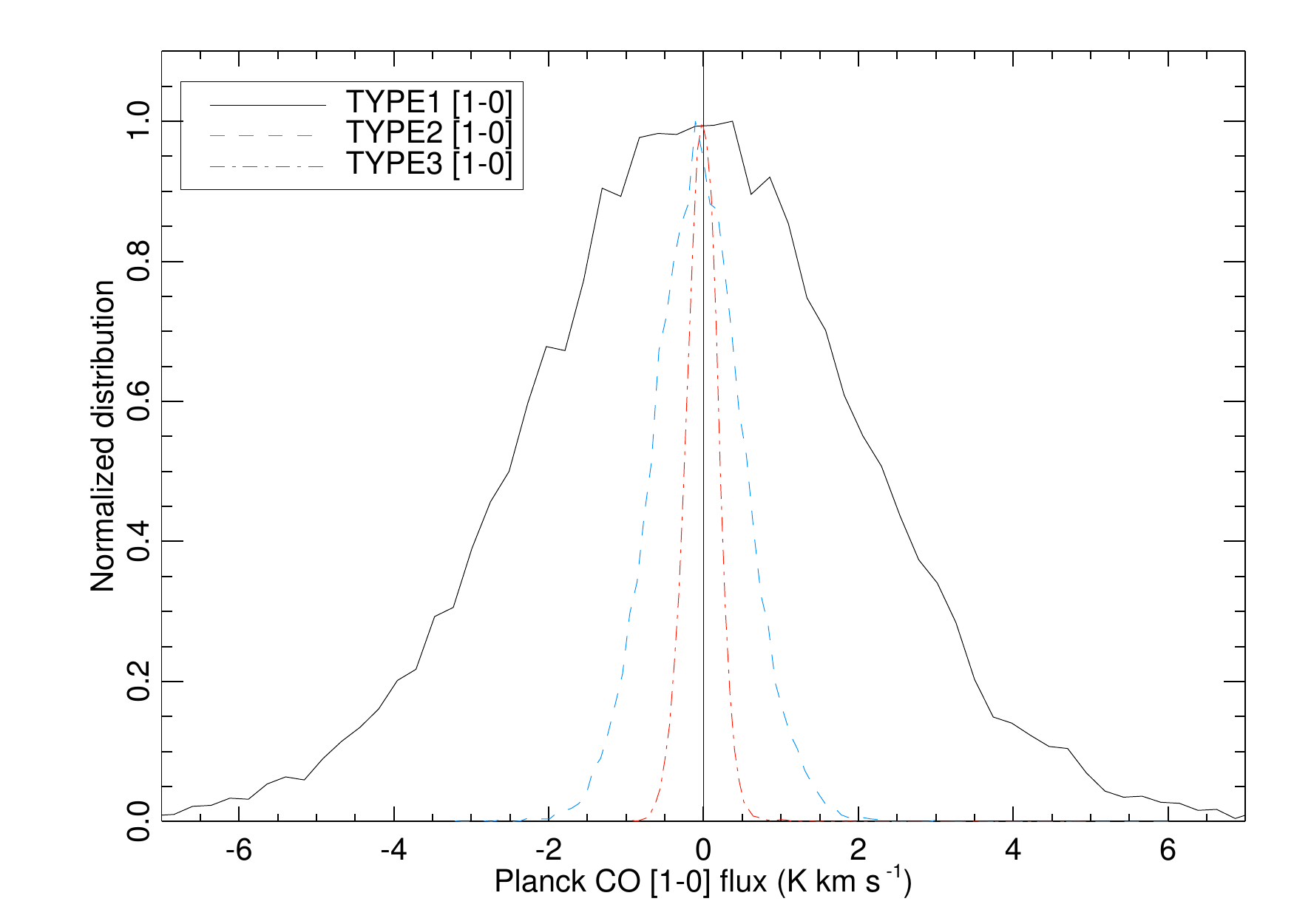}
\caption{Comparison of the \Planck\ \cooz\ maps with the high Galactic
  latitudes CO detections and non detections from MHHST00. \emph{Top}:
  correlation between the CO detections in MHHST00 and the
  \Planck\ \typetwo\ and \typethree\ maps. The latter was first
  smoothed to 15\arcm. \emph{Bottom}: distribution of all
  \Planck\ \cooz\ maps (at 15\arcm\ resolution) at the non-detection
  positions of MHHST00. The width of the distribution is compatible
  with the noise in the maps. Differences with respect to the values in Table~\ref{tab:prod_summary}
  can be explained by the inhomogeneities of the noise in the \Planck\ maps.
\label{fig:Planck_fluxes_on_cfa_nodetections}}
\end{figure}

In order to achieve a deep, high-Galactic-latitude statistical
comparison of the \Planck\ CO maps with ground-based observations,
we use the observations performed at CfA and described in
\citet{1998ApJ...492..205H} (HMT98) and  \citet{2000ApJ...535..167M} (MHHST00) for the
North and South Galactic hemisphere respectively. These data consist of 8\arcm\
beam observations of the  $|b|>30^\circ$ sky, on a regularly-spaced
grid, observable from Cambridge
(MA). In total, 10,443 points in the north Galactic hemisphere and
4,934 in the south were observed. A detection threshold
of~0.3 K$_{\rm RJ}\cdot$km$\cdot$s$^{-1}$ (3$\sigma$) was considered.
In total 26 detections were reported in the North and 133 in the South.
 
First, for each of these detection points, we compute the
corresponding \Planck\ \typetwo\ and \typethree\ CO [1-0] fluxes (in a
10\arcmin\ diameter disc). The \typeone\ map was not selected for this
analysis as its noise level makes it unsuitable for high Galactic latitudes. The
top panel in Fig. \ref{fig:Planck_fluxes_on_cfa_nodetections} shows
the correlation of the \Planck\ \typetwo\ and \typethree\ fluxes as a function of the 159
HMT98 and MHHST00 fluxes. A reasonable agreement and compatible
absolute calibration is reached in these faint CO-emitting regions.
The bottom panel presents the histogram of all three types of
\Planck\ CO maps fluxes at the 15,218 locations where no detection is
expected from the CfA surveys. The Gaussian distributions obtained are
all centred on zero and have a standard deviation compatible with the
noise level of the map. Discrepancies with respect to the values in Table~\ref{tab:prod_summary}
can be explained the inhomogeneities of the noise in the \Planck\ maps.
This comparison to the HMT98 and MHHST00
highlights the high quality of the \Planck\ CO maps in the faintest CO regions.

 \section{Conclusions}
 \label{sec:conclusion}
The \Planck-HFI maps contain a non-negligible component that is attributed
to the emission, in the HFI spectral bands, of the three first CO rotation
transitions coming from the interstellar medium. Several component separation methods
are used to isolate this component from the others (CMB and diffuse
Galactic emission, mainly thermal dust), by using the spectral
diversity and large coverage of 
\Planck-HFI. 

This paper presents all-sky maps of the velocity-integrated emission of
CO in the \joz, \jto, and \jtt\ transitions. Three different types
of maps are produced: i) \typeone\ maps obtained from a single-channel
analysis for the three transitions; ii) \typetwo\ maps
available for the first two transitions only and coming from a
multi-channel analysis; and iii) a single \typethree\ map relying on
a multi-line approach.
\\
We have characterized the maps in terms of resolution, noise and
systematic effects by a cross-comparison and internal validation of the
maps. Comparison to external CO data set was also performed and 
a general agreement is found, always within the estimated calibration
uncertainties of the data. Note however that the \typeone\ \jtt\ map
could not be as thoroughly tested as the others because of the few
external \cott\ data sets available. 

In summary, \typeone\ maps are the
noisiest but suffer from the least systematic errors and foreground
contamination. The \typetwo\ CO products provide high S/N single-line
maps at the price of larger systematics and foreground
contamination, and are of lower resolution.
Finally, the \typethree\ map gives a high S/N and high resolution CO map but
relies on the assumption of constant line ratios across the sky. Both
\typetwo\ and \typethree\ maps are more suitable for a use at intermediate and
high-Galactic latitudes.  The \typeone\ maps are,
by contrast, more reliable in terms of foreground contamination in the Galactic plane
and on high CO intensity regions. In all these maps the $^{13}$CO contribution
is not corrected and needs to be accounted for in any further scientific analysis.

The \Planck\ CO maps are used within the \Planck\ collaboration for the definition
of regions of the sky where CMB, thermal SZ and cosmic infrared
background cosmological studies can safely be performed. 
The \Planck\ CO maps presented in this paper are delivered as part of the \Planck\ nominal mission public archive.

As discussed in~\citet{planck2013-coscience}, Planck multi-line CO survey reveals that the bulk of the Galactic CO emission is subthermally excited and originates in a diffuse component. The highest sensitivity CO maps prove that CO emission extends well beyond the known boundaries of molecular cloud. 

\section{Acknowledgements}
The development of \Planck\ has been supported by: ESA; CNES and
CNRS/INSU-IN2P3-INP (France); ASI, CNR, and INAF (Italy); NASA and DoE
(USA); STFC and UKSA (UK); CSIC, MICINN, JA and RES (Spain); Tekes,
AoF and CSC (Finland); DLR and MPG (Germany); CSA (Canada); DTU Space
(Denmark); SER/SSO (Switzerland); RCN (Norway); SFI (Ireland);
FCT/MCTES (Portugal); and PRACE (EU). A description of the \Planck\
Collaboration and a list of its members, including the technical or
scientific activities in which they have been involved, can be found
at \url{http://www.sciops.esa.int/index.php?project=Planck\&page=Planck\_Collaboration}.
We acknowledge the use of the {\tt HEALPix} software.

\bibliographystyle{aa} 
\bibliography{biblio,Planck_bib} 

\appendix

\section{Description of the \textit{Planck} CO products}
\label{app:coprod}
The characteristics of the released maps are the following. We provide {\tt HEALPix} pixelization all-sky maps with N$_{\rm side}$=2048. For one transition, the CO velocity-integrated line signal map is given in K$.km \ s^{-1}$ units
at the CO transition frequency. A conversion factor from this unit to the native unit of HFI maps ($\rm K_{CMB}$) is provided in the header. Four maps are given in a single file per transition
(1-0), (2-1) and (3-2) and per \typeone, \typetwo, and \typethree, 

\begin{itemize}
\item a  CO signal map;
\item a CO null test map;
\item a standard deviation map;
\item a mask for which  a value of 1 corresponds to a valid pixel.
\end{itemize} 

\typeone\ products  have the native HFI resolution 9\parcm65, 4\parcm99 and 4\parcm82
for the CO [1$\rightarrow$0], [2$\rightarrow$1] and [3$\rightarrow$2] transitions respectively. \typetwo\
products have 15\arcm\ resolution and are given
for the two first transitions. The \typethree\ product has a 5\parcm5
effective resolution and corresponds to a combined map of the three CO transitions
normalized at the ($J=1\rightarrow 0$) transition frequency. These maps are shown in
the left column of  Figs.,~\ref{fig:fullskyCO10},
\ref{fig:fullskyCO21} and \ref{fig:fullskyCO32}. The corresponding
standard deviation maps are displayed in the right column.


\section{Validation of CO extraction methods on simulations}
\label{sec:simuvalidation}
\Planck\ component separation methods \citep[see][]{planck2013-p06} are generally validated using the {\tt FFP6} set of simulations \citep{planck2013-p28}, which aim
at providing a complete realization of the \Planck\ mission. They rely of the \Planck\ Sky Model
(PSM)  fully described in \citet{delabrouille2012}. {\tt FFP6} maps can be used for the multi-channel
and multi-line approaches, but they do not provide the individual
bolometer maps required to validate the single-channel CO extraction
method. For the latter, CO-tailored simulations are needed, as
described below. 

\subsection{Single-channel approach: \typeone\ maps}
We use here specific simulations of the \Planck\ bolometer maps at 100~GHz.
For each bolometer in the \Planck\ 100~GHz channel, we construct a map of the sky
emission considering CMB and foreground components (Galactic diffuse emission from synchrotron, thermal dust and free-free, and Galactic
and extraGalactic point sources). We use the same \Planck\ Sky Model (PSM) model as for the {\tt FFP6} simulations (see below for a detailed description). 
For the CO emission we just use the \citet{2001ApJ...547..792D} map as a template and applied the CO bandpass conversion coefficients 
from Table~\ref{tab:co_coeff} to convert to  $\mathrm{K}_{\mathrm{CMB}}$ units. The per bolometer null maps
were added to account for the noise contribution. The MILCA algorithm was then applied to these maps
to reconstruct the CO signal using the single channel approach discussed in Sect.~\ref{subsec:multibolo}. We found that when the $^{12}$CO conversion
coefficients are perfectly known the CO emission is well reconstructed. The contribution from foregrounds was estimated to be less than 0.5 \% of the CO signal overall
and the rms of the residuals consistent with the noise at about the 1--$\sigma$ level.

We also tested the sensitivity of the {\tt MILCA} algorithm to uncertainties on the CO conversion coefficients.
These uncertainties translate into noisier final CO maps and an overall calibration bias.
For 1\% and 5 \% uncertainties on the CO conversion coefficients we observe a combined
signal to-noise-reduction of  6\% and 60\%, respectively.

\subsection{Multi-channel approach:  \typetwo\ maps}
The multi-channel approach may be validated using \Planck\ {\tt FFP6} set
channel intensity maps. Relevant to our purpose here,
it includes in particular a $\nu^{-2.14}$ free-free emission component, a
two-component dust emission from model~7 of \citet{1999ApJ...524..867F}, and CO emission using the CO map of
\citet{2001ApJ...547..792D} as the template.

We apply the multi-channel CO extraction method described in Sect.~\ref{subsec:multifreq} to the set of {\tt FFP6}
intensity maps and characterize the reconstructed \cooz\ map with the
input map of the {\tt FFP6} simulations. As for some of the other tests done
throughout the paper, the comparison is performed in the Orion,
Taurus, and Polaris molecular clouds.

A linear correlation (${\rm CO}_{\rm type2}=a\,{\rm CO}_{\rm {\tt FFP6}}+b$) between the two maps is found in the
three regions, with slopes of $0.99\pm 0.01$, $1.013\pm 0.001$,
$0.999\pm 0.002$ for Taurus, Orion, and Polaris respectively. We also
checked the standard deviation of the residual maps ${\rm CO}_{\rm {\tt FFP6}}-{\rm CO}_{\rm type2}$
against the average $\sigma$ in each region obtained from the
\typetwo\ \cooz\ standard deviation map (see
Fig.~\ref{fig:fullskyCO10}). In Polaris, the residual is within the
$\pm$1--$\sigma$ level, while it is at about 1.5-$\sigma$ for Taurus and
Orion. In these two latter regions, the slightly poorer behaviour of
the residual is attributed to the simple modified blackbody dust modelling of the
multi-channel approach that cannot properly capture the dust of the
FPP6 simulations. With these caveats, the multi-channel
approach is successfully validated on {\tt FFP6} simulations.

\subsection{Multi-line approach:  \typethree\ map}
The quality of the reconstruction of the \typethree\ CO map was also validated using the {\tt FFP6} simulations. 
As described in Sect.~\ref{subsec:combined}, the \typethree\ CO map is a byproduct of the CMB oriented
component separation procedure and therefore, validation on
simulations is presented in the companion CMB-oriented 
component separation paper~\citet{planck2013-p06}.

\end{document}

%% file: Planck.tex
\def\setsymbol#1#2{\expandafter\def\csname #1\endcsname{#2}}
\def\getsymbol#1{\csname #1\endcsname}

\def\Planck{{\it Planck\/}}

\def\HeJT{$^4$He-JT}

\def\allearlypapers{\nocite{planck2011-1.1, planck2011-1.3, planck2011-1.4, planck2011-1.5, planck2011-1.6, planck2011-1.7, planck2011-1.10, planck2011-1.10sup, planck2011-5.1a, planck2011-5.1b, planck2011-5.2a, planck2011-5.2b, planck2011-5.2c, planck2011-6.1, planck2011-6.2, planck2011-6.3a, planck2011-6.4a, planck2011-6.4b, planck2011-6.6, planck2011-7.0, planck2011-7.2, planck2011-7.3, planck2011-7.7a, planck2011-7.7b, planck2011-7.12, planck2011-7.13}}

\newbox\tablebox    \newdimen\tablewidth
\def\leaderfil{\leaders\hbox to 5pt{\hss.\hss}\hfil}
%
%
\def\endPlancktable{\tablewidth=\columnwidth 
    $$\hss\copy\tablebox\hss$$
    \vskip-\lastskip\vskip -2pt}
\def\endPlancktablewide{\tablewidth=\textwidth 
    $$\hss\copy\tablebox\hss$$
    \vskip-\lastskip\vskip -2pt}
\def\tablenote#1 #2\par{\begingroup \parindent=0.8em
    \abovedisplayshortskip=0pt\belowdisplayshortskip=0pt
    \noindent
    $$\hss\vbox{\hsize\tablewidth \hangindent=\parindent \hangafter=1 \noindent
    \hbox to \parindent{$^#1$\hss}\strut#2\strut\par}\hss$$
    \endgroup}
\def\doubleline{\vskip 3pt\hrule \vskip 1.5pt \hrule \vskip 5pt}

%
\def\L2{\ifmmode L_2\else $L_2$\fi}
\def\dtt{\Delta T/T}
\def\DeltaT{\ifmmode \Delta T\else $\Delta T$\fi}
\def\deltat{\ifmmode \Delta t\else $\Delta t$\fi}
\def\fknee{\ifmmode f_{\rm knee}\else $f_{\rm knee}$\fi}
\def\Fmax{\ifmmode F_{\rm max}\else $F_{\rm max}$\fi}
\def\solar{\ifmmode{\rm M}_{\mathord\odot}\else${\rm M}_{\mathord\odot}$\fi}
\def\Msolar{\ifmmode{\rm M}_{\mathord\odot}\else${\rm M}_{\mathord\odot}$\fi}
\def\Lsolar{\ifmmode{\rm L}_{\mathord\odot}\else${\rm L}_{\mathord\odot}$\fi}
\def\mag{\sup{m}}
\def\inv{\ifmmode^{-1}\else$^{-1}$\fi}
\def\mo{\ifmmode^{-1}\else$^{-1}$\fi}
\def\sup#1{\ifmmode ^{\rm #1}\else $^{\rm #1}$\fi}
\def\expo#1{\ifmmode \times 10^{#1}\else $\times 10^{#1}$\fi}
\def\,{\thinspace}
\def\lsim{\mathrel{\raise .4ex\hbox{\rlap{$<$}\lower 1.2ex\hbox{$\sim$}}}}
\def\gsim{\mathrel{\raise .4ex\hbox{\rlap{$>$}\lower 1.2ex\hbox{$\sim$}}}}
\let\lea=\lsim
\let\gea=\gsim
\def\simprop{\mathrel{\raise .4ex\hbox{\rlap{$\propto$}\lower 1.2ex\hbox{$\sim$}}}}
\def\deg{\ifmmode^\circ\else$^\circ$\fi}
\def\pdeg{\ifmmode $\setbox0=\hbox{$^{\circ}$}\rlap{\hskip.11\wd0 .}$^{\circ}
          \else \setbox0=\hbox{$^{\circ}$}\rlap{\hskip.11\wd0 .}$^{\circ}$\fi}
\def\arcs{\ifmmode {^{\scriptstyle\prime\prime}}
          \else $^{\scriptstyle\prime\prime}$\fi}
\def\arcm{\ifmmode {^{\scriptstyle\prime}}
          \else $^{\scriptstyle\prime}$\fi}
\newdimen\sa  \newdimen\sb
\def\parcs{\sa=.07em \sb=.03em
     \ifmmode \hbox{\rlap{.}}^{\scriptstyle\prime\kern -\sb\prime}\hbox{\kern -\sa}
     \else \rlap{.}$^{\scriptstyle\prime\kern -\sb\prime}$\kern -\sa\fi}
\def\parcm{\sa=.08em \sb=.03em
     \ifmmode \hbox{\rlap{.}\kern\sa}^{\scriptstyle\prime}\hbox{\kern-\sb}
     \else \rlap{.}\kern\sa$^{\scriptstyle\prime}$\kern-\sb\fi}
\def\ra[#1 #2 #3.#4]{#1\sup{h}#2\sup{m}#3\sup{s}\llap.#4}
\def\dec[#1 #2 #3.#4]{#1\deg#2\arcm#3\arcs\llap.#4}
\def\deco[#1 #2 #3]{#1\deg#2\arcm#3\arcs}
\def\rra[#1 #2]{#1\sup{h}#2\sup{m}}
\def\page{\vfill\eject}
\def\dots{\relax\ifmmode \ldots\else $\ldots$\fi}
%
%
\def\WHzsr{\ifmmode $W\,Hz\mo\,sr\mo$\else W\,Hz\mo\,sr\mo\fi}
\def\mHz{\ifmmode $\,mHz$\else \,mHz\fi}
\def\GHz{\ifmmode $\,GHz$\else \,GHz\fi}
\def\mKs{\ifmmode $\,mK\,s$^{1/2}\else \,mK\,s$^{1/2}$\fi}
\def\muKs{\ifmmode \,\mu$K\,s$^{1/2}\else \,$\mu$K\,s$^{1/2}$\fi}
\def\muKRJs{\ifmmode \,\mu$K$_{\rm RJ}$\,s$^{1/2}\else \,$\mu$K$_{\rm RJ}$\,s$^{1/2}$\fi}
\def\muKHz{\ifmmode \,\mu$K\,Hz$^{-1/2}\else \,$\mu$K\,Hz$^{-1/2}$\fi}
\def\MJysr{\ifmmode \,$MJy\,sr\mo$\else \,MJy\,sr\mo\fi}
\def\MJysrmK{\ifmmode \,$MJy\,sr\mo$\,mK$_{\rm CMB}\mo\else \,MJy\,sr\mo\,mK$_{\rm CMB}\mo$\fi}
\def\microns{\ifmmode \,\mu$m$\else \,$\mu$m\fi}
\def\micron{\microns}
\def\muK{\ifmmode \,\mu$K$\else \,$\mu$\hbox{K}\fi}
\def\microK{\ifmmode \,\mu$K$\else \,$\mu$\hbox{K}\fi}
\def\muW{\ifmmode \,\mu$W$\else \,$\mu$\hbox{W}\fi}
\def\kms{\ifmmode $\,km\,s$^{-1}\else \,km\,s$^{-1}$\fi}
\def\kmsMpc{\ifmmode $\,\kms\,Mpc\mo$\else \,\kms\,Mpc\mo\fi}
%
%


\setsymbol{LFI:center:frequency:70GHz:units}{70.3\,GHz}
\setsymbol{LFI:center:frequency:44GHz:units}{44.1\,GHz}
\setsymbol{LFI:center:frequency:30GHz:units}{28.5\,GHz}

\setsymbol{LFI:center:frequency:70GHz}{70.3}
\setsymbol{LFI:center:frequency:44GHz}{44.1}
\setsymbol{LFI:center:frequency:30GHz}{28.5}

\setsymbol{LFI:center:frequency:LFI18:Rad:M:units}{71.7\GHz}
\setsymbol{LFI:center:frequency:LFI19:Rad:M:units}{67.5\GHz}
\setsymbol{LFI:center:frequency:LFI20:Rad:M:units}{69.2\GHz}
\setsymbol{LFI:center:frequency:LFI21:Rad:M:units}{70.4\GHz}
\setsymbol{LFI:center:frequency:LFI22:Rad:M:units}{71.5\GHz}
\setsymbol{LFI:center:frequency:LFI23:Rad:M:units}{70.8\GHz}
\setsymbol{LFI:center:frequency:LFI24:Rad:M:units}{44.4\GHz}
\setsymbol{LFI:center:frequency:LFI25:Rad:M:units}{44.0\GHz}
\setsymbol{LFI:center:frequency:LFI26:Rad:M:units}{43.9\GHz}
\setsymbol{LFI:center:frequency:LFI27:Rad:M:units}{28.3\GHz}
\setsymbol{LFI:center:frequency:LFI28:Rad:M:units}{28.8\GHz}
\setsymbol{LFI:center:frequency:LFI18:Rad:S:units}{70.1\GHz}
\setsymbol{LFI:center:frequency:LFI19:Rad:S:units}{69.6\GHz}
\setsymbol{LFI:center:frequency:LFI20:Rad:S:units}{69.5\GHz}
\setsymbol{LFI:center:frequency:LFI21:Rad:S:units}{69.5\GHz}
\setsymbol{LFI:center:frequency:LFI22:Rad:S:units}{72.8\GHz}
\setsymbol{LFI:center:frequency:LFI23:Rad:S:units}{71.3\GHz}
\setsymbol{LFI:center:frequency:LFI24:Rad:S:units}{44.1\GHz}
\setsymbol{LFI:center:frequency:LFI25:Rad:S:units}{44.1\GHz}
\setsymbol{LFI:center:frequency:LFI26:Rad:S:units}{44.1\GHz}
\setsymbol{LFI:center:frequency:LFI27:Rad:S:units}{28.5\GHz}
\setsymbol{LFI:center:frequency:LFI28:Rad:S:units}{28.2\GHz}

\setsymbol{LFI:center:frequency:LFI18:Rad:M}{71.7}
\setsymbol{LFI:center:frequency:LFI19:Rad:M}{67.5}
\setsymbol{LFI:center:frequency:LFI20:Rad:M}{69.2}
\setsymbol{LFI:center:frequency:LFI21:Rad:M}{70.4}
\setsymbol{LFI:center:frequency:LFI22:Rad:M}{71.5}
\setsymbol{LFI:center:frequency:LFI23:Rad:M}{70.8}
\setsymbol{LFI:center:frequency:LFI24:Rad:M}{44.4}
\setsymbol{LFI:center:frequency:LFI25:Rad:M}{44.0}
\setsymbol{LFI:center:frequency:LFI26:Rad:M}{43.9}
\setsymbol{LFI:center:frequency:LFI27:Rad:M}{28.3}
\setsymbol{LFI:center:frequency:LFI28:Rad:M}{28.8}
\setsymbol{LFI:center:frequency:LFI18:Rad:S}{70.1}
\setsymbol{LFI:center:frequency:LFI19:Rad:S}{69.6}
\setsymbol{LFI:center:frequency:LFI20:Rad:S}{69.5}
\setsymbol{LFI:center:frequency:LFI21:Rad:S}{69.5}
\setsymbol{LFI:center:frequency:LFI22:Rad:S}{72.8}
\setsymbol{LFI:center:frequency:LFI23:Rad:S}{71.3}
\setsymbol{LFI:center:frequency:LFI24:Rad:S}{44.1}
\setsymbol{LFI:center:frequency:LFI25:Rad:S}{44.1}
\setsymbol{LFI:center:frequency:LFI26:Rad:S}{44.1}
\setsymbol{LFI:center:frequency:LFI27:Rad:S}{28.5}
\setsymbol{LFI:center:frequency:LFI28:Rad:S}{28.2}


\setsymbol{LFI:white:noise:sensitivity:70GHz:units}{134.7\muKs}
\setsymbol{LFI:white:noise:sensitivity:44GHz:units}{164.7\muKs}
\setsymbol{LFI:white:noise:sensitivity:30GHz:units}{143.4\muKs}

\setsymbol{LFI:white:noise:sensitivity:70GHz}{134.7}
\setsymbol{LFI:white:noise:sensitivity:44GHz}{164.7}
\setsymbol{LFI:white:noise:sensitivity:30GHz}{143.4}


\setsymbol{LFI:white:noise:sensitivity:LFI18:Rad:M:units}{512.0\muKs}
\setsymbol{LFI:white:noise:sensitivity:LFI19:Rad:M:units}{581.4\muKs}
\setsymbol{LFI:white:noise:sensitivity:LFI20:Rad:M:units}{590.8\muKs}
\setsymbol{LFI:white:noise:sensitivity:LFI21:Rad:M:units}{455.2\muKs}
\setsymbol{LFI:white:noise:sensitivity:LFI22:Rad:M:units}{492.0\muKs}
\setsymbol{LFI:white:noise:sensitivity:LFI23:Rad:M:units}{507.7\muKs}
\setsymbol{LFI:white:noise:sensitivity:LFI24:Rad:M:units}{462.2\muKs}
\setsymbol{LFI:white:noise:sensitivity:LFI25:Rad:M:units}{413.6\muKs}
\setsymbol{LFI:white:noise:sensitivity:LFI26:Rad:M:units}{478.6\muKs}
\setsymbol{LFI:white:noise:sensitivity:LFI27:Rad:M:units}{277.7\muKs}
\setsymbol{LFI:white:noise:sensitivity:LFI28:Rad:M:units}{312.3\muKs}
\setsymbol{LFI:white:noise:sensitivity:LFI18:Rad:S:units}{465.7\muKs}
\setsymbol{LFI:white:noise:sensitivity:LFI19:Rad:S:units}{555.6\muKs}
\setsymbol{LFI:white:noise:sensitivity:LFI20:Rad:S:units}{623.2\muKs}
\setsymbol{LFI:white:noise:sensitivity:LFI21:Rad:S:units}{564.1\muKs}
\setsymbol{LFI:white:noise:sensitivity:LFI22:Rad:S:units}{534.4\muKs}
\setsymbol{LFI:white:noise:sensitivity:LFI23:Rad:S:units}{542.4\muKs}
\setsymbol{LFI:white:noise:sensitivity:LFI24:Rad:S:units}{399.2\muKs}
\setsymbol{LFI:white:noise:sensitivity:LFI25:Rad:S:units}{392.6\muKs}
\setsymbol{LFI:white:noise:sensitivity:LFI26:Rad:S:units}{418.6\muKs}
\setsymbol{LFI:white:noise:sensitivity:LFI27:Rad:S:units}{302.9\muKs}
\setsymbol{LFI:white:noise:sensitivity:LFI28:Rad:S:units}{285.3\muKs}

\setsymbol{LFI:white:noise:sensitivity:LFI18:Rad:M}{512.0}
\setsymbol{LFI:white:noise:sensitivity:LFI19:Rad:M}{581.4}
\setsymbol{LFI:white:noise:sensitivity:LFI20:Rad:M}{590.8}
\setsymbol{LFI:white:noise:sensitivity:LFI21:Rad:M}{455.2}
\setsymbol{LFI:white:noise:sensitivity:LFI22:Rad:M}{492.0}
\setsymbol{LFI:white:noise:sensitivity:LFI23:Rad:M}{507.7}
\setsymbol{LFI:white:noise:sensitivity:LFI24:Rad:M}{462.2}
\setsymbol{LFI:white:noise:sensitivity:LFI25:Rad:M}{413.6}
\setsymbol{LFI:white:noise:sensitivity:LFI26:Rad:M}{478.6}
\setsymbol{LFI:white:noise:sensitivity:LFI27:Rad:M}{277.7}
\setsymbol{LFI:white:noise:sensitivity:LFI28:Rad:M}{312.3}
\setsymbol{LFI:white:noise:sensitivity:LFI18:Rad:S}{465.7}
\setsymbol{LFI:white:noise:sensitivity:LFI19:Rad:S}{555.6}
\setsymbol{LFI:white:noise:sensitivity:LFI20:Rad:S}{623.2}
\setsymbol{LFI:white:noise:sensitivity:LFI21:Rad:S}{564.1}
\setsymbol{LFI:white:noise:sensitivity:LFI22:Rad:S}{534.4}
\setsymbol{LFI:white:noise:sensitivity:LFI23:Rad:S}{542.4}
\setsymbol{LFI:white:noise:sensitivity:LFI24:Rad:S}{399.2}
\setsymbol{LFI:white:noise:sensitivity:LFI25:Rad:S}{392.6}
\setsymbol{LFI:white:noise:sensitivity:LFI26:Rad:S}{418.6}
\setsymbol{LFI:white:noise:sensitivity:LFI27:Rad:S}{302.9}
\setsymbol{LFI:white:noise:sensitivity:LFI28:Rad:S}{285.3}


\setsymbol{LFI:knee:frequency:70GHz:units}{29.5\mHz}
\setsymbol{LFI:knee:frequency:44GHz:units}{56.2\mHz}
\setsymbol{LFI:knee:frequency:30GHz:units}{113.7\mHz}

\setsymbol{LFI:knee:frequency:70GHz}{29.5}
\setsymbol{LFI:knee:frequency:44GHz}{56.2}
\setsymbol{LFI:knee:frequency:30GHz}{113.7}

\setsymbol{LFI:knee:frequency:LFI18:Rad:M:units}{16.3\mHz}
\setsymbol{LFI:knee:frequency:LFI19:Rad:M:units}{15.1\mHz}
\setsymbol{LFI:knee:frequency:LFI20:Rad:M:units}{18.7\mHz}
\setsymbol{LFI:knee:frequency:LFI21:Rad:M:units}{37.2\mHz}
\setsymbol{LFI:knee:frequency:LFI22:Rad:M:units}{12.7\mHz}
\setsymbol{LFI:knee:frequency:LFI23:Rad:M:units}{34.6\mHz}
\setsymbol{LFI:knee:frequency:LFI24:Rad:M:units}{46.2\mHz}
\setsymbol{LFI:knee:frequency:LFI25:Rad:M:units}{24.9\mHz}
\setsymbol{LFI:knee:frequency:LFI26:Rad:M:units}{67.6\mHz}
\setsymbol{LFI:knee:frequency:LFI27:Rad:M:units}{187.4\mHz}
\setsymbol{LFI:knee:frequency:LFI28:Rad:M:units}{122.2\mHz}
\setsymbol{LFI:knee:frequency:LFI18:Rad:S:units}{17.7\mHz}
\setsymbol{LFI:knee:frequency:LFI19:Rad:S:units}{22.0\mHz}
\setsymbol{LFI:knee:frequency:LFI20:Rad:S:units}{8.7\mHz}
\setsymbol{LFI:knee:frequency:LFI21:Rad:S:units}{25.9\mHz}
\setsymbol{LFI:knee:frequency:LFI22:Rad:S:units}{15.8\mHz}
\setsymbol{LFI:knee:frequency:LFI23:Rad:S:units}{129.8\mHz}
\setsymbol{LFI:knee:frequency:LFI24:Rad:S:units}{100.9\mHz}
\setsymbol{LFI:knee:frequency:LFI25:Rad:S:units}{38.9\mHz}
\setsymbol{LFI:knee:frequency:LFI26:Rad:S:units}{58.9\mHz}
\setsymbol{LFI:knee:frequency:LFI27:Rad:S:units}{104.4\mHz}
\setsymbol{LFI:knee:frequency:LFI28:Rad:S:units}{40.7\mHz}

\setsymbol{LFI:knee:frequency:LFI18:Rad:M}{16.3}
\setsymbol{LFI:knee:frequency:LFI19:Rad:M}{15.1}
\setsymbol{LFI:knee:frequency:LFI20:Rad:M}{18.7}
\setsymbol{LFI:knee:frequency:LFI21:Rad:M}{37.2}
\setsymbol{LFI:knee:frequency:LFI22:Rad:M}{12.7}
\setsymbol{LFI:knee:frequency:LFI23:Rad:M}{34.6}
\setsymbol{LFI:knee:frequency:LFI24:Rad:M}{46.2}
\setsymbol{LFI:knee:frequency:LFI25:Rad:M}{24.9}
\setsymbol{LFI:knee:frequency:LFI26:Rad:M}{67.6}
\setsymbol{LFI:knee:frequency:LFI27:Rad:M}{187.4}
\setsymbol{LFI:knee:frequency:LFI28:Rad:M}{122.2}
\setsymbol{LFI:knee:frequency:LFI18:Rad:S}{17.7}
\setsymbol{LFI:knee:frequency:LFI19:Rad:S}{22.0}
\setsymbol{LFI:knee:frequency:LFI20:Rad:S}{8.7}
\setsymbol{LFI:knee:frequency:LFI21:Rad:S}{25.9}
\setsymbol{LFI:knee:frequency:LFI22:Rad:S}{15.8}
\setsymbol{LFI:knee:frequency:LFI23:Rad:S}{129.8}
\setsymbol{LFI:knee:frequency:LFI24:Rad:S}{100.9}
\setsymbol{LFI:knee:frequency:LFI25:Rad:S}{38.9}
\setsymbol{LFI:knee:frequency:LFI26:Rad:S}{58.9}
\setsymbol{LFI:knee:frequency:LFI27:Rad:S}{104.4}
\setsymbol{LFI:knee:frequency:LFI28:Rad:S}{40.7}


\setsymbol{LFI:slope:70GHz:units}{$-1.03$\mHz}
\setsymbol{LFI:slope:44GHz:units}{$-0.89$\mHz}
\setsymbol{LFI:slope:30GHz:units}{$-0.87$\mHz}

\setsymbol{LFI:slope:70GHz}{$-1.03$}
\setsymbol{LFI:slope:44GHz}{$-0.89$}
\setsymbol{LFI:slope:30GHz}{$-0.87$}

\setsymbol{LFI:slope:LFI18:Rad:M:units}{$-1.04$\mHz}
\setsymbol{LFI:slope:LFI19:Rad:M:units}{$-1.09$\mHz}
\setsymbol{LFI:slope:LFI20:Rad:M:units}{$-0.69$\mHz}
\setsymbol{LFI:slope:LFI21:Rad:M:units}{$-1.56$\mHz}
\setsymbol{LFI:slope:LFI22:Rad:M:units}{$-1.01$\mHz}
\setsymbol{LFI:slope:LFI23:Rad:M:units}{$-0.96$\mHz}
\setsymbol{LFI:slope:LFI24:Rad:M:units}{$-0.83$\mHz}
\setsymbol{LFI:slope:LFI25:Rad:M:units}{$-0.91$\mHz}
\setsymbol{LFI:slope:LFI26:Rad:M:units}{$-0.95$\mHz}
\setsymbol{LFI:slope:LFI27:Rad:M:units}{$-0.87$\mHz}
\setsymbol{LFI:slope:LFI28:Rad:M:units}{$-0.88$\mHz}
\setsymbol{LFI:slope:LFI18:Rad:S:units}{$-1.15$\mHz}
\setsymbol{LFI:slope:LFI19:Rad:S:units}{$-1.00$\mHz}
\setsymbol{LFI:slope:LFI20:Rad:S:units}{$-0.95$\mHz}
\setsymbol{LFI:slope:LFI21:Rad:S:units}{$-0.92$\mHz}
\setsymbol{LFI:slope:LFI22:Rad:S:units}{$-1.01$\mHz}
\setsymbol{LFI:slope:LFI23:Rad:S:units}{$-0.95$\mHz}
\setsymbol{LFI:slope:LFI24:Rad:S:units}{$-0.73$\mHz}
\setsymbol{LFI:slope:LFI25:Rad:S:units}{$-1.16$\mHz}
\setsymbol{LFI:slope:LFI26:Rad:S:units}{$-0.79$\mHz}
\setsymbol{LFI:slope:LFI27:Rad:S:units}{$-0.82$\mHz}
\setsymbol{LFI:slope:LFI28:Rad:S:units}{$-0.91$\mHz}

\setsymbol{LFI:slope:LFI18:Rad:M}{$-1.04$}
\setsymbol{LFI:slope:LFI19:Rad:M}{$-1.09$}
\setsymbol{LFI:slope:LFI20:Rad:M}{$-0.69$}
\setsymbol{LFI:slope:LFI21:Rad:M}{$-1.56$}
\setsymbol{LFI:slope:LFI22:Rad:M}{$-1.01$}
\setsymbol{LFI:slope:LFI23:Rad:M}{$-0.96$}
\setsymbol{LFI:slope:LFI24:Rad:M}{$-0.83$}
\setsymbol{LFI:slope:LFI25:Rad:M}{$-0.91$}
\setsymbol{LFI:slope:LFI26:Rad:M}{$-0.95$}
\setsymbol{LFI:slope:LFI27:Rad:M}{$-0.87$}
\setsymbol{LFI:slope:LFI28:Rad:M}{$-0.88$}
\setsymbol{LFI:slope:LFI18:Rad:S}{$-1.15$}
\setsymbol{LFI:slope:LFI19:Rad:S}{$-1.00$}
\setsymbol{LFI:slope:LFI20:Rad:S}{$-0.95$}
\setsymbol{LFI:slope:LFI21:Rad:S}{$-0.92$}
\setsymbol{LFI:slope:LFI22:Rad:S}{$-1.01$}
\setsymbol{LFI:slope:LFI23:Rad:S}{$-0.95$}
\setsymbol{LFI:slope:LFI24:Rad:S}{$-0.73$}
\setsymbol{LFI:slope:LFI25:Rad:S}{$-1.16$}
\setsymbol{LFI:slope:LFI26:Rad:S}{$-0.79$}
\setsymbol{LFI:slope:LFI27:Rad:S}{$-0.82$}
\setsymbol{LFI:slope:LFI28:Rad:S}{$-0.91$}


\setsymbol{LFI:FWHM:70GHz:units}{13\parcm01}
\setsymbol{LFI:FWHM:44GHz:units}{27\parcm92}
\setsymbol{LFI:FWHM:30GHz:units}{32\parcm65}

\setsymbol{LFI:FWHM:70GHz}{13.01}
\setsymbol{LFI:FWHM:44GHz}{27.92}
\setsymbol{LFI:FWHM:30GHz}{32.65}

\setsymbol{LFI:FWHM:LFI18:units}{13\parcm39}
\setsymbol{LFI:FWHM:LFI19:units}{13\parcm01}
\setsymbol{LFI:FWHM:LFI20:units}{12\parcm75}
\setsymbol{LFI:FWHM:LFI21:units}{12\parcm74}
\setsymbol{LFI:FWHM:LFI22:units}{12\parcm87}
\setsymbol{LFI:FWHM:LFI23:units}{13\parcm27}
\setsymbol{LFI:FWHM:LFI24:units}{22\parcm98}
\setsymbol{LFI:FWHM:LFI25:units}{30\parcm46}
\setsymbol{LFI:FWHM:LFI26:units}{30\parcm31}
\setsymbol{LFI:FWHM:LFI27:units}{32\parcm65}
\setsymbol{LFI:FWHM:LFI28:units}{32\parcm66}

\setsymbol{LFI:FWHM:LFI18}{13.39}
\setsymbol{LFI:FWHM:LFI19}{13.01}
\setsymbol{LFI:FWHM:LFI20}{12.75}
\setsymbol{LFI:FWHM:LFI21}{12.74}
\setsymbol{LFI:FWHM:LFI22}{12.87}
\setsymbol{LFI:FWHM:LFI23}{13.27}
\setsymbol{LFI:FWHM:LFI24}{22.98}
\setsymbol{LFI:FWHM:LFI25}{30.46}
\setsymbol{LFI:FWHM:LFI26}{30.31}
\setsymbol{LFI:FWHM:LFI27}{32.65}
\setsymbol{LFI:FWHM:LFI28}{32.66}



\setsymbol{LFI:FWHM:uncertainty:LFI18:units}{0.170\arcm}
\setsymbol{LFI:FWHM:uncertainty:LFI19:units}{0.174\arcm}
\setsymbol{LFI:FWHM:uncertainty:LFI20:units}{0.170\arcm}
\setsymbol{LFI:FWHM:uncertainty:LFI21:units}{0.156\arcm}
\setsymbol{LFI:FWHM:uncertainty:LFI22:units}{0.164\arcm}
\setsymbol{LFI:FWHM:uncertainty:LFI23:units}{0.171\arcm}
\setsymbol{LFI:FWHM:uncertainty:LFI24:units}{0.652\arcm}
\setsymbol{LFI:FWHM:uncertainty:LFI25:units}{1.075\arcm}
\setsymbol{LFI:FWHM:uncertainty:LFI26:units}{1.131\arcm}
\setsymbol{LFI:FWHM:uncertainty:LFI27:units}{1.266\arcm}
\setsymbol{LFI:FWHM:uncertainty:LFI28:units}{1.287\arcm}

\setsymbol{LFI:FWHM:uncertainty:LFI18}{0.170}
\setsymbol{LFI:FWHM:uncertainty:LFI19}{0.174}
\setsymbol{LFI:FWHM:uncertainty:LFI20}{0.170}
\setsymbol{LFI:FWHM:uncertainty:LFI21}{0.156}
\setsymbol{LFI:FWHM:uncertainty:LFI22}{0.164}
\setsymbol{LFI:FWHM:uncertainty:LFI23}{0.171}
\setsymbol{LFI:FWHM:uncertainty:LFI24}{0.652}
\setsymbol{LFI:FWHM:uncertainty:LFI25}{1.075}
\setsymbol{LFI:FWHM:uncertainty:LFI26}{1.131}
\setsymbol{LFI:FWHM:uncertainty:LFI27}{1.266}
\setsymbol{LFI:FWHM:uncertainty:LFI28}{1.287}


\setsymbol{HFI:center:frequency:100GHz:units}{100\,GHz}
\setsymbol{HFI:center:frequency:143GHz:units}{143\,GHz}
\setsymbol{HFI:center:frequency:217GHz:units}{217\,GHz}
\setsymbol{HFI:center:frequency:353GHz:units}{353\,GHz}
\setsymbol{HFI:center:frequency:545GHz:units}{545\,GHz}
\setsymbol{HFI:center:frequency:857GHz:units}{857\,GHz}

\setsymbol{HFI:center:frequency:100GHz}{100}
\setsymbol{HFI:center:frequency:143GHz}{143}
\setsymbol{HFI:center:frequency:217GHz}{217}
\setsymbol{HFI:center:frequency:353GHz}{353}
\setsymbol{HFI:center:frequency:545GHz}{545}
\setsymbol{HFI:center:frequency:857GHz}{857}


\setsymbol{HFI:Ndetectors:100GHz}{8}
\setsymbol{HFI:Ndetectors:143GHz}{11}
\setsymbol{HFI:Ndetectors:217GHz}{12}
\setsymbol{HFI:Ndetectors:353GHz}{12}
\setsymbol{HFI:Ndetectors:545GHz}{3}
\setsymbol{HFI:Ndetectors:857GHz}{4}


\setsymbol{HFI:FWHM:Maps:100GHz:units}{9\parcm88}
\setsymbol{HFI:FWHM:Maps:143GHz:units}{7\parcm18}
\setsymbol{HFI:FWHM:Maps:217GHz:units}{4\parcm87}
\setsymbol{HFI:FWHM:Maps:353GHz:units}{4\parcm65}
\setsymbol{HFI:FWHM:Maps:545GHz:units}{4\parcm72}
\setsymbol{HFI:FWHM:Maps:857GHz:units}{4\parcm39}
\setsymbol{HFI:FWHM:Maps:100GHz}{9.88}
\setsymbol{HFI:FWHM:Maps:143GHz}{7.18}
\setsymbol{HFI:FWHM:Maps:217GHz}{4.87}
\setsymbol{HFI:FWHM:Maps:353GHz}{4.65}
\setsymbol{HFI:FWHM:Maps:545GHz}{4.72}
\setsymbol{HFI:FWHM:Maps:857GHz}{4.39}


\setsymbol{HFI:beam:ellipticity:Maps:100GHz}{1.15}
\setsymbol{HFI:beam:ellipticity:Maps:143GHz}{1.01}
\setsymbol{HFI:beam:ellipticity:Maps:217GHz}{1.06}
\setsymbol{HFI:beam:ellipticity:Maps:353GHz}{1.05}
\setsymbol{HFI:beam:ellipticity:Maps:545GHz}{1.14}
\setsymbol{HFI:beam:ellipticity:Maps:857GHz}{1.19}


\setsymbol{HFI:FWHM:Mars:100GHz:units}{9\parcm37}
\setsymbol{HFI:FWHM:Mars:143GHz:units}{7\parcm04}
\setsymbol{HFI:FWHM:Mars:217GHz:units}{4\parcm68}
\setsymbol{HFI:FWHM:Mars:353GHz:units}{4\parcm43}
\setsymbol{HFI:FWHM:Mars:545GHz:units}{3\parcm80}
\setsymbol{HFI:FWHM:Mars:857GHz:units}{3\parcm67}

\setsymbol{HFI:FWHM:Mars:100GHz}{9.37}
\setsymbol{HFI:FWHM:Mars:143GHz}{7.04}
\setsymbol{HFI:FWHM:Mars:217GHz}{4.68}
\setsymbol{HFI:FWHM:Mars:353GHz}{4.43}
\setsymbol{HFI:FWHM:Mars:545GHz}{3.80}
\setsymbol{HFI:FWHM:Mars:857GHz}{3.67}


\setsymbol{HFI:beam:ellipticity:Mars:100GHz}{1.18}
\setsymbol{HFI:beam:ellipticity:Mars:143GHz}{1.03}
\setsymbol{HFI:beam:ellipticity:Mars:217GHz}{1.14}
\setsymbol{HFI:beam:ellipticity:Mars:353GHz}{1.09}
\setsymbol{HFI:beam:ellipticity:Mars:545GHz}{1.25}
\setsymbol{HFI:beam:ellipticity:Mars:857GHz}{1.03}


\setsymbol{HFI:CMB:relative:calibration:100GHz}{$\lsim 1\%$}
\setsymbol{HFI:CMB:relative:calibration:143GHz}{$\lsim 1\%$}
\setsymbol{HFI:CMB:relative:calibration:217GHz}{$\lsim 1\%$}
\setsymbol{HFI:CMB:relative:calibration:353GHz}{$\lsim 1\%$}
\setsymbol{HFI:CMB:relative:calibration:545GHz}{}
\setsymbol{HFI:CMB:relative:calibration:857GHz}{}


\setsymbol{HFI:CMB:absolute:calibration:100GHz}{$\lsim 2\%$}
\setsymbol{HFI:CMB:absolute:calibration:143GHz}{$\lsim 2\%$}
\setsymbol{HFI:CMB:absolute:calibration:217GHz}{$\lsim 2\%$}
\setsymbol{HFI:CMB:absolute:calibration:353GHz}{$\lsim 2\%$}
\setsymbol{HFI:CMB:absolute:calibration:545GHz}{}
\setsymbol{HFI:CMB:absolute:calibration:857GHz}{}


\setsymbol{HFI:FIRAS:gain:calibration:accuracy:statistical:100GHz}{}
\setsymbol{HFI:FIRAS:gain:calibration:accuracy:statistical:143GHz}{}
\setsymbol{HFI:FIRAS:gain:calibration:accuracy:statistical:217GHz}{}
\setsymbol{HFI:FIRAS:gain:calibration:accuracy:statistical:353GHz}{2.5\%}
\setsymbol{HFI:FIRAS:gain:calibration:accuracy:statistical:545GHz}{1\%}
\setsymbol{HFI:FIRAS:gain:calibration:accuracy:statistical:857GHz}{0.5\%}


\setsymbol{HFI:FIRAS:gain:calibration:accuracy:systematic:100GHz}{}
\setsymbol{HFI:FIRAS:gain:calibration:accuracy:systematic:143GHz}{}
\setsymbol{HFI:FIRAS:gain:calibration:accuracy:systematic:217GHz}{}
\setsymbol{HFI:FIRAS:gain:calibration:accuracy:systematic:353GHz}{}
\setsymbol{HFI:FIRAS:gain:calibration:accuracy:systematic:545GHz}{7\%}
\setsymbol{HFI:FIRAS:gain:calibration:accuracy:systematic:857GHz}{7\%}


\setsymbol{HFI:FIRAS:zero:point:accuracy:100GHz:units}{0.8\MJysr}
\setsymbol{HFI:FIRAS:zero:point:accuracy:143GHz:units}{}
\setsymbol{HFI:FIRAS:zero:point:accuracy:217GHz:units}{}
\setsymbol{HFI:FIRAS:zero:point:accuracy:353GHz:units}{1.4\MJysr}
\setsymbol{HFI:FIRAS:zero:point:accuracy:545GHz:units}{2.2\MJysr}
\setsymbol{HFI:FIRAS:zero:point:accuracy:857GHz:units}{1.7\MJysr}

\setsymbol{HFI:FIRAS:zero:point:accuracy:100GHz}{0.8}
\setsymbol{HFI:FIRAS:zero:point:accuracy:143GHz}{}
\setsymbol{HFI:FIRAS:zero:point:accuracy:217GHz}{}
\setsymbol{HFI:FIRAS:zero:point:accuracy:353GHz}{1.4}
\setsymbol{HFI:FIRAS:zero:point:accuracy:545GHz}{2.2}
\setsymbol{HFI:FIRAS:zero:point:accuracy:857GHz}{1.7}


\setsymbol{HFI:unit:conversion:100GHz:units}{0.2415\MJysrmK}
\setsymbol{HFI:unit:conversion:143GHz:units}{0.3694\MJysrmK}
\setsymbol{HFI:unit:conversion:217GHz:units}{0.4811\MJysrmK}
\setsymbol{HFI:unit:conversion:353GHz:units}{0.2883\MJysrmK}
\setsymbol{HFI:unit:conversion:545GHz:units}{0.05826\MJysrmK}
\setsymbol{HFI:unit:conversion:857GHz:units}{0.002238\MJysrmK}

\setsymbol{HFI:unit:conversion:100GHz}{0.2415}
\setsymbol{HFI:unit:conversion:143GHz}{0.3694}
\setsymbol{HFI:unit:conversion:217GHz}{0.4811}
\setsymbol{HFI:unit:conversion:353GHz}{0.2883}
\setsymbol{HFI:unit:conversion:545GHz}{0.05826}
\setsymbol{HFI:unit:conversion:857GHz}{0.002238}


\setsymbol{HFI:colour:correction:alpha=-2:V1.01:100GHz}{0.9893}
\setsymbol{HFI:colour:correction:alpha=-2:V1.01:143GHz}{0.9759}
\setsymbol{HFI:colour:correction:alpha=-2:V1.01:217GHz}{1.0007}
\setsymbol{HFI:colour:correction:alpha=-2:V1.01:353GHz}{1.0028}
\setsymbol{HFI:colour:correction:alpha=-2:V1.01:545GHz}{1.0019}
\setsymbol{HFI:colour:correction:alpha=-2:V1.01:857GHz}{0.9889}


\setsymbol{HFI:colour:correction:alpha=0:V1.01:100GHz}{1.0008}
\setsymbol{HFI:colour:correction:alpha=0:V1.01:143GHz}{1.0148}
\setsymbol{HFI:colour:correction:alpha=0:V1.01:217GHz}{0.9909}
\setsymbol{HFI:colour:correction:alpha=0:V1.01:353GHz}{0.9888}
\setsymbol{HFI:colour:correction:alpha=0:V1.01:545GHz}{0.9878}
\setsymbol{HFI:colour:correction:alpha=0:V1.01:857GHz}{1.0014}

%% file: AuthorList_P03a_CO_authors_and_institutes.tex
\author{\small
Planck Collaboration:
P.~A.~R.~Ade\inst{90}
\and
N.~Aghanim\inst{60}
\and
M.~I.~R.~Alves\inst{60}
\and
C.~Armitage-Caplan\inst{95}
\and
M.~Arnaud\inst{75}
\and
M.~Ashdown\inst{72, 6}
\and
F.~Atrio-Barandela\inst{19}
\and
J.~Aumont\inst{60}
\and
C.~Baccigalupi\inst{89}
\and
A.~J.~Banday\inst{98, 10}
% ...
}
\institute{\small
APC, AstroParticule et Cosmologie, Universit\'{e} Paris Diderot, CNRS/IN2P3, CEA/lrfu, Observatoire de Paris, Sorbonne Paris Cit\'{e}, 10, rue Alice Domon et L\'{e}onie Duquet, 75205 Paris Cedex 13, France\\
\and
Aalto University Mets\"{a}hovi Radio Observatory, Mets\"{a}hovintie 114, FIN-02540 Kylm\"{a}l\"{a}, Finland\\
\and
African Institute for Mathematical Sciences, 6-8 Melrose Road, Muizenberg, Cape Town, South Africa\\
\and
Agenzia Spaziale Italiana Science Data Center, c/o ESRIN, via Galileo Galilei, Frascati, Italy\\
\and
Agenzia Spaziale Italiana, Viale Liegi 26, Roma, Italy\\
\and
Astrophysics Group, Cavendish Laboratory, University of Cambridge, J J Thomson Avenue, Cambridge CB3 0HE, U.K.\\
\and
Astrophysics Research Institute, Liverpool John Moores University, Twelve Quays House, Egerton Wharf, Birkenhead CH41 1LD, U.K.\\
\and
Astrophysics \& Cosmology Research Unit, School of Mathematics, Statistics \& Computer Science, University of KwaZulu-Natal, Westville Campus, Private Bag X54001, Durban 4000, South Africa\\
\and
CITA, University of Toronto, 60 St. George St., Toronto, ON M5S 3H8, Canada\\
\and
CNRS, IRAP, 9 Av. colonel Roche, BP 44346, F-31028 Toulouse cedex 4, France\\
\and
California Institute of Technology, Pasadena, California, U.S.A.\\
\and
Centre for Theoretical Cosmology, DAMTP, University of Cambridge, Wilberforce Road, Cambridge CB3 0WA U.K.\\
\and
Centro de Estudios de F\'{i}sica del Cosmos de Arag\'{o}n (CEFCA), Plaza San Juan, 1, planta 2, E-44001, Teruel, Spain\\
\and
Computational Cosmology Center, Lawrence Berkeley National Laboratory, Berkeley, California, U.S.A.\\
\and
Consejo Superior de Investigaciones Cient\'{\i}ficas (CSIC), Madrid, Spain\\
\and
DSM/Irfu/SPP, CEA-Saclay, F-91191 Gif-sur-Yvette Cedex, France\\
\and
DTU Space, National Space Institute, Technical University of Denmark, Elektrovej 327, DK-2800 Kgs. Lyngby, Denmark\\
\and
D\'{e}partement de Physique Th\'{e}orique, Universit\'{e} de Gen\`{e}ve, 24, Quai E. Ansermet,1211 Gen\`{e}ve 4, Switzerland\\
\and
Departamento de F\'{\i}sica Fundamental, Facultad de Ciencias, Universidad de Salamanca, 37008 Salamanca, Spain\\
\and
Departamento de F\'{\i}sica, Universidad de Oviedo, Avda. Calvo Sotelo s/n, Oviedo, Spain\\
\and
Department of Astronomy and Astrophysics, University of Toronto, 50 Saint George Street, Toronto, Ontario, Canada\\
\and
Department of Astrophysics/IMAPP, Radboud University Nijmegen, P.O. Box 9010, 6500 GL Nijmegen, The Netherlands\\
\and
Department of Electrical Engineering and Computer Sciences, University of California, Berkeley, California, U.S.A.\\
\and
Department of Physics \& Astronomy, University of British Columbia, 6224 Agricultural Road, Vancouver, British Columbia, Canada\\
\and
Department of Physics and Astronomy, Dana and David Dornsife College of Letter, Arts and Sciences, University of Southern California, Los Angeles, CA 90089, U.S.A.\\
\and
Department of Physics and Astronomy, University College London, London WC1E 6BT, U.K.\\
\and
Department of Physics, Gustaf H\"{a}llstr\"{o}min katu 2a, University of Helsinki, Helsinki, Finland\\
\and
Department of Physics, Nagoya University, Chikusa-ku, Nagoya, 464-8602, Japan\\
\and
Department of Physics, Princeton University, Princeton, New Jersey, U.S.A.\\
\and
Department of Physics, University of California, One Shields Avenue, Davis, California, U.S.A.\\
\and
Department of Physics, University of California, Santa Barbara, California, U.S.A.\\
\and
Department of Physics, University of Illinois at Urbana-Champaign, 1110 West Green Street, Urbana, Illinois, U.S.A.\\
\and
Dipartimento di Fisica e Astronomia G. Galilei, Universit\`{a} degli Studi di Padova, via Marzolo 8, 35131 Padova, Italy\\
\and
Dipartimento di Fisica e Scienze della Terra, Universit\`{a} di Ferrara, Via Saragat 1, 44122 Ferrara, Italy\\
\and
Dipartimento di Fisica, Universit\`{a} La Sapienza, P. le A. Moro 2, Roma, Italy\\
\and
Dipartimento di Fisica, Universit\`{a} degli Studi di Milano, Via Celoria, 16, Milano, Italy\\
\and
Dipartimento di Fisica, Universit\`{a} degli Studi di Trieste, via A. Valerio 2, Trieste, Italy\\
\and
Dipartimento di Fisica, Universit\`{a} di Roma Tor Vergata, Via della Ricerca Scientifica, 1, Roma, Italy\\
\and
Discovery Center, Niels Bohr Institute, Blegdamsvej 17, Copenhagen, Denmark\\
\and
Dpto. Astrof\'{i}sica, Universidad de La Laguna (ULL), E-38206 La Laguna, Tenerife, Spain\\
\and
European Space Agency, ESAC, Planck Science Office, Camino bajo del Castillo, s/n, Urbanizaci\'{o}n Villafranca del Castillo, Villanueva de la Ca\~{n}ada, Madrid, Spain\\
\and
European Space Agency, ESTEC, Keplerlaan 1, 2201 AZ Noordwijk, The Netherlands\\
\and
Finnish Centre for Astronomy with ESO (FINCA), University of Turku, V\"{a}is\"{a}l\"{a}ntie 20, FIN-21500, Piikki\"{o}, Finland\\
\and
Graduate School of Science and Engineering, Kagoshima University, 1-21-35 Korimoto, Kagoshima, Kagoshima 890-0065, Japan\\
\and
Helsinki Institute of Physics, Gustaf H\"{a}llstr\"{o}min katu 2, University of Helsinki, Helsinki, Finland\\
\and
INAF - Osservatorio Astrofisico di Catania, Via S. Sofia 78, Catania, Italy\\
\and
INAF - Osservatorio Astronomico di Padova, Vicolo dell'Osservatorio 5, Padova, Italy\\
\and
INAF - Osservatorio Astronomico di Roma, via di Frascati 33, Monte Porzio Catone, Italy\\
\and
INAF - Osservatorio Astronomico di Trieste, Via G.B. Tiepolo 11, Trieste, Italy\\
\and
INAF/IASF Bologna, Via Gobetti 101, Bologna, Italy\\
\and
INAF/IASF Milano, Via E. Bassini 15, Milano, Italy\\
\and
INFN, Sezione di Bologna, Via Irnerio 46, I-40126, Bologna, Italy\\
\and
INFN, Sezione di Roma 1, Universit\`{a} di Roma Sapienza, Piazzale Aldo Moro 2, 00185, Roma, Italy\\
\and
IPAG: Institut de Plan\'{e}tologie et d'Astrophysique de Grenoble, Universit\'{e} Joseph Fourier, Grenoble 1 / CNRS-INSU, UMR 5274, Grenoble, F-38041, France\\
\and
IUCAA, Post Bag 4, Ganeshkhind, Pune University Campus, Pune 411 007, India\\
\and
Imperial College London, Astrophysics group, Blackett Laboratory, Prince Consort Road, London, SW7 2AZ, U.K.\\
\and
Infrared Processing and Analysis Center, California Institute of Technology, Pasadena, CA 91125, U.S.A.\\
\and
Institut N\'{e}el, CNRS, Universit\'{e} Joseph Fourier Grenoble I, 25 rue des Martyrs, Grenoble, France\\
\and
Institut Universitaire de France, 103, bd Saint-Michel, 75005, Paris, France\\
\and
Institut d'Astrophysique Spatiale, CNRS (UMR8617) Universit\'{e} Paris-Sud 11, B\^{a}timent 121, Orsay, France\\
\and
Institut d'Astrophysique de Paris, CNRS (UMR7095), 98 bis Boulevard Arago, F-75014, Paris, France\\
\and
Institute for Space Sciences, Bucharest-Magurale, Romania\\
\and
Institute of Astronomy and Astrophysics, Academia Sinica, Taipei, Taiwan\\
\and
Institute of Astronomy, University of Cambridge, Madingley Road, Cambridge CB3 0HA, U.K.\\
\and
Institute of Astronomy, University of Tokyo, 2-21-1 Osawa, Mitaka, Tokyo, Japan\\
\and
Institute of Theoretical Astrophysics, University of Oslo, Blindern, Oslo, Norway\\
\and
Instituto de Astrof\'{\i}sica de Canarias, C/V\'{\i}a L\'{a}ctea s/n, La Laguna, Tenerife, Spain\\
\and
Instituto de F\'{\i}sica de Cantabria (CSIC-Universidad de Cantabria), Avda. de los Castros s/n, Santander, Spain\\
\and
Jet Propulsion Laboratory, California Institute of Technology, 4800 Oak Grove Drive, Pasadena, California, U.S.A.\\
\and
Jodrell Bank Centre for Astrophysics, Alan Turing Building, School of Physics and Astronomy, The University of Manchester, Oxford Road, Manchester, M13 9PL, U.K.\\
\and
Joint Astronomy Centre, 660 N. Aohoku Place, University Park, Hilo, Hawaii 96720, U.S.A.\\
\and
Kavli Institute for Cosmology Cambridge, Madingley Road, Cambridge, CB3 0HA, U.K.\\
\and
LAL, Universit\'{e} Paris-Sud, CNRS/IN2P3, Orsay, France\\
\and
LERMA, CNRS, Observatoire de Paris, 61 Avenue de l'Observatoire, Paris, France\\
\and
Laboratoire AIM, IRFU/Service d'Astrophysique - CEA/DSM - CNRS - Universit\'{e} Paris Diderot, B\^{a}t. 709, CEA-Saclay, F-91191 Gif-sur-Yvette Cedex, France\\
\and
Laboratoire Traitement et Communication de l'Information, CNRS (UMR 5141) and T\'{e}l\'{e}com ParisTech, 46 rue Barrault F-75634 Paris Cedex 13, France\\
\and
Laboratoire de Physique Subatomique et de Cosmologie, Universit\'{e} Joseph Fourier Grenoble I, CNRS/IN2P3, Institut National Polytechnique de Grenoble, 53 rue des Martyrs, 38026 Grenoble cedex, France\\
\and
Laboratoire de Physique Th\'{e}orique, Universit\'{e} Paris-Sud 11 \& CNRS, B\^{a}timent 210, 91405 Orsay, France\\
\and
Lawrence Berkeley National Laboratory, Berkeley, California, U.S.A.\\
\and
Max-Planck-Institut f\"{u}r Astrophysik, Karl-Schwarzschild-Str. 1, 85741 Garching, Germany\\
\and
McGill Physics, Ernest Rutherford Physics Building, McGill University, 3600 rue University, Montr\'{e}al, QC, H3A 2T8, Canada\\
\and
MilliLab, VTT Technical Research Centre of Finland, Tietotie 3, Espoo, Finland\\
\and
National Astronomical Observatory of Japan, 2-21-1 Osawa, Mitaka, Tokyo 181-8588, Japan\\
\and
Niels Bohr Institute, Blegdamsvej 17, Copenhagen, Denmark\\
\and
Nobeyama Radio Observatory, National Astronomical Observatory of Japan, 462-2 Nobeyama, Minamimaki, Minamisaku, Nagano 384-1305, Japan\\
\and
Observational Cosmology, Mail Stop 367-17, California Institute of Technology, Pasadena, CA, 91125, U.S.A.\\
\and
Optical Science Laboratory, University College London, Gower Street, London, U.K.\\
\and
SB-ITP-LPPC, EPFL, CH-1015, Lausanne, Switzerland\\
\and
SISSA, Astrophysics Sector, via Bonomea 265, 34136, Trieste, Italy\\
\and
School of Physics and Astronomy, Cardiff University, Queens Buildings, The Parade, Cardiff, CF24 3AA, U.K.\\
\and
Space Research Institute (IKI), Russian Academy of Sciences, Profsoyuznaya Str, 84/32, Moscow, 117997, Russia\\
\and
Space Sciences Laboratory, University of California, Berkeley, California, U.S.A.\\
\and
Special Astrophysical Observatory, Russian Academy of Sciences, Nizhnij Arkhyz, Zelenchukskiy region, Karachai-Cherkessian Republic, 369167, Russia\\
\and
Stanford University, Dept of Physics, Varian Physics Bldg, 382 Via Pueblo Mall, Stanford, California, U.S.A.\\
\and
Sub-Department of Astrophysics, University of Oxford, Keble Road, Oxford OX1 3RH, U.K.\\
\and
Theory Division, PH-TH, CERN, CH-1211, Geneva 23, Switzerland\\
\and
UPMC Univ Paris 06, UMR7095, 98 bis Boulevard Arago, F-75014, Paris, France\\
\and
Universit\'{e} de Toulouse, UPS-OMP, IRAP, F-31028 Toulouse cedex 4, France\\
% ...
}